\documentstyle[preprint,epsf,aps,floats,tighten]{revtex}
\input epsf
\newcommand{\ie}{\mbox{\rm i.e.}}
\newcommand{\eg}{\mbox{\rm e.g.}}
\newcommand{\ifm}[1]{\relax\ifmmode #1\else $#1$\fi}
\newcommand{\etal}{{\it et al.}}
\newcommand{\Dzero}{D\O}
\newcommand{\uaone}{UA1 Collaboration}
\newcommand{\uatwo}{UA2 Collaboration}
\newcommand{\cdf}{CDF Collaboration}
\newcommand{\dzero}{\Dzero\ Collaboration}
\newcommand{\alephc}{ALEPH Collaboration}
\newcommand{\opalc}{OPAL Collaboration}
\newcommand{\PRL}{Phys. Rev. Lett.}
\newcommand{\PL}{Phys. Lett.}
\newcommand{\PR}{Phys. Rev.}
\newcommand{\NP}{Nucl. Phys.}
\newcommand{\NIM}{Nucl. Instrum. Methods in Phys. Res.}
\newcommand{\ZP}{Z.~Phys.}
\newcommand{\GEAN}{{\sc geant}}

\newcommand{\HERW}{{\sc herwig}}
\newcommand{\PDFL}{{\sc pdflib}}

\newcommand{\abseta}{\ifm{|\eta|}}

\newcommand{\qt}{\ifm{q_T}}
\newcommand{\px}{\ifm{p_x}}
\newcommand{\py}{\ifm{p_y}}
\newcommand{\pz}{\ifm{p_z}}
\newcommand{\mt}{\ifm{m_T}}

\newcommand{\pt}{\ifm{p_T}}
\renewcommand{\pt}{\ifm{p_T}}

\newcommand{\mpt}{\mbox{$\rlap{\kern0.1em/}\pt$}}
\newcommand{\mptv}{\mbox{$\rlap{\kern0.1em/}\vec\pt$}}
\newcommand{\PM}{\ifm{\pm}}
\newcommand{\lt}{\ifm{<}}
\newcommand{\gt}{\ifm{>}}

\newcommand{\Eg}{\ifm{E(\gamma)}}

\newcommand{\ut}{\ifm{u_T}}
\newcommand{\utv}{\ifm{\vec\ut}}
\newcommand{\upar}{\ifm{u_{\parallel}}}
\newcommand{\uper}{\ifm{u_{\perp}}}

\newcommand{\wb}{\ifm{W}}

\newcommand{\wev}{\ifm{W\to e\nu}}
\newcommand{\wegam}{\ifm{\wev\gamma}}

\newcommand{\wte}{\ifm{W\to \tau\nu\to e\nu\overline\nu\nu}}

\newcommand{\mw}{\ifm{M_{W}}}
\newcommand{\wwidth}{\ifm{\Gamma_{W}}}
\newcommand{\ptw}{\ifm{\pt(W)}}

\newcommand{\zb}{\ifm{Z}}
\newcommand{\zee}{\ifm{Z\to ee}}
\newcommand{\zeegam}{\ifm{\zee\gamma}}

\newcommand{\mz}{\ifm{M_{Z}}}

\newcommand{\ptee}{\ifm{\pt(ee)}}

\newcommand{\pee}{\ifm{p(ee)}}
\newcommand{\peev}{\ifm{\vec\pee}}
\newcommand{\mee}{\ifm{m(ee)}}

\newcommand{\tev}{\ifm{\tau\to e\nu\overline\nu}}
\newcommand{\qqbar}{\ifm{q\overline{q}}}

\newcommand{\ee}{\ifm{e^+e^-}}
\newcommand{\ppbar}{\ifm{p\overline{p}}}

\newcommand{\GeV}{\mbox{GeV}}

\newcommand{\Ee}{\ifm{E(e)}}

\newcommand{\pe}{\ifm{p(e)}}
\newcommand{\pev}{\ifm{\vec\pe}}
\newcommand{\pte}{\ifm{\pt(e)}}
\newcommand{\ptev}{\ifm{\vec\pte}}
\newcommand{\peonev}{\ifm{\vec p(e_1)}}
\newcommand{\petwov}{\ifm{\vec p(e_2)}}
\newcommand{\phie}{\ifm{\phi(e)}}
\newcommand{\te}{\ifm{\theta(e)}}
\newcommand{\etae}{\ifm{\eta(e)}}

\newcommand{\rcal}{\ifm{r_{\rm cal}}}

\newcommand{\rtrk}{\ifm{r_{\rm trk}}}
\newcommand{\xcal}{\ifm{x_{\rm cal}}}
\newcommand{\ycal}{\ifm{y_{\rm cal}}}
\newcommand{\zcal}{\ifm{z_{\rm cal}}}
\newcommand{\xtrk}{\ifm{x_{\rm trk}}}
\newcommand{\ytrk}{\ifm{y_{\rm trk}}}
\newcommand{\ztrk}{\ifm{z_{\rm trk}}}
\newcommand{\xvtx}{\ifm{x_{\rm vtx}}}
\newcommand{\yvtx}{\ifm{y_{\rm vtx}}}
\newcommand{\zvtx}{\ifm{z_{\rm vtx}}}
\newcommand{\fiso}{\ifm{f_{\rm iso}}}

\newcommand{\sigm}{\ifm{\sigma_{\rm trk}}}

\newcommand{\pnuz}{\ifm{\pz(\nu)}}

\newcommand{\ptnu}{\ifm{\pt(\nu)}}
\newcommand{\ptnuv}{\ifm{\vec\ptnu}}
\newcommand{\phinu}{\ifm{\phi(\nu)}}

\newcommand{\rrec}{\ifm{{\rm R_{rec}}}}
\newcommand{\sigrec}{\ifm{\sigma_{\rm rec}}}
\newcommand{\dupar}{\ifm{\Delta \upar}}
\newcommand{\mdupar}{\ifm{\langle \dupar/\delta \eta \delta \phi \rangle}}
\newcommand{\alphaccem}{\ifm{\alpha_{\rm CC}}}
\newcommand{\alphaecem}{\ifm{\alpha_{\rm EC}}}
\newcommand{\deltaccem}{\ifm{\delta_{\rm CC}}}
\newcommand{\deltaecem}{\ifm{\delta_{\rm EC}}}
\newcommand{\alphamb}{\ifm{\alpha_{\rm mb}}}
\newcommand{\srec}{\ifm{s_{\rm rec}}}
\newcommand{\alpharec}{\ifm{\alpha_{\rm rec}}}
\newcommand{\betarec}{\ifm{\beta_{\rm rec}}}

\newcommand{\regam}{\ifm{\Delta R(e\gamma)}}
\newcommand{\rzero}{\ifm{R_0}}

\newcommand{\betacdc}{\ifm{\beta_{\rm CDC}}}

\newcommand{\betaec}{\ifm{\beta_{\rm EC}}}
\newcommand{\betafdc}{\ifm{\beta_{\rm FDC}}}

\newcommand{\cem}{\ifm{c_{\rm EM}}}
\newcommand{\cccem}{\ifm{c_{\rm CC}}}
\newcommand{\cecem}{\ifm{c_{\rm EC}}}
\newcommand{\sem}{\ifm{s_{\rm EM}}}
\newcommand{\nem}{\ifm{n_{\rm EM}}}

\begin{document}
\pagestyle{myheadings}
\onecolumn

\title{ A measurement of the \wb\ boson mass using large rapidity electrons}

%
\author{                                                                      
B.~Abbott,$^{45}$                                                             
M.~Abolins,$^{42}$                                                            
V.~Abramov,$^{18}$                                                            
B.S.~Acharya,$^{11}$                                                          
I.~Adam,$^{44}$                                                               
D.L.~Adams,$^{54}$                                                            
M.~Adams,$^{28}$                                                              
S.~Ahn,$^{27}$                                                                
V.~Akimov,$^{16}$                                                             
G.A.~Alves,$^{2}$
% ...
\\                                                                            
\vskip 0.30cm     
\centerline{(D\O\ Collaboration)}                    
\vskip 0.30cm     
}                                                                             
\address{                                                                     
\centerline{$^{1}$Universidad de Buenos Aires, Buenos Aires, Argentina}       
\centerline{$^{2}$LAFEX, Centro Brasileiro de Pesquisas F{\'\i}sicas,         
                  Rio de Janeiro, Brazil}                                     
\centerline{$^{3}$Universidade do Estado do Rio de Janeiro,                   
                  Rio de Janeiro, Brazil}                                     
\centerline{$^{4}$Institute of High Energy Physics, Beijing,                  
                  People's Republic of China}                                 
\centerline{$^{5}$Universidad de los Andes, Bogot\'{a}, Colombia}             
\centerline{$^{6}$Universidad San Francisco de Quito, Quito, Ecuador}         
\centerline{$^{7}$Institut des Sciences Nucl\'eaires, IN2P3-CNRS,             
                  Universite de Grenoble 1, Grenoble, France}                 
\centerline{$^{8}$DAPNIA/Service de Physique des Particules, CEA, Saclay,     
                  France}                                                     
\centerline{$^{9}$Panjab University, Chandigarh, India}                       
\centerline{$^{10}$Delhi University, Delhi, India}                            
\centerline{$^{11}$Tata Institute of Fundamental Research, Mumbai, India}     
\centerline{$^{12}$Kyungsung University, Pusan, Korea}                        
\centerline{$^{13}$Seoul National University, Seoul, Korea}                   
\centerline{$^{14}$CINVESTAV, Mexico City, Mexico}                            
\centerline{$^{15}$Institute of Nuclear Physics, Krak\'ow, Poland}            
\centerline{$^{16}$Institute for Theoretical and Experimental Physics,        
                   Moscow, Russia}                                            
\centerline{$^{17}$Moscow State University, Moscow, Russia}                   
\centerline{$^{18}$Institute for High Energy Physics, Protvino, Russia}       
\centerline{$^{19}$Lancaster University, Lancaster, United Kingdom}           
\centerline{$^{20}$University of Arizona, Tucson, Arizona 85721}              
\centerline{$^{21}$Lawrence Berkeley National Laboratory and University of    
                   California, Berkeley, California 94720}                    
\centerline{$^{22}$University of California, Davis, California 95616}         
\centerline{$^{23}$University of California, Irvine, California 92697}        
\centerline{$^{24}$University of California, Riverside, California 92521}     
\centerline{$^{25}$Florida State University, Tallahassee, Florida 32306}      
\centerline{$^{26}$University of Hawaii, Honolulu, Hawaii 96822}              
\centerline{$^{27}$Fermi National Accelerator Laboratory, Batavia,            
                   Illinois 60510}                                            
\centerline{$^{28}$University of Illinois at Chicago, Chicago,                
                   Illinois 60607}                                            
\centerline{$^{29}$Northern Illinois University, DeKalb, Illinois 60115}      
\centerline{$^{30}$Northwestern University, Evanston, Illinois 60208}         
\centerline{$^{31}$Indiana University, Bloomington, Indiana 47405}            
\centerline{$^{32}$University of Notre Dame, Notre Dame, Indiana 46556}       
\centerline{$^{33}$Purdue University, West Lafayette, Indiana 47907}          
\centerline{$^{34}$Iowa State University, Ames, Iowa 50011}                   
\centerline{$^{35}$University of Kansas, Lawrence, Kansas 66045}              
\centerline{$^{36}$Kansas State University, Manhattan, Kansas 66506}          
\centerline{$^{37}$Louisiana Tech University, Ruston, Louisiana 71272}        
\centerline{$^{38}$University of Maryland, College Park, Maryland 20742}      
\centerline{$^{39}$Boston University, Boston, Massachusetts 02215}            
\centerline{$^{40}$Northeastern University, Boston, Massachusetts 02115}      
\centerline{$^{41}$University of Michigan, Ann Arbor, Michigan 48109}         
\centerline{$^{42}$Michigan State University, East Lansing, Michigan 48824}   
\centerline{$^{43}$University of Nebraska, Lincoln, Nebraska 68588}           
\centerline{$^{44}$Columbia University, New York, New York 10027}             
\centerline{$^{45}$New York University, New York, New York 10003}             
\centerline{$^{46}$University of Rochester, Rochester, New York 14627}        
\centerline{$^{47}$State University of New York, Stony Brook,                 
                   New York 11794}                                            
\centerline{$^{48}$Brookhaven National Laboratory, Upton, New York 11973}     
\centerline{$^{49}$Langston University, Langston, Oklahoma 73050}             
\centerline{$^{50}$University of Oklahoma, Norman, Oklahoma 73019}            
\centerline{$^{51}$Brown University, Providence, Rhode Island 02912}          
\centerline{$^{52}$University of Texas, Arlington, Texas 76019}               
\centerline{$^{53}$Texas A\&M University, College Station, Texas 77843}       
\centerline{$^{54}$Rice University, Houston, Texas 77005}                     
}                                                                             

\maketitle

\vskip 0.30cm     
\centerline{\today } 

\begin{abstract}
We present a measurement of the \wb\ boson mass using data collected by the
\Dzero\ experiment at the Fermilab Tevatron during 1994--1995. We identify \wb\
bosons by their decays to $e\nu$ final states where the electron is detected
 in a forward calorimeter. We extract the \wb\ boson mass, \mw, by
fitting the transverse mass and transverse electron and neutrino
 momentum spectra
from a sample of 11{,}089 \wev\ decay candidates. We use a sample of
1{,}687 dielectron events, mostly due to \zee\ decays,
 to constrain our model of
the detector response.  Using the forward calorimeter data,
  we measure $\mw = 80.691
\pm0.227$~GeV. Combining the forward calorimeter measurements 
  with our previously published central calorimeter 
results, 
 we obtain $\mw = 80.482\pm0.091$~GeV.
\end{abstract}

\pacs{ PACS numbers: 14.70.Fm, 12.15.Ji, 13.38.Be, 13.85.Qk }


\section{ Introduction }
\label{sec-intro}
In this article we describe the first measurement \cite{ecmwprl} 
of the mass of
the \wb\ boson using electrons detected at large rapidities (i.e. between
 1.5 and 2.5). 
 We use data collected in 1994--1995 with the \Dzero\ detector \cite{d0nim} at
the Fermilab Tevatron \ppbar\ collider.
 This measurement performed with the D\O\ forward calorimeters \cite{d0ec} 
 complements our previous
 measurements with central electrons \cite{wmass1bcc,wmass1acc}
  and the more complete combined 
 rapidity coverage gives useful constraints on model parameters that permit
 reduction of the systematic error, in addition to increasing the
  statistical
 precision. 

The study of the properties of the \wb\ boson began in  1983 with its discovery
by the UA1 \cite{UA1_W_discovery} and UA2 \cite{UA2_W_discovery} collaborations
at the CERN \ppbar\ collider. Together with the discovery of the \zb\ boson in
the same year \cite{UA1_Z_discovery,UA2_Z_discovery}, it provided a direct
confirmation of the unified description
  of the weak and electromagnetic interactions
\cite{SM}, which --- together with the theory of the strong interaction, 
 quantum chromodynamics (QCD) --- now constitutes the standard model.

Since the \wb\ and \zb\ bosons are carriers of the weak force, their properties
are intimately coupled to the structure of the model. The properties of the
\zb\ boson have been studied in great detail in \ee\ collisions \cite{mz}. 
The study of the \wb\ boson 
has proven to be significantly more difficult, since
it is charged and so cannot be resonantly produced in \ee\ collisions.
Until recently its direct study has therefore been the realm of experiments at 
\ppbar\ colliders \cite{wmass1bcc,wmass1acc,UA2,CDF}. 
 Direct measurements of the \wb\
boson mass have also been carried out at LEP2 \cite{L3,ALEPH,OPAL,DELPHI} using
nonresonant \wb\ pair production. A summary of these
measurements can be found in Table~\ref{tab:mw} at the end of this article.

\begin{figure}[htb]
\vspace{-0.5cm}
\epsfxsize=3.0in
\centerline{\epsfbox{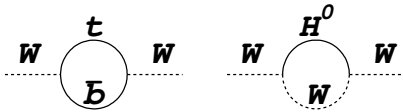}}
\vspace{-1cm}
\caption{ Loop diagrams contributing to the \wb\ boson mass.}
\label{fig:loop}
\end{figure}

The standard model links the \wb\ boson mass to other parameters,
\begin{equation}
\mw^2 = \left({\pi\alpha(\mz^2)\over \sqrt{2} G_F}\right)
{\mz^2\over(\mz^2-\mw^2)({1-\Delta r_{EW}})} \quad 
\label{eq:mw1}
\end{equation}
in the ``on shell'' scheme \cite{onshell}. Aside from the radiative corrections
$\Delta r_{EW}$, the \wb\ boson mass is thus
determined by three precisely measured quantities, the mass of the \zb\ boson
\mz \cite{mz}, the Fermi constant $G_F$ \cite{PDG},
and the electromagnetic coupling constant $\alpha$ evaluated at $Q^2=\mz^2$
\cite{PDG}:
\begin {eqnarray}
\mz & = & 91.1867\pm0.0021\ \hbox{GeV}, \label{eq:mz} \\
G_F & = & (1.16639\pm0.00001)\times10^{-5}\ \hbox{GeV}^{-2} \quad , \\
\alpha & = & (128.88\pm0.09)^{-1} \quad .
\end{eqnarray}
From the measured \wb\ boson mass, we can derive the size of
the radiative corrections $\Delta r_{EW}$. Within
the framework of the standard model, these corrections are dominated
by loops involving the top quark and the Higgs boson (see
Fig.~\ref{fig:loop}). 
The correction from the $t\overline b$ loop is substantial
because of the large mass difference between the two quarks. It is proportional
to $m_t^2$ for large values of the top quark mass $m_t$. Since $m_t$ has
been measured \cite{mtop_lj,cdf_mt}, this
contribution can be calculated within the standard model. For a large Higgs
boson mass, $m_H$, the correction from the Higgs loop is proportional to 
 $\ln(m_H)$. In extensions to the standard model, new particles may give rise
to additional corrections to the value of \mw. In the minimal supersymmetric
extension of the standard model (MSSM), for example,
additional corrections can increase the predicted
\wb\ mass by up to 250~MeV~\cite{susy}.

A measurement of the \wb\ boson mass therefore constitutes a test of
the standard model.
In conjunction with a measurement of the top quark mass,
the standard model predicts \mw\ up to a 200~MeV uncertainty due to the
unknown Higgs boson mass. 
By comparing the standard model calculation
 to the measured value of the \wb\ boson mass,
we can constrain the mass of the Higgs boson,
the agent of the electroweak symmetry breaking in the standard model
 that has up to now eluded
experimental detection.
A discrepancy with the range allowed by the standard model could indicate new
physics.
The experimental challenge is thus to measure the \wb\
boson mass to sufficient precision, about 0.1\%, to be sensitive to these
corrections.

\section{Overview}
\label{sec-overview}
\subsection {Conventions}
We use a Cartesian coordinate system with the $z$-axis defined by
the direction of the proton beam, the $x$-axis pointing radially out of the
Tevatron ring, and the $y$-axis pointing up. A vector $\vec p$ is then defined
in terms of its projections on these three axes, \px, \py, \pz.  Since
protons and antiprotons in the Tevatron are unpolarized,
all physical processes are invariant
with respect to rotations around the beam direction.
It is therefore
convenient to use a cylindrical coordinate system, in which the same
vector is given by the magnitude of its component transverse to the
beam direction, \pt, its azimuth $\phi$, and \pz. In \ppbar\ collisions,
the center-of-mass frame of the parton-parton collisions is approximately at
rest in the plane transverse to the beam direction but has an undetermined
motion along the beam direction. Therefore the plane transverse to the beam
direction is of special importance, and sometimes we work with two-dimensional
vectors defined in the $x$-$y$ plane. They are written with a subscript
$T$, \eg\ $\vec \pt$. We also use spherical coordinates by replacing
\pz\ with the polar angle
$\theta$ (as measured between \pz\ and the $z$-axis)
or the pseudorapidity $\eta=-\ln\tan\left(\theta/2\right)$. The origin
of the coordinate system is in general the reconstructed 
 position of the \ppbar\
interaction when describing the interaction, and the geometrical
center of the detector when describing the detector.
For convenience, we use units in which $c=\hbar=1$.

\subsection {\boldmath \wb\ \unboldmath and \boldmath \zb\ \unboldmath 
 Boson Production and Decay}

\begin{figure}[ht]
\vspace{ -2.9cm}
\epsfxsize=3.0in
\centerline{\epsfbox{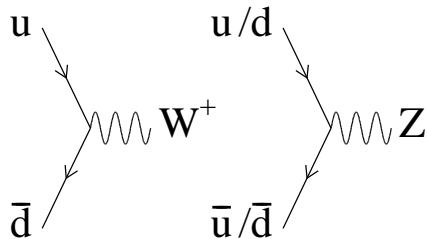}}
\vspace{-.8cm }
\caption{ Lowest order diagrams for \wb\ and \zb\ boson production.}
\label{fig:wzproduction}
\end{figure}

In \ppbar\ collisions at $\sqrt{s}=1.8$~TeV, \wb\ and \zb\ bosons are produced
predominantly through quark-antiquark annihilation. Figure
\ref{fig:wzproduction} shows the lowest-order diagrams. The quarks in the
initial state may radiate gluons which are usually
very soft but may sometimes be energetic enough to give rise to hadron jets in
the detector. In the reaction, the initial proton and antiproton
break up and the fragments hadronize. We refer to everything except the
vector boson and its decay products collectively as the underlying event. Since
the initial proton and antiproton momentum vectors add to zero, the same must be
true for the vector sum of all final state momenta and therefore the vector
boson recoils against all particles in the underlying event. The sum of the
transverse momenta of the recoiling particles must balance the transverse
momentum of the boson, which is typically small compared to its mass
but has a long tail to large values.

We identify \wb\ and
\zb\ bosons by their leptonic decays. The \Dzero\ detector
(Sec.~\ref{sec-exper})
is best suited for a precision measurement of electrons and
positrons\footnote{In the following we use ``electron'' generically for
both electrons and positrons.}, and we therefore use the decay channel
\wev\ to measure the \wb\ boson mass. \zee\ decays serve as an important
calibration sample. About 11\% of the \wb\ bosons decay to $e\nu$ and about
3.3\% of the \zb\ bosons decay to $ee$. The leptons typically have
transverse momenta  of about half the mass of the decaying boson and are
well isolated from other large energy deposits in the calorimeter. Gauge 
vector boson
decays are the dominant source of isolated high-\pt\ leptons at the Tevatron,
and therefore these decays allow us to select clean samples of \wb\ and \zb\
boson decays.

\subsection {Event Characteristics}

In events due to the process $\ppbar\to(\wev)+X$, where $X$ stands for the
underlying event, we detect the electron and all
particles recoiling against the \wb\ boson 
 with pseudorapidity $-4<\eta<4$. The
neutrino escapes undetected.
In the calorimeter we cannot
resolve individual recoil particles, but we measure their energies summed over
detector segments. Recoil particles with $|\eta| \gt 4$ escape unmeasured
through
the beampipe, possibly carrying away substantial momentum along the beam
direction. This means that we cannot measure the sum of the $z$-components of
the recoil momenta, $u_z$, precisely. Since these particles escape at a very
small angle
with respect to the beam, their transverse momenta are typically small and  neglecting
them in the sum of the transverse recoil momenta, \utv\ causes a small amount
 of smearing of \utv. We measure \utv\
by summing the observed energy flow vectorially over all detector segments.
Thus, we reduce the reconstruction of every candidate event to a measurement of
the electron momentum \pev\ and \utv.

Since the neutrino escapes undetected, the sum of all measured final state
transverse momenta does not add to zero. The missing transverse momentum \mptv,
required to balance the transverse momentum sum, is a measure of the transverse
momentum of the neutrino.
The neutrino momentum component along the beam direction cannot be determined,
because $u_z$ is not measured well.
The signature of a \wev\ decay is therefore an isolated
high-\pt\ electron and large missing transverse momentum.

In the case of \zee\ decays, the signature consists of two isolated high-\pt\
electrons and we measure the momenta of both leptons, \peonev\ and \petwov, and
\utv\ in the detector.

\subsection {Mass Measurement Strategy}

Since \pnuz\ is unknown, we cannot reconstruct the $e\nu$ invariant mass for
\wev\ candidate events and therefore must resort to other kinematic variables
for the mass measurement.

For recent measurements \cite{UA2,CDF,wmass1acc,wmass1bcc} the transverse mass,
\begin{eqnarray}
\mt = \sqrt{2 \pte \ptnu \left(1-\cos\left(\phie-\phinu\right)\right)} \quad ,
\label{eq:mtdefn}
\end{eqnarray}
was used. This variable has the advantage that its spectrum is relatively
insensitive to the production dynamics of the \wb\ boson. Corrections to \mt\
due to the motion
of the \wb\ are of order $\left(\qt/\mw\right)^2$,
where \qt\ is the transverse momentum of the \wb\ boson. It is also insensitive
to selection biases that prefer certain event topologies
(Sec.~\ref{sec-elec-upar}). However, it makes use of
the inferred neutrino \pt\ and is therefore sensitive to the response of the
detector to the recoil particles.

The electron \pt\ spectrum provides an alternative measurement of 
 the \wb\ mass.
It is measured with better resolution than the neutrino \pt\ and is insensitive
to the recoil momentum measurement. However, its shape is sensitive to the
motion
of the \wb\ boson
 and receives corrections of order $\qt/\mw$. It thus requires a
better understanding of the \wb\ boson production dynamics than the \mt\
spectrum does.

\begin{figure}[ht]
\vskip -0cm
\epsfxsize=3.0in
\centerline{\epsfbox{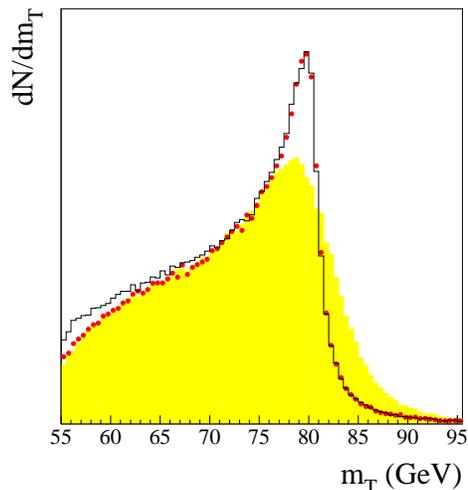}}
\caption{The \mt\ spectrum for \wb\ bosons with $q_T=0$ (------), with the
correct
  $q_T$ distribution ($\bullet$), and with detector resolutions (shaded).}
\label{fig:sensitivities1}
\end{figure}

\begin{figure}[ht]
\vskip -0cm
\epsfxsize=3.0in
\centerline{\epsfbox{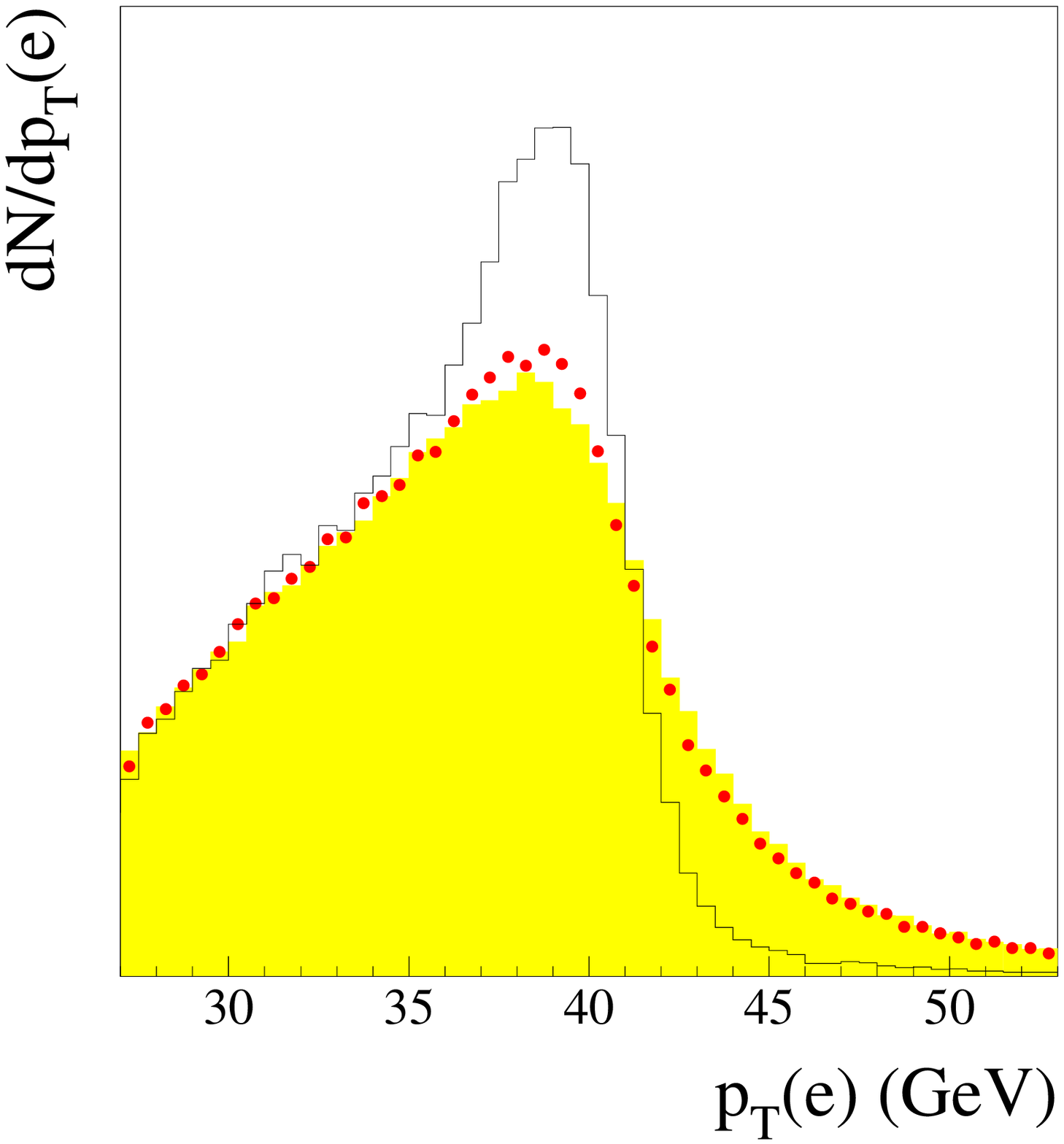}}
\caption{The \pte\ spectrum for \wb\ bosons with $q_T=0$ (------), with the
correct $q_T$ distribution
  ($\bullet$), and with detector resolutions (shaded).}
\label{fig:sensitivities2}
\end{figure}

These effects are  illustrated in Figs.~\ref{fig:sensitivities1} and
\ref{fig:sensitivities2}, which show the effect of the motion of the
\wb\ bosons and the detector resolutions on the shapes of the
 \mt\ and \pte\ spectra.
The solid line shows the shape of the distribution before the detector
simulation and with \qt=0. The points show the shape after \qt\ is added to
the system, and the shaded histogram also includes the detector simulation.
We observe that the shape of the \mt\ spectrum is dominated by detector
resolutions and the
shape of the \pte\ spectrum by the motion of the \wb\ boson. 

The shape of the neutrino \pt\ spectrum is sensitive to both the \wb\ boson
 production dynamics and the recoil momentum measurement. 
By performing the
measurement using all three spectra, we provide a powerful cross check with
complementary systematics.

All three spectra are equally sensitive to the electron energy response of the
detector. We calibrate this response by forcing the observed dielectron mass
peak in the \zee\ sample to agree with the known \zb\ mass\cite{mz}
(Sec.~\ref{sec-elec}). This means that we effectively measure the ratio
of \wb\ and \zb\ masses, which is equivalent to a measurement of the \wb\ mass
because the \zb\ mass is known precisely.

To carry out these measurements, we perform a maximum likelihood fit to the
spectra. Since the shape of the spectra, including all
the experimental effects, cannot be computed analytically, we
 need a Monte Carlo
simulation program that can predict the shape of the spectra as a function of
the \wb\ mass.  To measure the \wb\ mass to a
precision of order 100~MeV, we wish to estimate individual systematic effects
with a statistical error of
  5~MeV. Our technique requires a Monte Carlo sample of
10 million accepted \wb\ bosons for each such effect. The program
therefore must be capable of generating large event
 samples in a reasonable time. We obtain the required Monte Carlo statistics
  by employing a parameterized model of the
detector response.

We next summarize the aspects of the accelerator and detector
that are important for our measurement (Sec.~\ref{sec-exper}). Then we
describe the data selection (Sec.~\ref{sec-data}) and
the fast Monte Carlo model (Sec.~\ref{sec-mc}). Most parameters in
the model are determined from our data. We describe the determination of
the various components of the Monte Carlo model in
Secs.~\ref{sec-elec}-\ref{sec-back}.
After tuning the model, we fit the
kinematic spectra (Sec.~\ref{sec-fit}),
perform some consistency checks (Sec.~\ref{sec-checks}),
and discuss the systematic uncertainties (Sec.~\ref{sec-syst}).
We present the error analysis in Sec.~\ref{erroranalysis}, and
  summarize the results and present
the conclusions in Sec.~\ref{sec-results}.

\section {Experimental Method}
\label{sec-exper}
\subsection{ Accelerator }
During the data run, 
the Fermilab Tevatron\cite{Tevatron} collided proton and antiproton
beams at a center-of-mass energy of $\sqrt{s}=1.8$~TeV. Six bunches each
of protons and antiprotons circulated around the ring in opposite directions.
Bunches crossed at the intersection regions every 3.5 $\mu$s. During the
1994--1995 running period, the accelerator reached a peak luminosity of
$2.5\times10^{31} \hbox{cm}^{-2} \hbox{s}^{-1}$ and
delivered an integrated luminosity of
about 100 pb$^{-1}$. The beam interaction region at D\O\ 
 was at the center of the 
 detector with an r.m.s. length of 27 cm. 

The Tevatron tunnel also housed a 150~GeV proton synchrotron, called the Main
Ring,  used as an injector for the Tevatron and 
 accelerated protons for antiproton production during collider operation.
Since the Main Ring beampipe passed through the outer section of the \Dzero\
calorimeter, passing proton bunches gave rise to backgrounds in the detector.
We eliminated this background using timing cuts based on the accelerator clock
signal.

\subsection{ Detector }
\label{sec-detect}

\subsubsection{Overview}
The \Dzero\ detector consists of three major subsystems: an inner tracking
detector, a calorimeter, and a muon spectrometer.
It is described in
detail in Ref.~\cite{d0nim}. We describe only the features that are most
important for this measurement.

\subsubsection{Inner Tracking Detector}
\begin{figure}[htpb!]
\vspace{0.5in}
\epsfxsize=3.0in
\centerline{\epsfbox{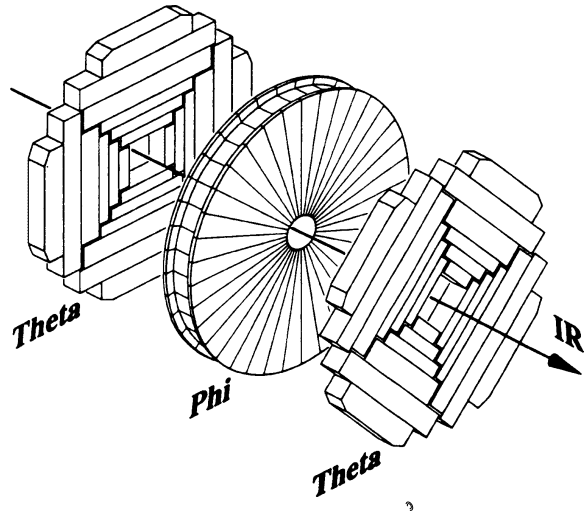}}
\vspace{0.02in}
\caption{ An exploded view of a \Dzero\ forward drift chamber (FDC).}
\label{fig:d0fdc}
\end{figure}

The inner tracking detector is designed to measure the trajectories of charged
particles.
It consists of a vertex drift chamber, a transition
radiation detector, a central drift chamber (CDC), and two forward drift
chambers (FDC). There is no central magnetic field.
The CDC covers the region $|\eta|<1.0$.
The FDC covers the region $1.4<\abseta<3.0$. Each FDC consists of
 three separate chambers: a $\Phi$ module, with radial wires which measures the
 $\phi$ coordinate, sandwiched between a pair of $\Theta$ modules which measure
 (approximately) the radial coordinate. Figure~\ref{fig:d0fdc} shows one of
 the two FDC detectors. 

\subsubsection{Calorimeter}
\label{sec-exper-cal}

\begin{figure}[htpb!]
\vspace{0.5in}
\epsfxsize=4.5in
\centerline{\epsfbox{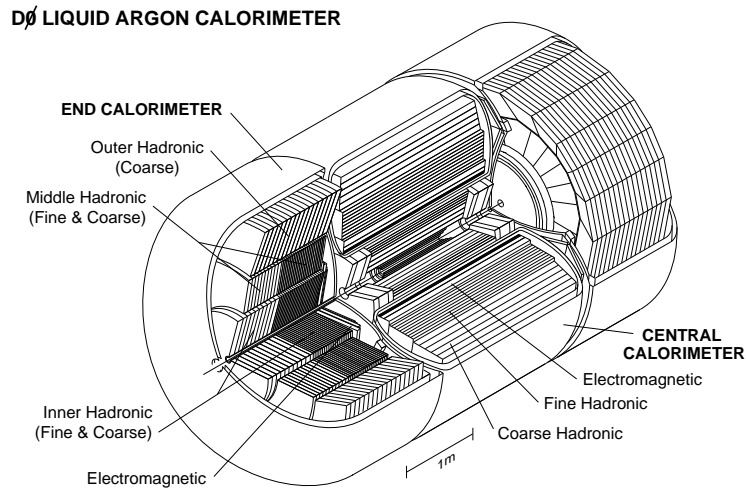}}
\vspace{-0.02in}
\caption{ A cutaway view of the \Dzero\ calorimeter and tracking system.}
\label{fig:d0cal}
\end{figure}

The uranium/liquid-argon sampling calorimeter (Fig.~\ref{fig:d0cal}) 
 is the most important part of the
detector for this measurement.
 There are three calorimeters: a central calorimeter
(CC) and two end calorimeters (EC), each housed in its own cryostat.
Each is segmented into an electromagnetic (EM) section, a fine
hadronic (FH) section, and a coarse hadronic (CH) section, with increasingly
coarser sampling. 

\begin{figure}[htpb!]
\vspace{0.5in}
\epsfxsize=3.0in
\centerline{\epsfbox{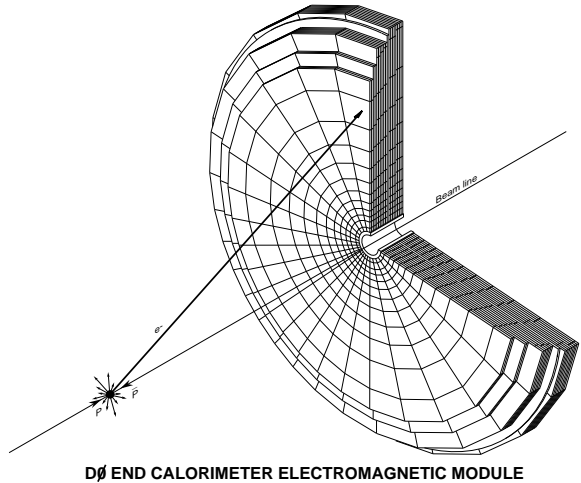}}
\vspace{-0.02in}
\caption{The ECEM section of an end calorimeter.}
\label{ecemcal}
\end{figure}

The ECEM section (Fig.~\ref{ecemcal}) has a monolithic construction of 
 alternating uranium plates, liquid-argon gaps, and multilayer
 printed-circuit readout 
 boards.
Each end calorimeter is divided into about 1000 pseudo-projective
towers, each covering 0.1$\times$0.1 in $\eta\times\phi$. The EM section
is segmented into four layers, 0.3, 2.6, 7.9, and 9.3 radiation lengths
thick. The third layer, in which electromagnetic showers typically reach their
maximum, is transversely segmented into cells covering 0.05$\times$0.05 in
$\eta\times\phi$. The EC hadronic section is segmented into
five layers. The entire calorimeter is 7--9 nuclear interaction lengths
thick. There are no projective cracks in the calorimeter and it provides
hermetic and almost uniform coverage for particles with $\abseta<4$.

The signals from arrays of 2$\times$2 calorimeter towers covering
0.2$\times$0.2 in $\eta\times\phi$ are added together
electronically for the EM section alone and for the EM and hadronic
 sections together, and shaped with a
fast rise time for use in the Level 1 trigger. We refer to these arrays of
2$\times$2 calorimeter towers as ``trigger towers."

The liquid argon has unit gain and the end calorimeter response was
 extremely stable during the entire run. The liquid-argon response 
 was monitored with radioactive sources of
$\alpha$ and $\beta$ particles throughout the run, as were the gains and
pedestals of all readout channels. Details can be found in Ref.~\cite{calmon}.
  
The ECEM calorimeter provides a measurement of energy and position of the
electrons from the \wb\ and \zb\ boson
 decays. Due to the fine segmentation of the
third layer, we can measure the position of the shower centroid with a
precision of about 1 mm in the azimuthal and radial directions.

\begin{figure}[htpb!]
\vspace{0.1in}
\epsfxsize=3.0in
\centerline{\epsfbox{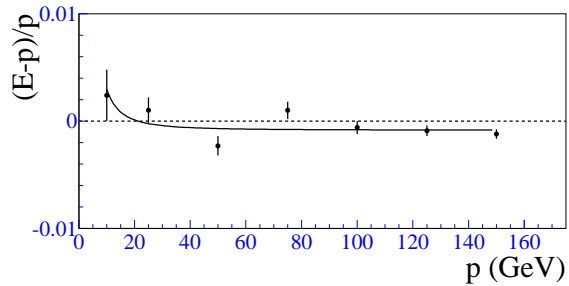}}
\vspace{0.1in}
\caption{ The fractional deviation of the reconstructed electron energy
from the beam momentum from beam tests of an ECEM module.}
\label{fig:tbresponse}
\end{figure}

We have
 studied the response of the ECEM calorimeter to electrons in beam tests
\cite{d0ec,testbeam}. To reconstruct the
electron energy we add the signals $a_i$ observed in each EM layer
($i=1, \ldots ,4$) and the first FH layer ($i=5$) of an array of  5$\times$5
calorimeter towers, centered on the most energetic tower,
weighted by a layer-dependent sampling weight $s_i$,
\begin{equation}
E = A \sum_{i=1}^5 s_i a_i - \deltaecem \quad .
\label{eq:emresponse}
\end{equation}
To determine the sampling weights we minimize
\begin{equation}
\chi^2 = \sum {(p-E)^2 \over \sigma_{\rm EM}^2} \quad ,
\end{equation}
where the sum runs 
over all events, $\sigma_{\rm EM}$ is the resolution given
in Eq.~\ref{eq:emresolution} and $p$ is the beam momentum.
We obtain $A=3.74$~MeV/ADC count, $\deltaecem =-300$~MeV, 
$s_1=1.47$, $s_2=1.00$,
$s_4=1.10$, and $s_5=1.67$. We arbitrarily fix $s_3=1$. The value of
 $\deltaecem$
depends on the amount of uninstrumented
  material in front of the calorimeter.  The
parameters $s_1$ to $s_4$ weight
the four EM layers and $s_5$ the first FH layer.
Figure~\ref{fig:tbresponse} shows the fractional deviation of $E$ as a
function of the beam momentum $p$. Above 20~GeV the non-linearity
 is less than 0.1\%.

The fractional energy resolution can
be parameterized as a function of electron energy using constant, sampling, and
noise terms as
\begin{equation}
\left({\sigma_{\rm EM}\over E}\right)^2 = \cem^2 +
\left({\sem \over\sqrt{E}}\right)^2 + \left({\nem \over E}\right)^2,
\label{eq:emresolution}
\end{equation}
with $\cem =0.003$, $\sem =0.157$~GeV$^{1/2}$
 and $\nem =0.29$~GeV in the end 
calorimeters, as measured in beam tests \cite{d0ec,testbeam}.

\subsubsection {Muon Spectrometer}
The D\O\ muon spectrometer consists of five separate solid-iron toroidal
 magnets, together with sets of proportional drift tube chambers to measure
 the track coordinates. The central toroid covers the region 
$\mid \! \eta  \! \mid \leq 1$, 
two end toroids cover $1 < \mid \! \eta  \! \mid \leq 2.5$, 
and the small-angle muon system covers $2.5 < \mid \! \eta  \! \mid \leq 3.6$.
There is one layer of chambers inside the toroids and two layers outside
 for detecting and  reconstructing the trajectory and the momentum of muons.  
\subsubsection {Luminosity Monitor}
Two arrays of scintillator hodoscopes, mounted in front of the EC cryostats,
register hits with a 220 ps time resolution. They serve to detect the
 occurance of 
an inelastic \ppbar\ interaction. The particles from the
breakup of the proton give rise to hits in the hodoscopes on one side of the
detector that are tightly clustered in time.  
For events with a single interaction,
the location of the interaction vertex can be determined with a resolution of 3
cm from the time difference between the hits on the two sides of the detector
for use in the Level 2 trigger.
This array is also called the Level 0 trigger because the detection of an
inelastic \ppbar\ interaction is required for most triggers.

\subsubsection{Trigger}
Readout of the detector is controlled by a two-level trigger system.
Level 1 consists of an and-or network that can be programmed to trigger on a
\ppbar\ crossing if a number of preselected conditions are satisfied.
 The Level 1
trigger decision is 
taken within the 3.5 $\mu$s time interval between crossings.
As an extension to Level 1, a trigger processor (Level 1.5) may be invoked
to execute simple algorithms on the limited information available at the time
of a Level 1 accept. For electrons, the processor uses the energy deposits in
each trigger tower as inputs. The detector cannot accept any triggers until the
Level 1.5 processor completes execution and accepts or rejects the event.

Level 2 of the trigger consists of a farm of 48 VAXstation 4000's.
At this level, the complete event is available.
More sophisticated algorithms refine the
trigger decisions and events are accepted based on preprogrammed conditions.
Events accepted by Level 2 are written to magnetic tape for offline
reconstruction.

\section{ Data Selection }
\label{sec-data}
\subsection{Trigger }
\label{sec:trigger}

The conditions required at trigger Level 1 for \wb\ and \zb\ boson
 candidates are:
\begin{itemize}
\item $\underline{\hbox{\ppbar\ interaction:}}$ Level 0 hodoscopes register
hits consistent with a \ppbar\ interaction. Using monitor trigger data, 
the efficiency of this condition has been measured to be 98.6\%.

\item $\underline{\hbox{Main Ring Veto:}}$ No Main Ring proton bunch passes
through the detector within 800 ns of the   \ppbar\ crossing
and no protons were injected into the Main Ring less than 400 ms before the
\ppbar\ crossing.

\item $\underline{\hbox{EM trigger towers:}}$ There are one or more EM trigger
towers with $E\sin\theta>T$, where $E$ is the energy measured in the tower,
$\theta$ is the polar angle of the tower with the beam 
 measured from the center of the detector, and
$T$ is 
 a programmable threshold. This requirement is fully efficient for electrons
with $\pt>2 T$.
\end{itemize}

The Level 1.5 processor recomputes the transverse electron energy by adding the
adjacent EM trigger tower with the largest signal to the EM trigger tower that
exceeded the Level 1 threshold. In addition, the signal in the EM trigger tower
that exceeded the Level 1 threshold must constitute at least 85\% of the signal
registered in this tower if the hadronic layers are also included. This EM
fraction requirement is fully efficient for electron candidates that pass our
offline selection (Sec.~\ref{sec-data_samples}).

Level 2 uses the EM trigger tower that exceeded the Level 1 threshold as a
starting point.
The Level 2 algorithm finds the most energetic of the four calorimeter
towers that make up the trigger tower, and sums the energy in the EM
sections of a 3$\times$3 array of calorimeter towers around it.
It checks the longitudinal shower shape by applying cuts on the fraction of the
energy in the different EM layers. The transverse shower shape is characterized
by the energy deposition pattern in the third EM layer. The difference between
the energies in concentric regions covering 0.25$\times$0.25 and
0.15$\times$0.15 in
$\eta\times\phi$ must be consistent with an electron.
Level 2 also imposes an isolation
condition requiring
\begin{eqnarray}
 \frac{\sum_iE_i\sin\theta_i - \pt}{\pt} \lt 0.15 \quad ,
\end{eqnarray}
where $E_i$ and $\theta_i$ are the energy and polar angle of cell $i$, 
 the sum runs over
all cells within a cone of radius $R=\sqrt{\Delta\phi^2+\Delta\eta^2}=0.4$
around the electron direction and \pt\ is the transverse momentum of the
electron~\cite{McKinley_thesis}.

\begin{figure}[htpb!]
\vspace{-0.3in}
\epsfxsize=3.0in
\centerline{\epsfbox{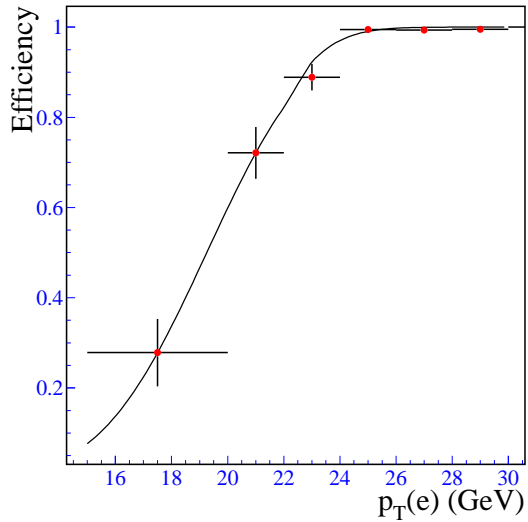}}
\vspace{0.1in}
\caption{ The relative efficiency of the Level 2 electron filter
for a threshold of 20~GeV for EC electrons, as a function of the 
 \pte\ computed offline for the $W$ boson mass analysis.  }
\label{fig:ptetrig}
\end{figure}

The \pt\ of the electron computed at Level 2 is based
on its energy and the $z$-position of
the interaction vertex measured by the Level 0 hodoscopes. Level 2 accepts
events that have a minimum number of EM clusters that satisfy the shape cuts
and have \pt\ above a preprogrammed threshold. Figure
\ref{fig:ptetrig} shows the measured relative efficiency of the Level 2 
 electron
filter for forward electrons
 versus electron \pt\ for a Level 2 \pt\ threshold of 20~GeV.
We determine this efficiency using \zb\ boson data taken with a lower
threshold value (16~GeV) for one electron.
The efficiency is the fraction of electrons above a Level 2 \pt\
threshold of 20~GeV.
The curve is the parameterization used in the fast Monte Carlo 
 (see Sec.~\ref{sec-mc}).

\begin{figure}[htpb!]
\vspace{-0.0in}
\epsfxsize=3.0in
\centerline{\epsfbox{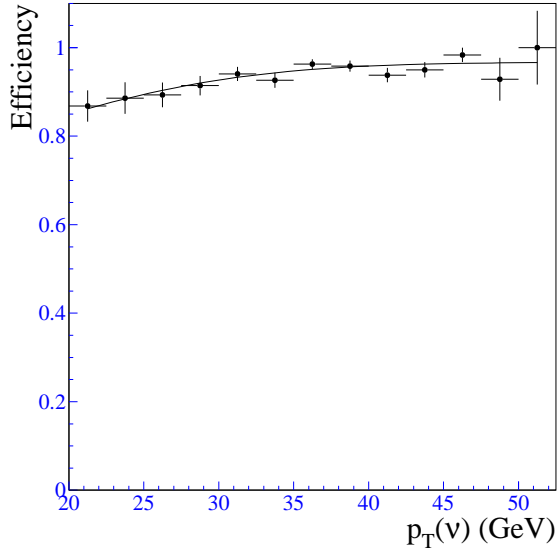}}
\vspace{0.1in}
\caption{ The efficiency of a 15~GeV Level 2 \mpt\ requirement for EC
 electrons, as a function of the \ptnu\ computed for the \wb\ boson mass
 analysis.  }
\label{fig:ptmisstrig}
\end{figure}

Level 2 also computes the missing transverse momentum based on the energy
registered in each calorimeter cell and the vertex $z$-position as measured
 by the Level 0 hodoscopes. The level 2 \wb\ boson trigger requires 
 minimum \mpt\ of 15 GeV. 
We determine the efficiency curve for a 15~GeV Level 2 \mpt\ requirement
from data taken without the Level 2 \mpt\ condition. Figure
\ref{fig:ptmisstrig} shows the measured efficiency versus \ptnu\ as computed
 for the \wb\ mass analysis, when the 
 electron is detected in the end calorimeters.
The curve is the parameterization used in the fast Monte Carlo.

\subsection{ Reconstruction }

\subsubsection{ Electron }
\label{sec-data-elec}

We identify electrons as clusters of adjacent calorimeter cells 
 with significant
energy deposits. Only clusters with at least 90\% of
their energy in the EM section and at least 60\% of their energy in the
most energetic calorimeter tower are considered as electron candidates.
For most electrons we also reconstruct a track in the CDC or FDC that points
towards the centroid of the cluster.

We compute the forward 
 electron energy \Ee\ from the signals in all cells of the EM
layers and the first FH layer whose centers lie within
  a projective cone of radius 20 cm
  and centered at the cluster centroid. 
 In the computation we use the sampling weights and
calibration constants determined using
the test-beam data (Sec.~\ref{sec-exper-cal}), except for the overall energy
 scale $A$ and the offset
$\deltaecem$, which we take from an {\em in situ} calibration
(Sec.~\ref{sec-EMresponse}).

The calorimeter shower centroid position (\xcal, \ycal, \zcal), the
 track coordinates (\xtrk, \ytrk, \ztrk), and the proton
beam trajectory define the electron angle. 
We determine the position of the electron shower centroid 
$\vec x_{\rm cal}=(\xcal,\ycal,\zcal)$ in the calorimeter
from the energy depositions in the third EM layer 
by computing the weighted mean
of the positions $\vec x_i$ of the cell centers,
\begin{equation}
\vec x_{\rm cal} = {\sum_i w_i \vec x_i\over \sum_i w_i} \quad .
\end{equation}
The weights are given by
\begin{equation}
w_i = \max\left(0,w_0 + \log\left({E_i\over\Ee}\right)\right) \quad ,
\end{equation}
where $E_i$ is the energy in cell $i$, $w_0$ is a parameter which 
depends upon \etae, and $E(e)$ is the energy of the electron.
The FDC track coordinates are reported at a fixed $z$ position using a 
 straight line fit to all the
drift chamber
  hits on the track. The calibration of the radial coordinates measured in 
 the cylindrical coordinate system 
contributes a systematic uncertainty to the \wb\ boson mass
measurement. Using tracks from many events reconstructed in the vertex
drift chamber, we measure the beam trajectory for every run.
The closest approach to the beam trajectory of the line through the
 shower centroid
and the track coordinates defines the $z$-position of the interaction
vertex (\zvtx). The beam trajectory provides (\xvtx,\yvtx). 
 In \zee\ events, we may have
two electron candidates with tracks. In this case we take the point determined 
from the more central electron as the
interaction vertex, because this gives better resolution. 
Using only the electron track to determine the position of
the interaction vertex, rather than all tracks in the event, makes the
resolution of this measurement less sensitive to the
luminosity and avoids confusion
between vertices in events with more than one \ppbar\ interaction.

We then define the azimuth \phie\ and the polar angle
  \te\ of the electron using
the vertex and the shower centroid positions
\begin{eqnarray}
\tan\phie & = & {\ycal - \yvtx \over \xcal - \xvtx } \quad , \\
\tan\te & = &   {\sqrt{\xcal^2+\ycal^2} - \sqrt{\xvtx^2+\yvtx^2}
                        \over \zcal - \zvtx } \quad .
\end{eqnarray}
Neglecting the electron mass, the momentum of the electron is given by
\begin{equation}
\pev = \Ee \left(\begin{array}{l} \sin\te\cos\phie \\
                                  \sin\te\sin\phie \\
                                  \cos\te            \end{array} \right).
\end{equation}

\subsubsection{ Recoil }
We reconstruct the transverse momentum of all particles recoiling against the
\wb\ or \zb\ boson by taking the vector sum
\begin{equation}
\utv = \sum_i E_i \sin\theta_i \left(
\begin{array}{c} \cos\phi_i \\ \sin\phi_i \end{array} \right),
\end{equation}
where the sum runs over all calorimeter cells that were read out, except those
that belong to electron cones. $E_i$ are the
cell energies, and $\phi_i$ and $\theta_i$ are the azimuth and 
polar angle of
the center of cell $i$ with respect to the interaction vertex.

\subsubsection{ Derived Quantities }
\begin{figure}[htb]
\vspace{-0.3in}
\epsfxsize=3.0in
\centerline{\epsfbox{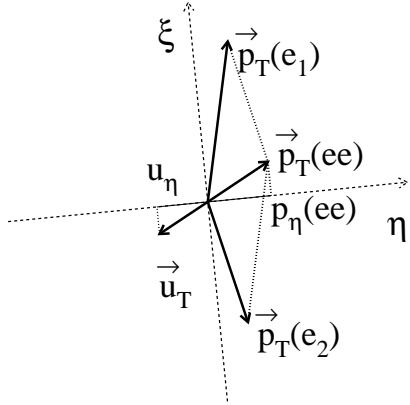}} 
\vspace{-0.2in}
\caption{ Illustration of momentum vectors in the transverse plane
for \zee\ candidates. The vectors drawn as thick lines are directly
measured.}
\label{fig:zdef}
\end{figure}

In the case of \zee\ decays, we define the dielectron momentum
\begin{equation}
\peev = \peonev + \petwov
\end{equation}
and the dielectron invariant mass
\begin{equation}
\mee = \sqrt{2 E(e_1) E(e_2) (1-\cos\omega)} \quad ,
\end{equation}
where $\omega$ is the opening angle between the two electrons. It is
useful to define a coordinate system in the plane transverse
to the beam that depends only on the electron directions. We follow the
conventions first introduced by UA2\cite{UA2} and call the axis
along the inner bisector of the transverse directions of the
two electrons the $\eta$-axis and the axis
perpendicular to that the $\xi$-axis. Projections on these axes are denoted
with subscripts $\eta$ or $\xi$. Figure~\ref{fig:zdef} illustrates these
definitions.

\begin{figure}[htb]
\vspace{-0.3in}
\epsfxsize=3.0in
\centerline{\epsfbox{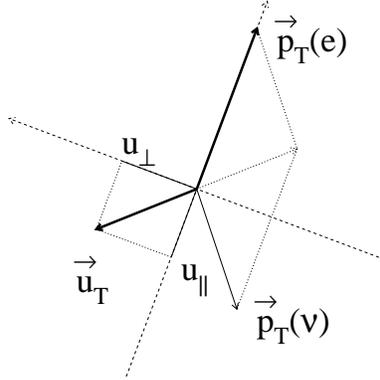}} 
\vspace{-0.2in}
\caption{ Illustration of momentum vectors in the transverse plane
for \wev\  candidates. The vectors drawn as thick lines are directly
measured.}
\label{fig:wdef}
\end{figure}

In the case of \wev\ decays, we define the transverse neutrino momentum
\begin{equation}
\ptnuv = -\ptev - \utv
\end{equation}
and the transverse mass (Eq.~\ref{eq:mtdefn}).
Useful quantities are the projection of the transverse recoil momentum on the
transverse component of the electron direction,
\begin{equation}
\upar = \utv\cdot\hat \pte \quad ,
\end{equation}
and the projection 
perpendicular to the transverse component of the electron direction,
\begin{equation}
\uper = \utv\cdot(\hat\pte\times\hat z) \quad .
\end{equation}
Figure~\ref{fig:wdef} illustrates these definitions.

\subsection{ Electron Identification }

\subsubsection{ Fiducial Cuts}
 Electrons in the ECEM are defined by the pseudorapidity $\eta$ of the cluster
 centroid position with respect to the center of the detector. We define
 forward electrons by $1.5 \leq \mid \eta_{\rm det} (e) \mid \leq 2.5$.  
\subsubsection{ Quality Variables }
We test how well the shape of a cluster agrees with that expected for an
electromagnetic shower by computing a
quality variable ($\chi^2$) for all cell energies using a
41-dimensional covariance matrix. The covariance matrix was determined from
\GEAN-based \cite{geant} simulations \cite{top_prd} that were tuned to agree
 with extensive test beam measurements.

To determine how well a track matches a cluster,
  we extrapolate the track to the
third EM layer in the end calorimeter and compute the distance between the
extrapolated track and the cluster centroid in the azimuthal direction,
$\Delta s$, and in the radial direction, $\Delta \rho$. The variable
\begin{equation}
\sigm^2 = \left({\Delta s\over \delta s}\right)^2 + \left({\Delta \rho \over
 \delta \rho }\right)^2,
\end{equation}
quantifies the quality of the match. 
The parameters $\delta s=0.25$ cm 
 and $\delta \rho =1.0$ cm are the resolutions with which
$\Delta s$ and $\Delta \rho $ are measured, as
determined using  the end calorimeter electrons from \wev\ decays.

 In the EC, electrons must have a matched track in the forward drift chamber
  to suppress background due to misidentification.  In
 the CC, we define ``tight'' and ``loose'' criteria. The tight criteria 
 require a
 matched track in the CDC, defined as the track with the smallest $\sigm$.
  The loose criteria do not require a matched track and
help increase the electron finding efficiency for \zee\ decays with at least
 one central electron.

The isolation fraction is defined as
\begin{equation}
    \fiso = {E_{\rm cone}-E_{\rm core}\over E_{\rm core}},
\end{equation}
where $E_{\rm cone}$ is the energy in a cone of
radius $R=0.4$ around the direction of
the electron, summed over the entire depth of the calorimeter, and
$E_{\rm core}$ is the energy in a cone of $R=0.2$, summed over the EM
calorimeter only.

 We use the $dE/dx$ information provided by the FDC 
 on the tracks associated with
 the EM calorimeter cluster. The $dE/dx$ 
information helps to distinguish between
  singly-ionizing electron tracks and 
doubly-ionizing tracks from photon conversions.

We identify electron candidates in the forward detectors by making loose cuts
 on the shower shape $\chi^2$, the track-cluster match quality, and the
 shower electromagnetic energy fraction. The electromagnetic energy fraction
 is the ratio of the cluster energy measured in the electromagnetic calorimeter
 to the total cluster energy (including the hadronic calorimeter), and is a 
 measure of the longitudinal shower profile. We then use a cut on 
 a 4-variable likelihood
 ratio $\lambda _4$ 
 which combines the information in these variables and  the track $dE/dx$
 into a single variable. The final cut on the likelihood ratio $\lambda _4$
 gives the maximum discrimination between
 electrons and jet background, i.e. gives the maximum background rejection
 for any given electron selection efficiency.  

\begin{figure}[htpb!]
\vspace{0.1in}
\epsfxsize=3.0in
\centerline{\epsfbox{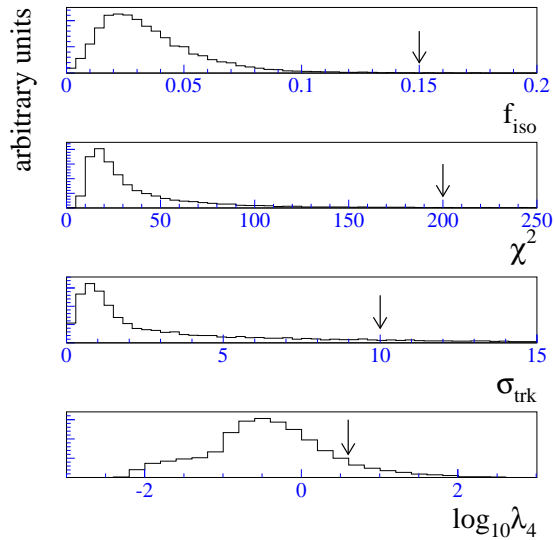}}
\vspace{0.1in}
\caption{ Distributions of the EC electron identification variables for
 \wev\ candidates in the data.
The arrows indicate the cut values. }
\label{fig:eid}
\end{figure}

Figure~\ref{fig:eid} shows the distributions of the quality variables
for electrons in the EC data; the arrows indicate the cut values.
Table~\ref{tab:e_selection} summarizes the electron selection criteria.

\begin{table}[ht]
\begin{center}
\caption{\small Electron selection criteria. $\Delta\phi_{\rm cal}$ is
 the difference in azimuthal angle between the cluster centroid and the 
 CC module edge. 
}
\medskip
\begin{tabular}{lccc}
variable      & CC (loose)          & CC (tight)          & EC (tight) \\ \hline
fiducial cuts & $|\Delta\phi_{\rm cal}|>0.02$ & $|\Delta\phi_{\rm cal}|>0.02$ & --- \\
              & $|\zcal|<108$ cm & $|\zcal|<108$ cm & $1.5\leq|\eta|\leq2.5$ \\
              & ---                 & $|\ztrk|<80$ cm     & --- \\
shower shape  & $\chi^2<100$        & $\chi^2<100$        & $\chi^2<200$ \\
isolation     & $\fiso <0.15$       & $\fiso <0.15$       & $\fiso <0.15$ \\
track match   & ---                 & $\sigm <5$          & $\sigm <10$ \\
4-variable   &                     &           &  \\
likelihood ratio   & ---            & ---          & $\lambda_4 <$4 \\
\end{tabular}
\label{tab:e_selection}
\end{center}
\end{table}

\subsection{ Data Samples }
\label{sec-data_samples}
The data were collected
  during the 1994--1995 Tevatron run. After the removal of
runs in which parts of the detector were not operating adequately, the data
 correspond 
to an integrated luminosity of 82 pb$^{-1}$.
We select \wb\ boson decay candidates by requiring:

\begin{tabular}{ll}
Level 1: & \ppbar\ interaction \\
         & Main Ring Veto \\
	 & EM trigger tower above 10~GeV \\
Level 1.5: & $\geq 1$ EM cluster above 15~GeV \\
Level 2: & electron candidate with $\pt>20$~GeV \\
         & momentum imbalance $\mpt>15$~GeV \\
offline: & $\geq 1$ tight electron candidate in EC \\
         & $\pte > 30$~GeV \\
         & $\ptnu > 30$~GeV \\
         & $\ut < 15$~GeV \\
\end{tabular}

\noindent
This selection gives us 11{,}089 \wb\ boson candidates. 
We select \zb\ boson decay candidates by requiring:

\begin{tabular}{ll}
Level 1: & \ppbar\ interaction \\
         & $\geq 2$ EM trigger towers above 7~GeV \\
Level 1.5: & $\geq 1$ EM cluster above 10~GeV \\
Level 2: & $\geq 2$ electron candidates with $\pt>20$~GeV \\
offline: & $\geq 2$ electron candidates \\
	 & $\pte > 30$~GeV (EC) \\ 
         & or $\pte > 25$~GeV (CC) \\
\end{tabular}

\noindent We accept \zee\ decays with at least one electron candidate in the
EC and the other in the CC or the EC.
EC candidates must pass the tight electron
selection criteria. A CC candidate may pass only the
loose criteria. We use the 1{,}687 events with at least one
 electron in the EC
(CC/EC + EC/EC \zb\ samples) to calibrate the
calorimeter response to electrons (Sec.~\ref{sec-elec}). These events need
not pass the Main Ring Veto cut because Main Ring background does not affect
the EM calorimeter. Of these events, 
those that do pass the Main Ring Veto have been 
used to calibrate the recoil momentum response.
The events for which both electrons are in the EC (EC/EC
 \zb\ sample) and
 which pass the Main Ring Veto serve to check the calibration of the recoil
 response (Sec.~\ref{sec-recoil}). Table~\ref{tab:sample} summarizes
the data samples.


\begin{table}[ht]
\begin{center}
\caption{\small Number of \wb\ and \zb\ boson candidate events.}
\medskip
\begin{tabular}{lccc}
channel & \multicolumn{2}{c}{\zee} & \wev \\
fiducial region of electrons   & CC/EC & EC/EC &    EC \\
            &   1265 &  422  & 11089 \\
\end{tabular}
\label{tab:sample}
\end{center}
\end{table}

\begin{figure}[htpb!]
\vspace{0.2in}
\epsfxsize=3.0in
\centerline{\epsfbox{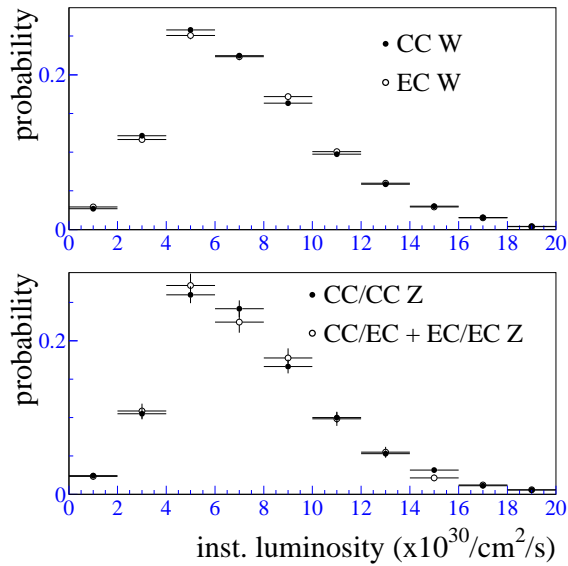}}
\vspace{0.1in}
\caption{ The instantaneous
 luminosity distribution of the \wb\ (top) and the \zb\
(bottom) boson samples.}
\label{fig:lum}
\end{figure}

Figure~\ref{fig:lum} shows the luminosity of the colliding beams
during the \wb\ and \zb\ boson data collection.

On several occasions we use a sample of 295{,}000 random \ppbar\ interaction
events for calibration purposes. We collected
these data concurrently with the \wb\ and
\zb\ signal data, requiring only a \ppbar\ interaction at
Level 1. We refer to these data as ``minimum bias events."

\section{ Fast Monte Carlo Model }
\label{sec-mc}
\subsection {Overview}

The fast Monte Carlo model consists of three parts. First we simulate the
production of the \wb\ or \zb\ boson by generating the boson four-momentum and
other characteristics of the event such as
  the $z$-position of the interaction
vertex and the luminosity.  The event luminosity is required for
 luminosity-dependent parametrizations in the detector simulation.  Then 
 we simulate the
decay of the boson. At this point
we know the true \pt\ of the boson and the momenta of its decay products. We
next  apply a parameterized detector model to these momenta to simulate
the observed transverse recoil momentum and the observed electron momenta.

Our fast Monte Carlo program is very similar to the one used in our published
 CC analysis \cite{wmass1bcc}, with some modifications in the simulation
 of forward electron events.
 
\subsection { Vector Boson Production }
\label{sec-vb_prod}

To specify the production dynamics of vector bosons
in \ppbar\ collisions completely, we need to know the differential
production cross section
in mass $Q$, rapidity $y$, and transverse momentum \qt\ of the
produced \wb\ bosons. To speed up the event generation, we factorize
this into
\begin{equation}
{d^3\sigma\over d \qt^2  d y d Q } \approx
\left.{d^2\sigma\over d \qt^2 d y}\right|_{Q^2=\mw^2} \times
{d\sigma\over d Q}
\label{eq:prod}
\end{equation}
to generate \qt, $y$, and $Q$ of the bosons.

For \ppbar\ collisions, the vector boson production
cross section is given by the parton cross section $\widetilde\sigma_{i,j}$
convoluted with the parton distribution functions (pdf) $f(x,Q^2)$ and summed
over parton flavors $i,j$:
\begin {eqnarray}
{d^2\sigma\over d \qt^2  d y} &=&
     \sum_{i,j}\int dx_1 \int dx_2 f_i(x_1,Q^2) f_j(x_2,Q^2) \nonumber \\
 & & \ \ \delta(sx_1x_2-Q^2){d^2\widetilde\sigma_{i,j}\over d \qt^2  d y} 
 \quad .
\label{eq:wprod}
\end {eqnarray}
The cross section $d^2\sigma / d \qt^2 d y |_{Q^2=\mw^2}$ has been
 computed by several authors~\cite{LY,AK} using a perturbative
calculation \cite{AR} for the high-\qt\ regime and the Collins-Soper
resummation formalism \cite{CSS,AEMG} for the low-\qt\ regime. We use the
code provided by the authors of Ref.~\cite{LY} and the MRST parton
distribution functions \cite{mrst} to compute the cross section. 
The production of $WW$, $WZ$ and $W \gamma$
  is suppressed by three orders of magnitude compared to inclusive W 
  production.

We use a Breit-Wigner curve with a mass-dependent
  width for the line shape of the
\wb\ boson. The intrinsic width of the \wb\ is $\wwidth=2.062 \pm 0.059 $~GeV
\cite{Wwidth}. The line shape is skewed due to the momentum
distribution of the quarks inside the proton and antiproton. The mass spectrum
is given by
\begin {eqnarray}
{d\sigma\over d Q}={\cal L}_{\qqbar}(Q)
{Q^2\over (Q^2-\mw^2)^2+{Q^4\Gamma_W^2\over\mw^2}} \quad .
\end {eqnarray}
We call
\begin {eqnarray}
{\cal L}_{\qqbar}(Q) = {2Q\over s} \sum_{i,j}\int_{Q^2/s}^1 {d x\over x}
f_i(x,Q^2)f_j(Q^2/sx,Q^2)
\end {eqnarray}
the parton luminosity. To evaluate it,
  we generate \wev\ events using the \HERW\
Monte Carlo event generator \cite{herwig}, interfaced 
 with \PDFL\ \cite{PDFLIB},
and select the events subject to the same fiducial cuts as for the
\wb\ and \zb\ boson samples with at least one electron in EC. 
 We plot the mass spectrum
divided by the intrinsic line shape of the \wb\ boson. The result is
proportional to the parton luminosity,  and we parameterize
the shape of the spectrum with the function~\cite{wmass1acc}
\begin{equation}
{\cal L}_{\qqbar}(Q) = {e^{-\beta Q}\over Q} \quad .
\label{eq:partlum}
\end{equation}
Table~\ref{tab:wprod} shows the parton luminosity slope
 $\beta$ for \wb\ and \zb\ events for the
 different topologies. The value of $\beta$ depends on the rapidity
distribution of the \wb\ and \zb\ bosons, which is restricted by the
fiducial cuts that we impose on the decay leptons. The values of $\beta$ given
in Table~\ref{tab:wprod} are for the rapidity distributions of \wb\ and \zb\
bosons that satisfy the fiducial cuts given in
Sec.~\ref{sec-data}.
The uncertainty in $\beta$ is about 0.001 GeV$^{-1}$, due to Monte Carlo
statistics and uncertainties in the acceptance.

\begin{table}[ht]
\begin{center}
\caption{\small Parton luminosity slope $\beta$
 in the \wb\ and \zb\ boson production model. The $\beta$ value is
given for \wev\ decays with the electron in the 
 EC and for \zee\ decays with at
 least one electron in the EC.}
\medskip
\begin{tabular}{lcc}
                    & \zb\ production & {\wb\ production} \\
                    & $\beta$ (GeV$^{-1}$) & $\beta$ (GeV$^{-1}$) \\
\hline
CC/EC  & $9.9\times10^{-3}$ & ---
\\ 
EC/EC & $19.9\times10^{-3}$ & ---
\\ 
EC & ---                & $16.9\times10^{-3}$ 
\\ 
\end{tabular}
\label{tab:wprod}
\end{center}
\end{table}

Bosons can be produced by the annihilation of two valence quarks, two sea 
quarks, or one valence quark and one sea quark. Using the \HERW\ events, we 
 evaluate
 the fraction $f_{\rm ss}$ of bosons produced by the annihilation of two sea
 quarks.  We find $f_{\rm ss}=0.207$, independent of the   boson topology. 

To generate the boson four-momenta, we treat $d\sigma/d Q$ and $d^2\sigma/
d \qt^2 d y$ as probability density functions and pick $Q$ from the former
and a pair of $y$ and \qt\ values from the latter. For a fraction $f_{\rm ss}$
 the
 boson  helicity is $+1$ or $-1$ with equal probability. The remaining \wb\
bosons always have helicity $-1$. 
Finally, we pick the
$z$-position of the interaction vertex from a Gaussian
distribution centered at $z=0$ with a standard deviation of 27 cm and
a luminosity for each event from the histogram in Fig.~\ref{fig:lum}.

\subsection{ Vector Boson Decay }

\begin{figure}[htpb!]
\vspace{-0.3in}
\epsfxsize=3.0in
\centerline{\epsfbox{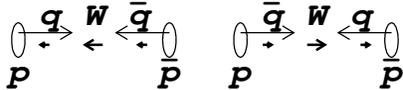}} 
\vspace{-0.02in}
\caption{ Polarization of the \wb\ boson
 produced in \ppbar\ collisions if the
quark comes from the proton (left) and if the antiquark comes from the proton
(right). The short 
 thick arrows indicate the orientations of the particle spins.}
\label{fig:wpol}
\end{figure}

At lowest order, the $W^\pm$ 
 boson is fully polarized along the beam direction due
to the $V \mp A$ coupling of the charged current. The resulting angular
distribution of the charged lepton in the \wb\ boson rest frame is given by
\begin{equation}
\label{eq:stnd_angle}
{d \sigma\over d \cos\theta^*} \propto (1-\lambda q\cos\theta^*)^2 \quad ,
\end{equation}
where $\lambda$ is the helicity of the \wb\ boson
 with respect to the proton
direction,
$q$ is the charge of the lepton, and $\theta^*$ 
is the angle between the charged
lepton and proton beam directions in the \wb\ rest frame.
The spin of the  \wb\ boson
 points along the direction of the incoming antiquark.
Most of the time, the quark comes from the proton and the antiquark from the
antiproton, so that $\lambda=-1$. Only if both quark and antiquark come from
the sea of the proton and antiproton, is there a 50\% chance that the quark
comes from the antiproton and the antiquark from the proton and in that case
$\lambda=1$ (see Fig.~\ref{fig:wpol}).

When ${\cal O}(\alpha_s)$ processes are included, the boson acquires finite
transverse momentum and Eq.~\ref{eq:stnd_angle} becomes~\cite{mirk}
\begin{eqnarray}
\label{eq:cos_angle}
{d\sigma\over d\cos\theta_{\rm CS}} \propto 
    \left( 1-\lambda q \alpha_1(\qt)\cos\theta_{\rm CS}+\alpha_2(\qt)\cos^2\theta_{\rm CS}
    \right)
\end{eqnarray}
for $W$ bosons after integration over $\phi$. The angle
$\theta_{\rm CS}$ in Eq.~\ref{eq:cos_angle} is now defined in the Collins-Soper
frame \cite{csframe}. The values of $\alpha_1$ and $\alpha_2$ as a function of
transverse boson momentum have been calculated at ${\cal O}(\alpha_s^2)$
\cite{mirk}. We have implemented the angular distribution given
in Eq.~\ref{eq:cos_angle} in the fast Monte Carlo.
 The angular distribution of the leptons from \zee\ decays is also 
generated according to Eq.~\ref{eq:cos_angle}, but with $\alpha_1$ and
$\alpha_2$ computed for \zee\ decays~\cite{mirk}.

Radiation from the decay electron or the \wb\ boson biases the mass
measurement. If the decay electron radiates a photon and the photon is 
 sufficiently 
  separated from the electron so that its energy is not included in the
electron energy, or if an on-shell \wb\ boson radiates a photon and therefore
 is
off-shell when it decays, the measured mass is biased low. We use the
calculation of Ref.~\cite{rad_decays_th} to generate \wegam\ and \zeegam\ 
decays. The calculation gives the fraction of events in which a photon with
energy $\Eg>E_0$ is radiated, and the angular distribution and energy
spectrum of the photons. Only radiation from the decay electron and the \wb\
boson, if the final state \wb\ is off-shell, is included to order $\alpha$.
Radiation by the initial quarks or the \wb\ boson,
  if the final \wb\ is on-shell,
does not affect the mass of the $e\nu$ pair from the \wb\ decay. We use a
minimum photon energy $E_0=50$~MeV, and calculate that in 30.6\% of all \wb\
decays a photon with $\Eg>50$~MeV is radiated. 
 Most of these photons are emitted
close to the electron direction and cannot be separated from the electron in
the calorimeter. For \zee\ decays, there is a 66\% probability for either
 of
the electrons to radiate a photon with $\Eg>50$~MeV.

If the photon and electron are close
together, they cannot be separated in the calorimeter.
The momentum of a photon
with $\regam\lt\rzero$ is therefore added to the electron momentum,
while for $\regam\geq\rzero$, a photon is considered separated
from the electron and its momentum is added to the recoil momentum. We use
$\rzero =20$ cm,
 which is the size of the cone in which the electron
energy is measured. We refer to \rzero\ as the photon coalescing radius. 

\wb\ boson decays through the channel $W\to\tau\nu\to e\nu\overline\nu\nu$ are
topologically indistinguishable from  \wev\ decays. We therefore include
these decays in the \wb\ decay model, properly accounting for the polarization
of the tau leptons in the decay angular distributions. In the standard model
 and neglecting small phase space effects, the fraction of \wb\
 boson decays to electrons  that proceed via tau decay 
  is $B(\tev)/\left(1+B(\tev)\right) = 0.151$.

\subsection{ Detector Model }

The detector simulation uses a parameterized model for detector
 response and resolution
to obtain a prediction for the distributions
  of the observed electron and recoil
momenta.

When simulating the detector response to an electron of energy $E_0$, we
compute the observed electron energy as
\begin{eqnarray}
E(e) = \alphaecem E_0 + \Delta E({\cal L},\eta,u_{||}) +
\sigma_{\rm EM} X \quad ,
\end{eqnarray}
where \alphaecem\ is the response of the end electromagnetic calorimeter,
$\Delta E$ is the energy due to particles from the underlying event within the
electron cone (parameterized as a function of luminosity ${\cal L}$, $\eta$
  and
$u_{||}$), $\sigma_{\rm EM}$ is the energy resolution of the electromagnetic
calorimeter, and $X$ is a random variable from a normal parent distribution
with zero mean and unit width.

The transverse energy measurement depends on the measurement of the electron
direction as well. We determine the shower centroid position by intersecting
  the
line defined by the event vertex and the electron direction with a plane
perpendicular to the beam
and located at $z = \pm$ 179 cm
 (the longitudinal center of the ECEM3 layer).  We
then smear the azimuthal and radial 
 coordinates of the intersection point by their
resolutions.  We determine the radial 
 coordinate of the
FDC track by intersecting the same line with a plane at $z= \pm 105$ cm, the
defined $z$ position of the FDC track centroid,
and smearing by the resolution. The measured angles are then obtained from the
smeared points as described in Section~\ref{sec-data-elec}.

The model for the particles recoiling against the \wb\ boson
 has two components: a
``hard'' component that models the $p_T$ of the \wb\ boson, 
and a ``soft'' component that models detector noise and pile-up. Pile-up refers
to the effects of additional \ppbar\ interactions in the same or previous beam
crossings. For the soft component we use the transverse momentum balance
\mptv\ measured in  minimum bias events recorded in the detector. The
 minimum bias events are weighted so that their luminosity distribution is
 the same as that of the $W$ sample. 
The observed recoil \pt\ is then given by
\begin{eqnarray}
\label{eq:ut}
\vec u_T &=& -\bigl(\rrec\qt+\sigrec X \bigr)\hat \qt \nonumber \\
         & & -\dupar({\cal L},\eta,\upar) \hat\pte \nonumber \\
         & & +\alphamb \mptv \quad , 
\end{eqnarray}
where \qt\ is the generated value of the boson transverse momentum,
\rrec\ is the (in general momentum-dependent) response, \sigrec\ is the 
resolution of the
calorimeter (parameterized as $\sigrec = \srec \sqrt{\ut}$),
 \dupar\ is the transverse energy flow into the electron window
(parameterized as a function of ${\cal L}$, $\eta$ and \upar), and
\alphamb\ is a correction factor that allows us to adjust the resolution to
the data, accounting for the difference between the data minimum bias events
 and the underlying spectator collisions in \wb\ events.
  The quantity \dupar\ is different from the transverse
 energy added to the
electron, $\Delta E_T$, because of the difference in the algorithms used to
 compute the electron $E_T$ and the recoil \pt.

We simulate selection biases due to the trigger requirements and the electron
isolation by accepting events with the estimated efficiencies.
Finally, we compute all the derived quantities from these observables
and apply fiducial and kinematic cuts.

\section{ Electron Measurement }
\label{sec-elec}

\subsection { Angular Calibrations }

\begin{figure}[htpb!]
\vskip 0.5 cm
\epsfxsize = 3.0in 
\centerline{\epsfbox{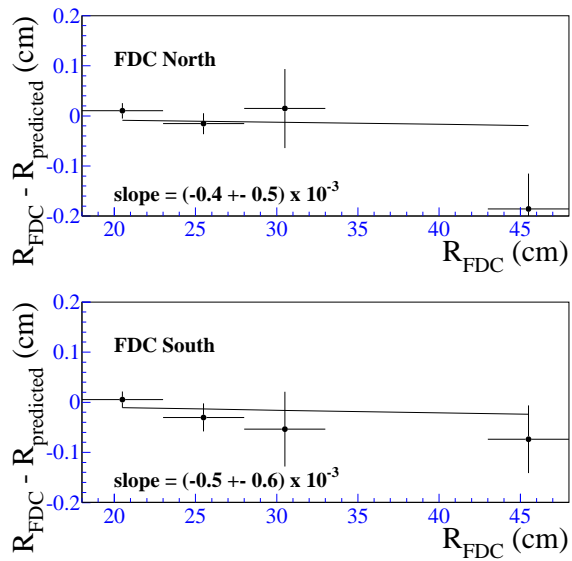}}
\vspace{0.1in}
\caption {Residue of the radial position of the FDC track centroid from the
 predicted 
 radial position of forward muon tracks at the FDC, as a function of the
 track radial position. 
The solid line is a fitted straight line constrained to pass through the 
origin. }
\label{fdcr}
\end{figure}

The FDC detectors 
have been studied and calibrated extensively in a test beam \cite{bantly}.
We use collider data muons which traverse the forward muon detectors and the
 FDC  to provide a cross-check of the test beam calibration of 
  the radial measurement of the track in the FDC.
We predict the trajectory
of the muon through the FDC by connecting the 
 hits in the innermost muon chambers with the reconstructed event vertex 
 by a straight line.
The  FDC track coordinate can then be compared relative to this
line. Figure~\ref{fdcr} shows the difference between the predicted
and the actual radial positions of the track.
These data are fit to a straight line constrained to pass through the
 origin.
We find the track position is consistent with the predicted position. 

\begin{figure}[htpb!]
\epsfxsize=3.0in
\centerline{\epsfbox{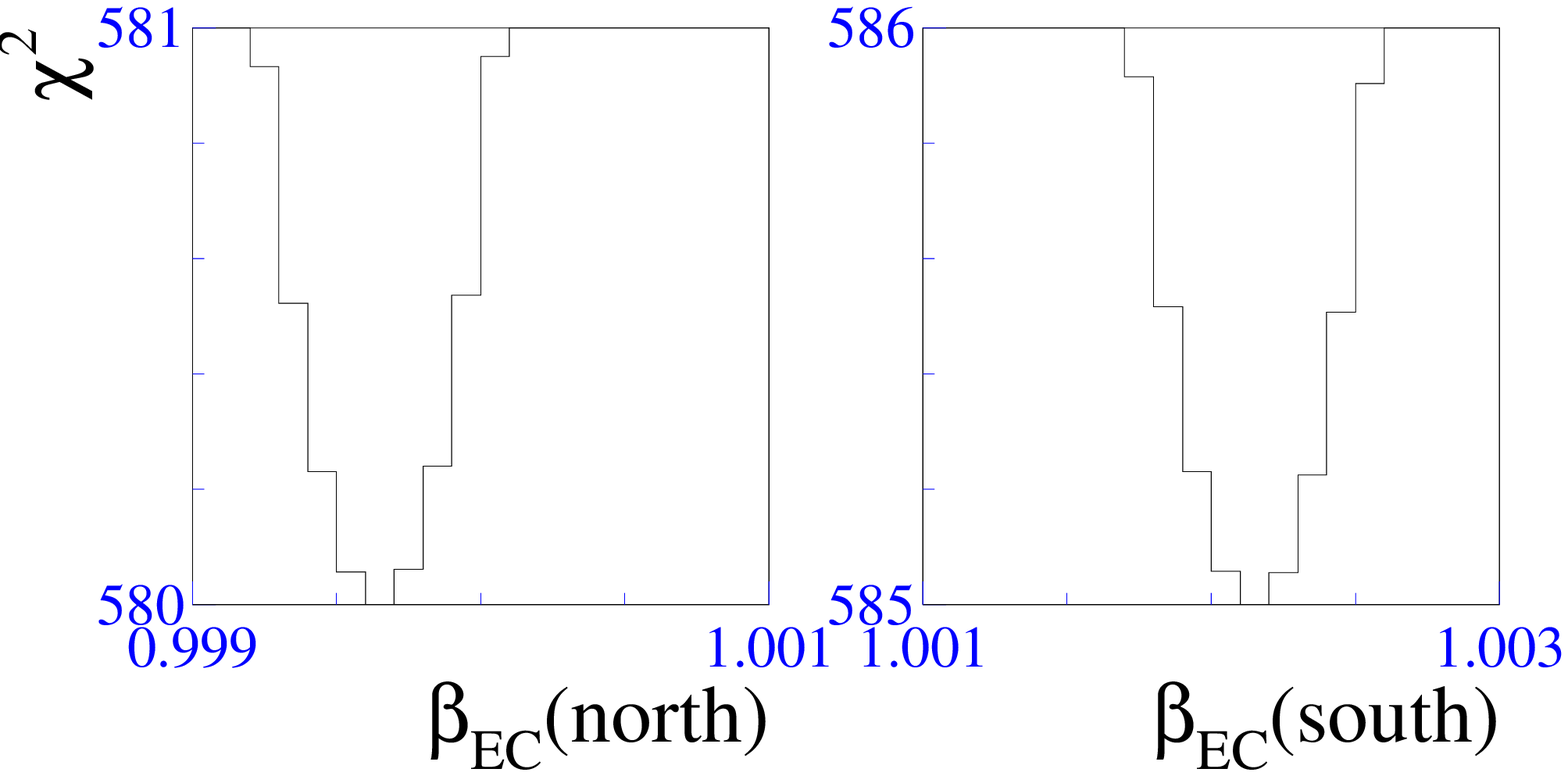}}
\vspace{0.1in}
\caption{The $\chi^2$ versus \betaec\ value.}
\label{fig:ecscale}
\end{figure}

We calibrate the shower centroid
 algorithm using Monte Carlo electrons simulated using \GEAN\
and electrons from the \zee\ data.
We apply a polynomial correction as a function of \rcal\ and the
 distance from the cell edges  based on
the Monte Carlo electrons.
We refine the calibration with the \zee\ data by exploiting the fact that 
both electrons originate from the same vertex. Using the algorithm 
described in Sec.~\ref{sec-data-elec}, we determine a
vertex for each electron from the shower centroid and the track coordinates.
 We minimize the difference between
the two vertex positions as a function of an \rcal\ scale factor \betaec\ 
 (see Fig.~\ref{fig:ecscale}). 
The correction factor is $\betaec = 0.9997 \pm 0.00044$ for EC North, and
 $\betaec = 1.00225 \pm 0.00044$ for EC South. We find no systematic
 radial dependence of these correction factors.   

We quantify the FDC and EC radial calibration uncertainty in terms of
scale factor uncertainties $\delta \betafdc= \pm 0.00054$ and $\delta 
 \betaec= \pm 0.0003$ for the radial coordinate. The
uncertainties in these scale factors lead to a 20 MeV
 uncertainty in the EC \wb\ boson 
mass measurement.

\subsection { Angular Resolutions }

\begin{figure}[htpb!]
\vspace{-0.3in}
\epsfxsize=3.0in
\centerline{\epsfbox{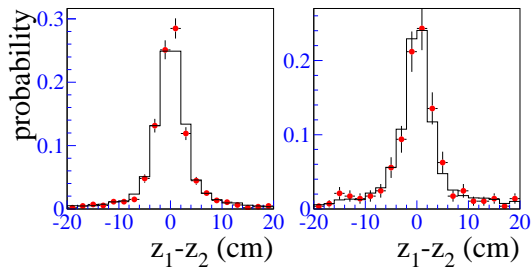}}
\vspace{0.1in}
\caption{ The distribution of $\zvtx(e_1) - \zvtx(e_2)$ for the
CC/EC (left) and EC/EC (right) \zee\ samples
  ($\bullet$) and the fast Monte Carlo simulation~\hbox{(------)}.}
\label{fig:rtrk}
\end{figure}

The resolution for the radial
 coordinate of the track, \rtrk, is
determined from the \zee\ sample. Both electrons originate from the same
interaction vertex and therefore the 
 difference between the interaction vertices
reconstructed from the two electrons separately, $\zvtx(e_1) -\zvtx(e_2)$, is a
measure of the resolution with which the electrons point back to the vertex. 
 The
points in Fig.~\ref{fig:rtrk} show the distribution
of $\zvtx(e_1) -\zvtx(e_2)$ observed in the CC/EC and EC/EC \zb\ samples 
with matching tracks required for both electrons.

A Monte Carlo study based on single electrons generated with a \GEAN\
simulation shows that the resolution of the
shower centroid algorithm is 0.1 cm in the EC, consistent with EC electron
 beam tests. We then tune the
resolution function for \rtrk\ in the fast Monte Carlo so that it
reproduces the shape of the $\zvtx(e_1) - \zvtx(e_2)$ distribution observed
in the data. We find that a resolution function consisting of two Gaussians
0.2 cm and 1.7 cm wide, with 20\% of the area under the wider Gaussian, fits
the data well. The histogram in Fig.~\ref{fig:rtrk} shows the Monte
Carlo prediction for the best fit, normalized to the same number of events as
the data. 

\subsection { Underlying Event Energy }
\label{sec-ue}

 We define a cone which is projective from the center of the detector,
 has a radius of 20 cm at the $z$ position of ECEM3 and is centered on the 
 electron cluster centroid. The cone extends over the four ECEM layers and the
 first ECFH layer. 
 This cone contains the entire energy deposited by the electron
shower plus some energy from other particles.
The energy in the window is  excluded from the computation of \utv. This causes
a bias in \upar, the component of \utv\ along the direction of the electron.
We call this bias \dupar. It is equal to the momentum flow observed
in the EM and first FH sections of a projective cone of radius 20 cm at ECEM3.

We use the \wb\ data sample to measure \dupar. For every
electron in the \wb\ sample, we compute the energy flow into an
azimuthally rotated position, keeping the cone radius and the radial position
 the same. For the rotated position
we compute the measured transverse energy. Since the $\eta \phi$ area
 of the cone increases as the electron $\eta$ increases, it is convenient to
 parameterize the transverse energy density, \dupar/$\delta \eta \delta \phi$.

\begin{figure}[htpb!]
\vspace{-0.3in}
\epsfxsize=3.0in
\centerline{\epsfbox{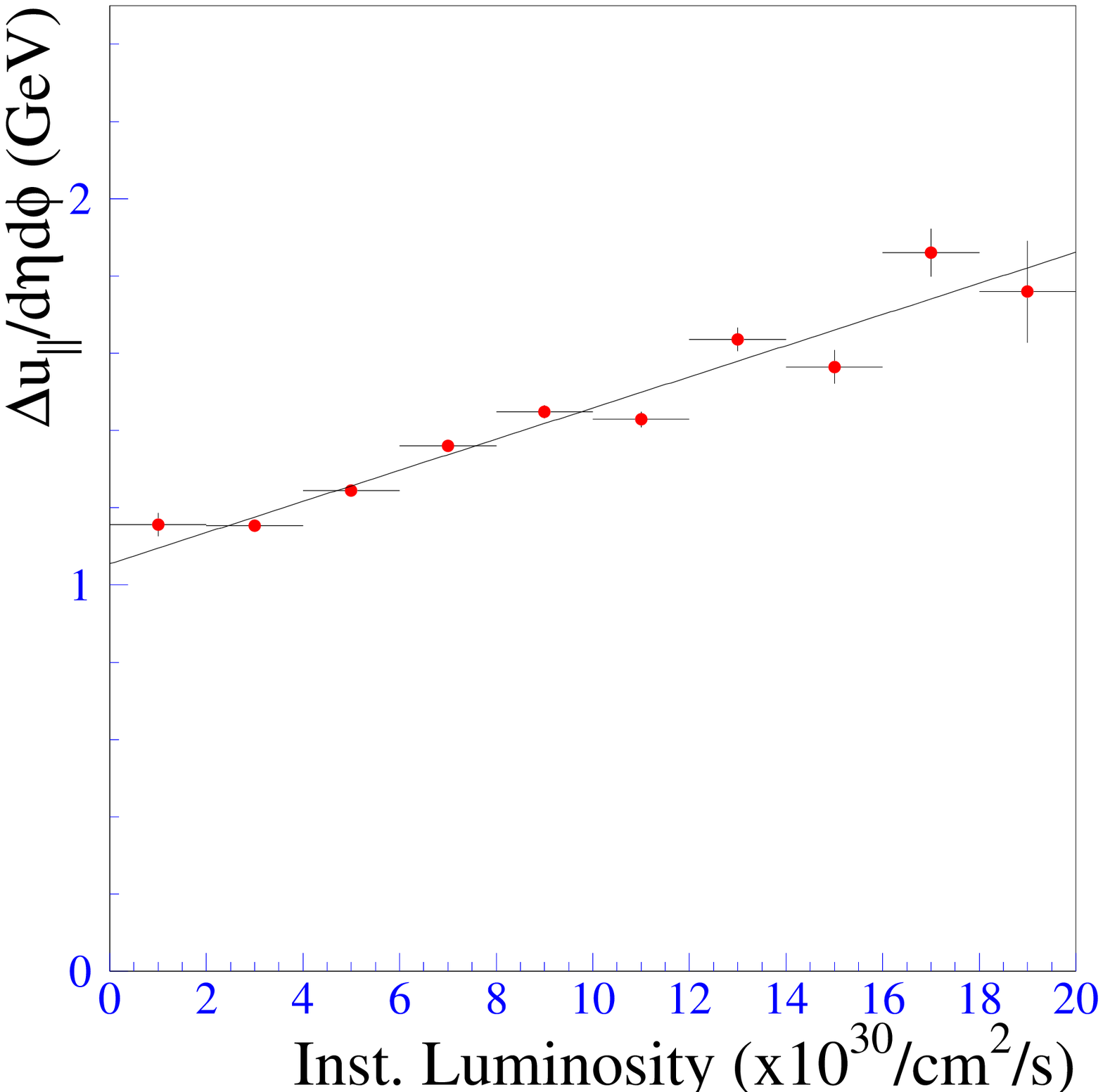}}
\vspace{0.1in}
\caption{ The instantaneous luminosity dependence of \mdupar.}
\label{fig:duparlum}
\vspace{0.4in}
\epsfxsize=3.0in
\centerline{\epsfbox{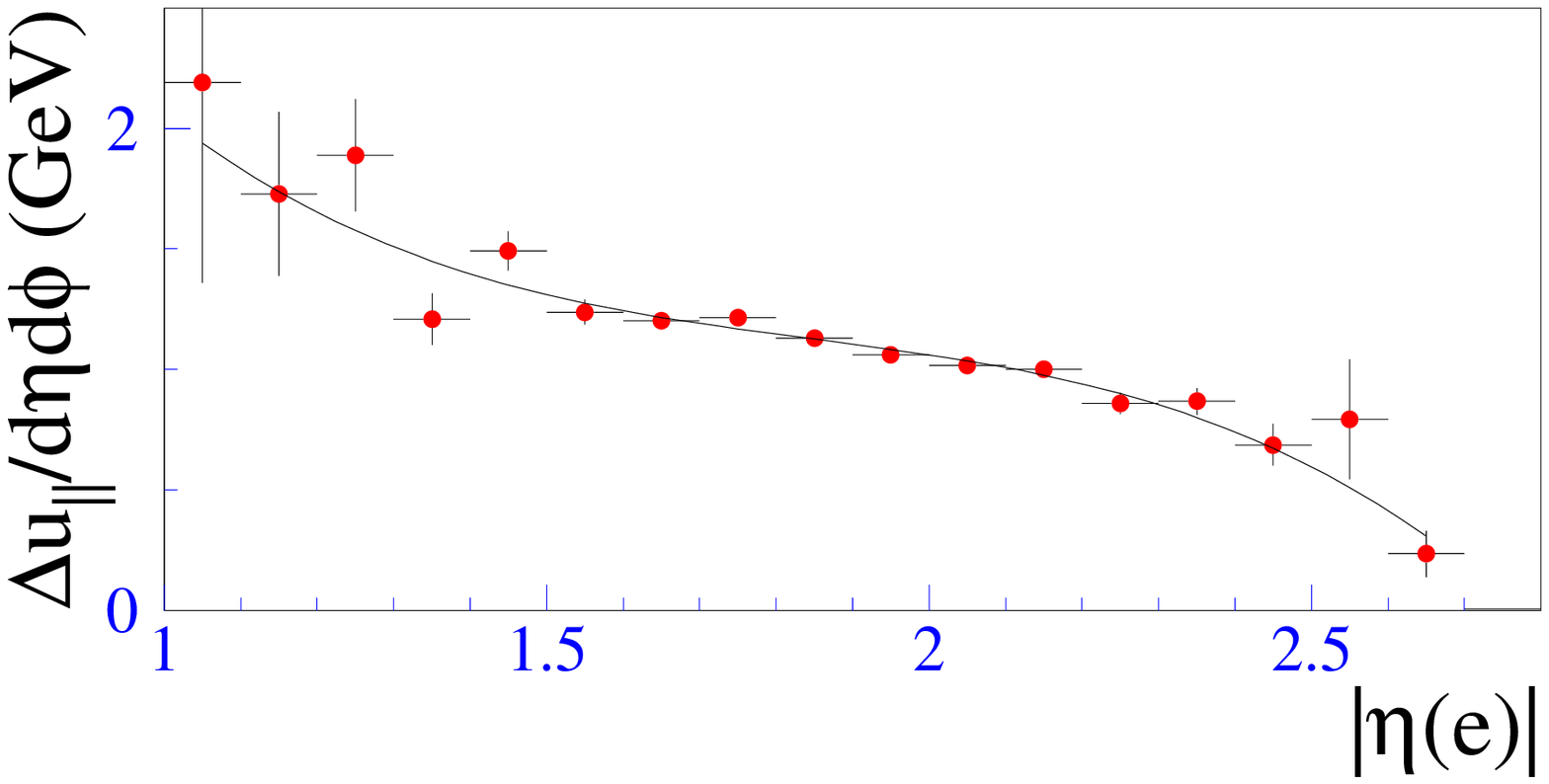}}
\vspace{0.1in}
\caption{ The variation of  \mdupar\ as a function of electron $\eta$.  }
\label{ue_ec_eta1}
\end{figure}

At higher luminosity the average number of interactions per event increases and
therefore \dupar/$\delta \eta \delta \phi$ increases (Fig.~\ref{fig:duparlum}).
The mean value of \dupar/$\delta \eta \delta \phi$ increases by 40~MeV per
10$^{30}$cm$^{-2}$s$^{-1}$. The underlying event energy flow into the electron
 cone depends on the electron $\eta$, as shown in Fig.~\ref{ue_ec_eta1},
 corrected back to zero luminosity.
 
\begin{figure}[htpb!]
\vspace{-0.3in}
\epsfxsize=3.0in
\centerline{\epsfbox{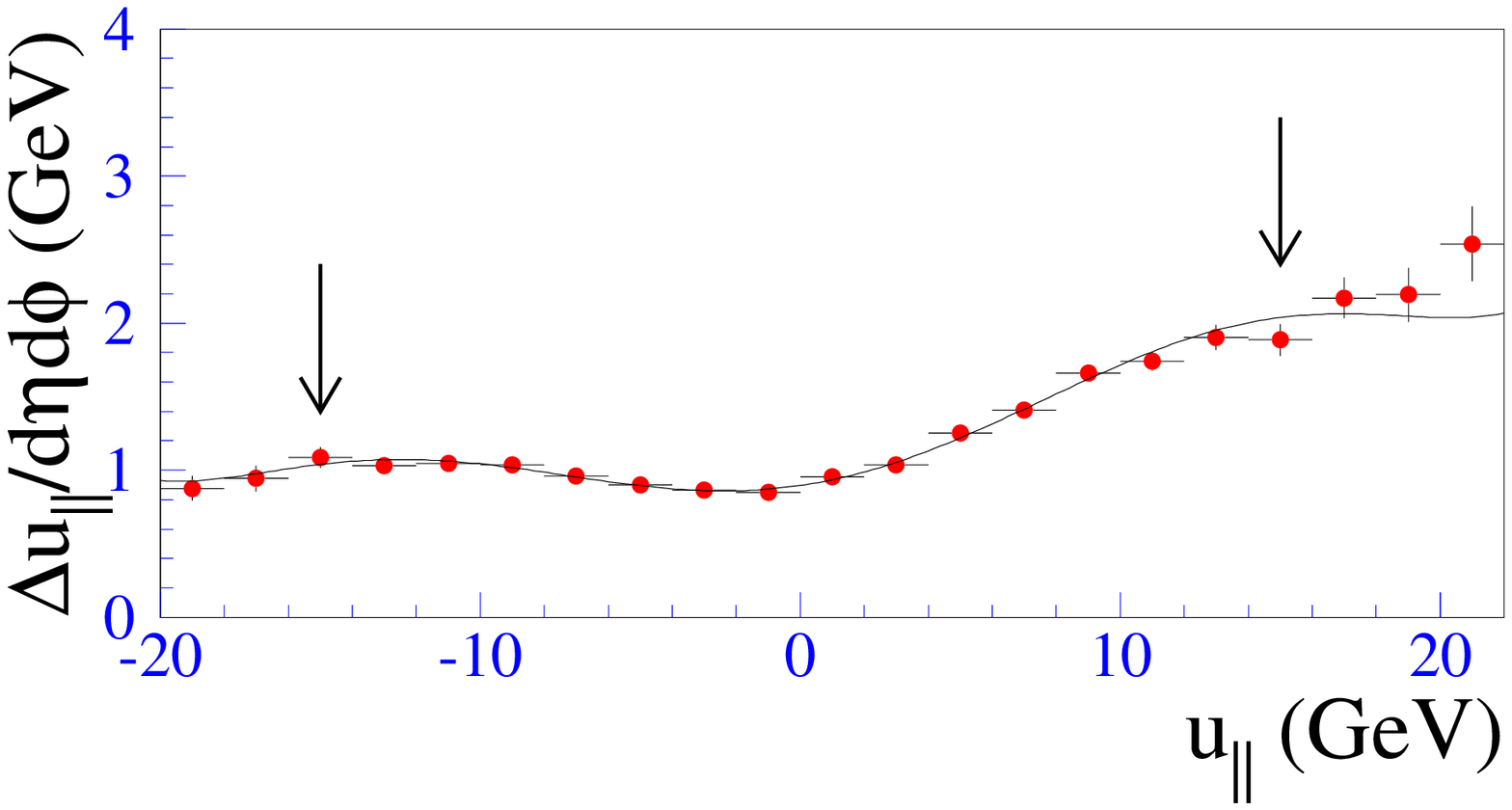}}
\vspace{0.1in}
\caption{ The variation of  \mdupar\ as a function of \upar.
The region between the arrows is populated by the \wb\ boson sample.  }
\label{fig:duupar}
\end{figure}

\begin{figure}[htpb!]
\vspace{-0.3in}
\epsfxsize=3.0in
\centerline{\epsfbox{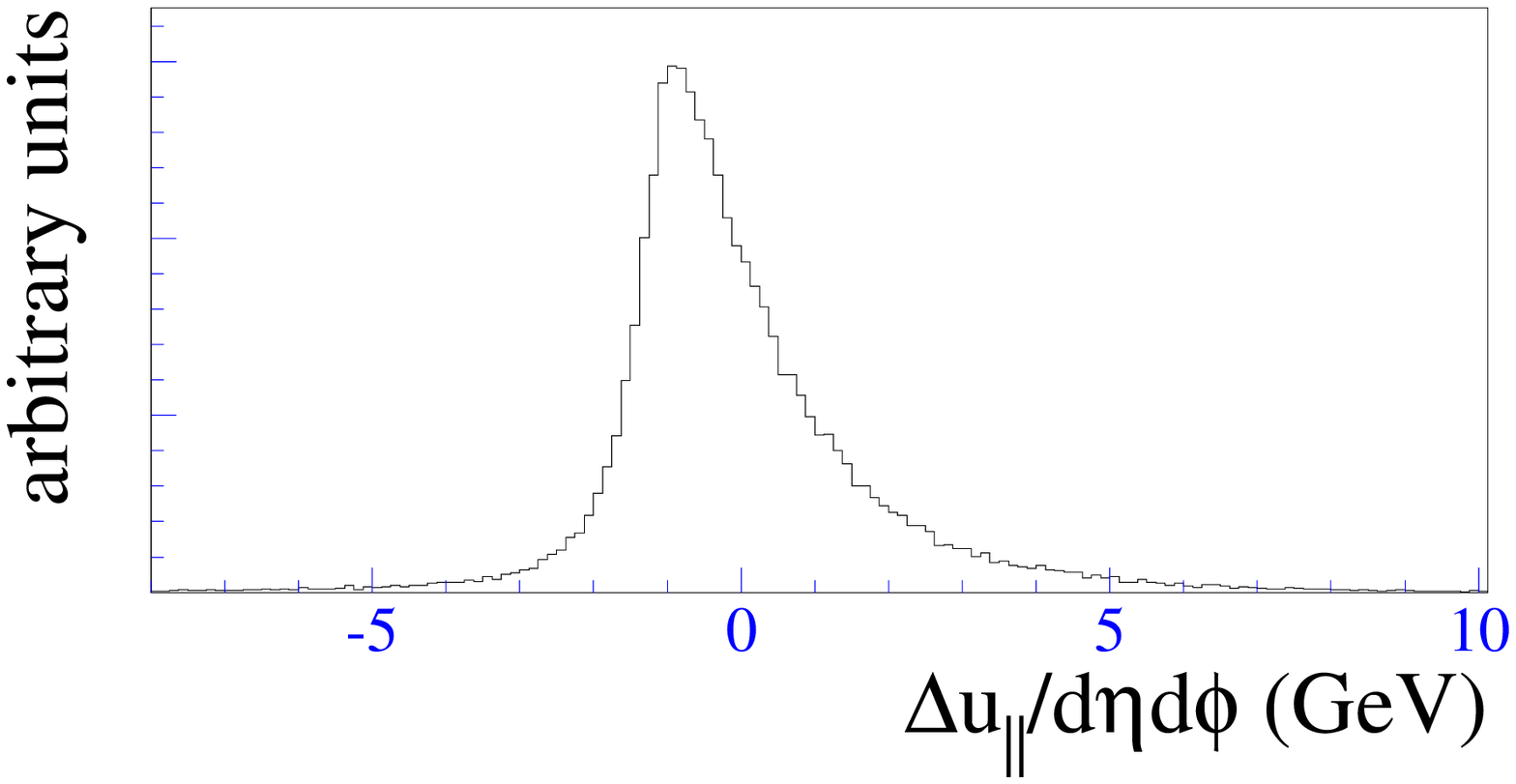}}
\vspace{0.1in}
\caption{ The distribution of \dupar/$\delta \eta \delta \phi$ in
the \wb\ signal sample, corrected to $\cal L$=0, $\mid \! \eta \! 
\mid = 2$, \upar =0 .}
\label{fig:deltaupar}
\end{figure}

The underlying event energy flow into the electron
cone also depends on the overlap between the recoil and the electron. 
We have found that the best measure of the recoil overlap is the component of
 the total recoil in the direction of the electron, which is \upar. 
Figure~\ref{fig:duupar} shows 
$\langle \dupar/\delta \eta \delta \phi ({\cal L}=0, \mid \! \eta \! 
 \mid = 2.0 ) \rangle$, the mean value for
\dupar/$\delta \eta \delta \phi$ 
 corrected  to zero luminosity and $\mid \! \eta \! \mid = 2.0$, as
a function of \upar. In the fast Monte Carlo model,
  a value \dupar/$\delta \eta \delta \phi$ is picked
from the distribution shown in Fig.~\ref{fig:deltaupar} for every event, 
corrected for \upar, $\eta$,  and luminosity dependences, and then scaled
 by the $\delta \eta \delta \phi$ area of a 20 cm cone at the electron $\eta$. 

\begin{figure}[htpb!]
\vspace{0.1in}
\epsfxsize=3.0in
\centerline{\epsfbox{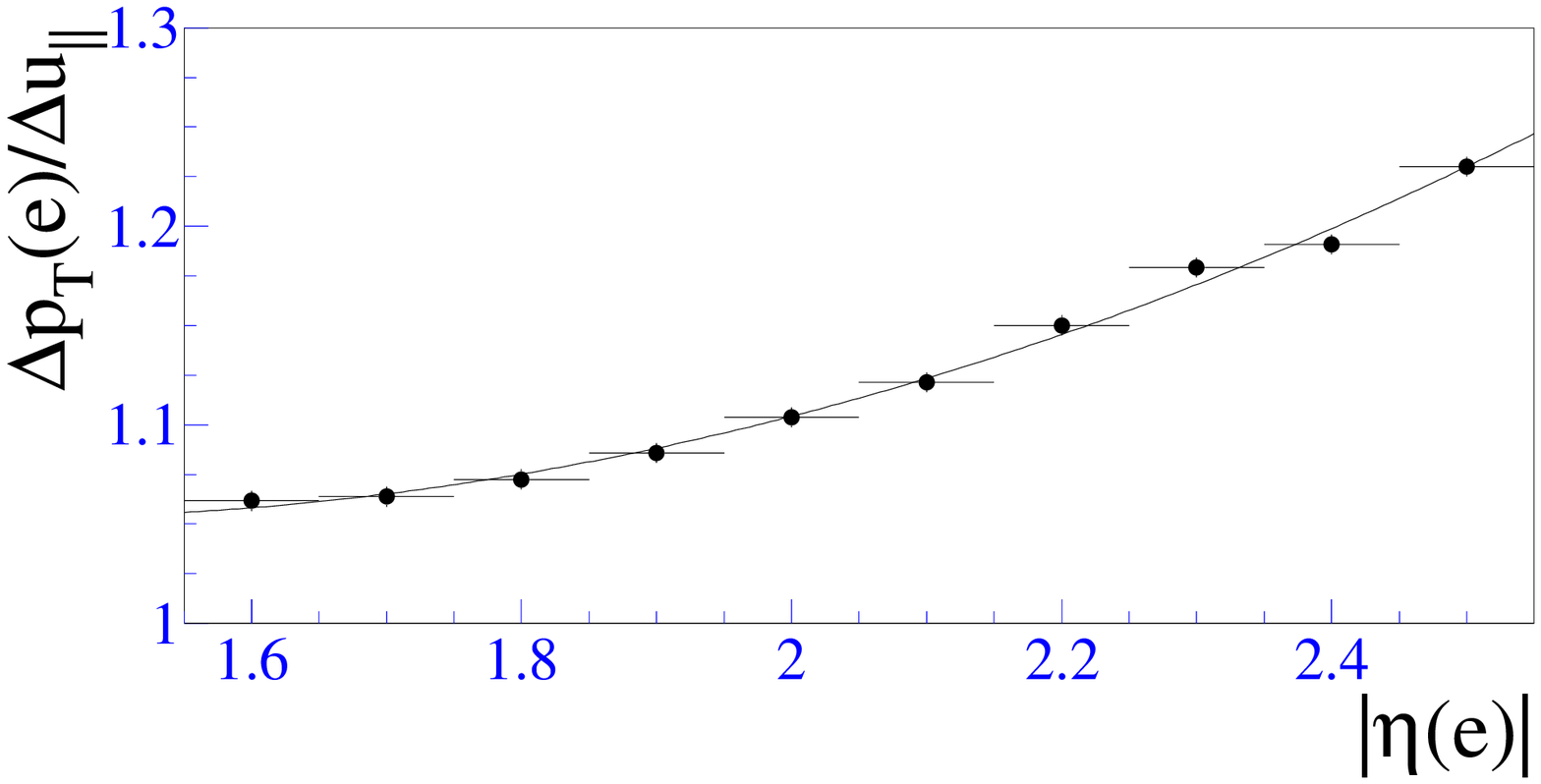}}
\vspace{0.1in}
\caption{ The ratio of the \mdupar\ corrections to the electron and the
 recoil as a function of electron $\eta$.  }
\label{ecue_elerec}
\end{figure}

 The measured electron transverse 
 energy is biased upwards  by the additional energy 
 $\Delta E_T$ in
 the window from the underlying event. $\Delta E_T$ is not equal to \dupar\ 
 because the electron $E_T$ is calculated by scaling the  sum of the
  cell energies by the electron angle, whereas \ut\  
 is obtained by summing the $E_T$ of each cell. The ratio of
 the two corrections as a function of electron $\eta$ is shown in 
 Fig.~\ref{ecue_elerec}.

 The uncertainty in the underlying event transverse energy density has a 
 statistical component (14 MeV) and a systematic component (24 MeV). 
 The systematic component
 is derived from the difference between the measurement close to the
 electron (where it is biased by the isolation requirement) and far from the
 electron (where it is not biased). The total uncertainty in the 
 underlying event transverse energy density is 28 MeV. 
\subsection { \boldmath  \upar\ \unboldmath Efficiency }
\label{sec-elec-upar}

The efficiency for electron identification depends on the 
 electron environment.
Well-isolated electrons are identified correctly more often than electrons near
other particles. Therefore \wb\ decays in which the electron is emitted in the
same direction as the particles recoiling against the \wb\ boson
 are selected
less often than \wb\ decays in which the electron is emitted in the direction
opposite the recoiling particles. This causes a bias in the lepton \pt\
distributions, shifting \pte\ to larger values and
\ptnu\ to lower values, whereas the \mt\ distribution is only slightly
affected.

\begin{figure}[htpb!]
\vspace{0.1in}
\epsfxsize=3.0in
\centerline{\epsfbox{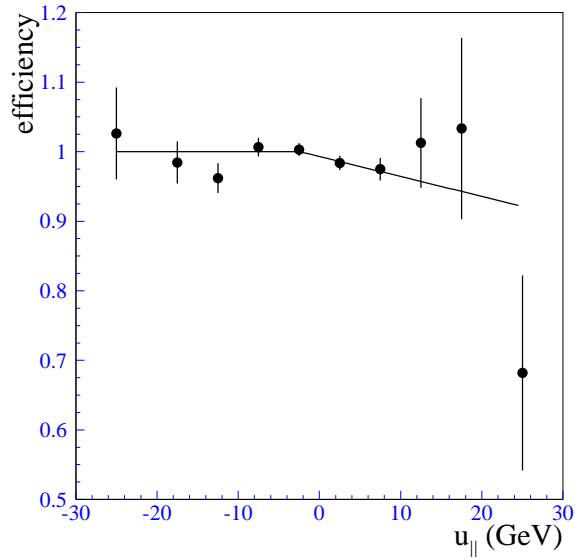}} 
\vspace{0.1in}
\caption{ The EC electron selection efficiency as a function of \upar.}
\label{fig:upareff}
\end{figure}

 We measure the electron finding efficiency as a function of \upar\
 using $Z \rightarrow ee$ events. The $Z$ event is tagged with one electron, 
 and the other electron provides an unbiased measurement of the efficiency.
 Following background subtraction, the measured  efficiency
is shown in Fig.~\ref{fig:upareff}. The line is a fit to a function
of the form
\begin{eqnarray}
\varepsilon(\upar)
= \varepsilon_0 \left\{ \begin{array}{ll} 1 & \hbox{for $\upar<u_0$} \\
                     1-s(\upar-u_0) & \hbox{otherwise.} \end{array} \right.
\end{eqnarray}
The parameter $\varepsilon_0$ is an overall efficiency which is
inconsequential for the \wb\ mass measurement, $u_0$ is the value of \upar\
at which the efficiency starts to decrease as a function of \upar, and $s$ is
the rate of decrease. We obtain the best fit for $u_0=-2.4$~GeV and
$s=0.0029$~GeV$^{-1}$. These two values are strongly anti-correlated. The
error on the slope 
 $\delta s=\pm 0.0012$~GeV$^{-1}$  accounts for the statistics
 of the $Z$ sample. 

\subsection { Electron Energy Response }
\label{sec-EMresponse}

Equation \ref{eq:emresponse} relates the reconstructed electron energy to the
recorded end calorimeter signals. Since the values for the constants were
determined in the test beam, we determine the offset \deltaecem\ and a
scale \alphaecem, 
which essentially modifies $A$, {\em in situ} with collider $Z 
\rightarrow ee$
data.

The electrons from \zb\ decays are not monoenergetic and therefore we can make
use of their energy spread to constrain \deltaecem. 
When both electrons are in the EC, 
 we can write
\begin{equation}
\mee = \alphaecem \mz + f_Z \deltaecem\
\end {equation}
for $\deltaecem \ll E(e_1)+E(e_2)$.
 $f_Z$ is a kinematic function related to the boost of the $Z$ boson,
  and is given by
 $f_Z=[E(e_1)+E(e_2)] (1-\cos\omega)/\mee$, where $\omega$ is the opening
 angle between the two electrons. When one electron is in the CC and one is
 in the EC, we can write
\begin{equation}
\mee = \sqrt{\alphaccem \alphaecem} \mz + f_Z \deltaecem,
\end {equation}
 where  $f_Z=E(e_2) (1-\cos\omega)/\mee$ and $e_2$ is the CC electron. 
 When we apply this formula, we have already corrected the CC electron for the
 corresponding CCEM offset, \deltaccem\ = $-0.16$ GeV, which was measured for
 our CC $W$ mass analysis \cite{wmass1bcc}. \alphaccem\ is the CC 
 electromagnetic energy scale, which is determined by fitting the \mee\ 
 spectrum of the CC/CC \zb\ sample.  

\begin{figure}[htpb!]
\vspace{-0.0in}
\epsfxsize=3.0in
\centerline{\epsfbox{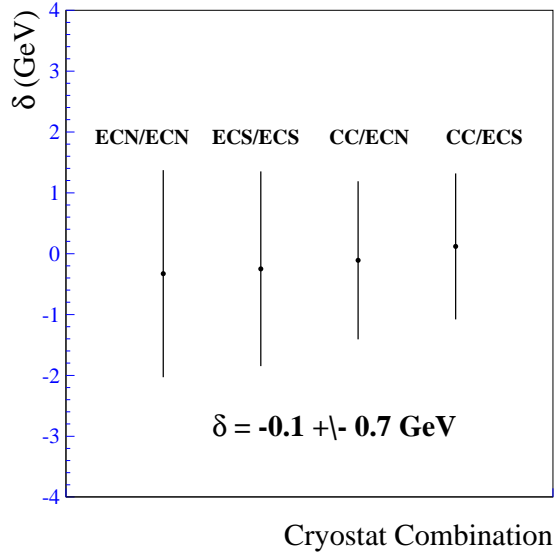}} 
\vspace{0.2in}
\caption{ The ECEM
 offset measurements using the CC/EC and EC/EC \zb\ samples. The labels 
 indicate the calorimeter cryostat in which each of the $Z$ decay electrons
 was detected. CC indicated the central calorimeter and ECN (ECS) indicates
 the north (south) end calorimeter respectively. }
\label{offset2}
\end{figure}

 We plot \mee\ versus $f_Z$
 and extract \deltaecem\ as the slope of the 
 fitted straight line. We use the fast Monte Carlo to correct for residual
 biases introduced by the kinematic cuts. The \deltaecem\ measurements from the
 CC/EC and EC/EC $Z$ samples are shown in Fig.~\ref{offset2} along with the
 statistical uncertainties. We obtain
 the average
 $\deltaecem =-0.1\pm0.7$~GeV. The uncertainty in this measurement of
\deltaecem\ is dominated by the statistical uncertainty due to the finite size
of the \zb\ sample. As Fig.~\ref{offset2} shows, the offsets measured in the
 north and south end calorimeters separately are completely consistent. 

\begin{figure}[htpb!]
\vspace{-0.0in}
\epsfxsize=3.0in
\centerline{\epsfbox{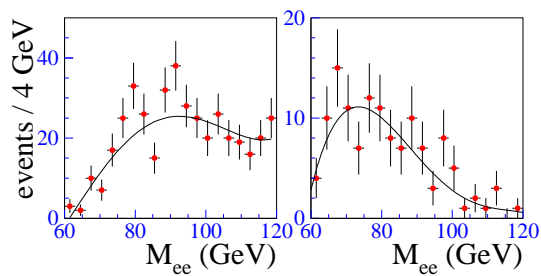}}
\vspace{0.1in}
\caption{ The dielectron mass spectrum from the CC/EC (left) and EC/EC (right)
 samples of 
 events with two EM clusters that fail the electron quality cuts.
The superimposed curves shows the fitted functions used to model the
 shape of the background in the \zb\ samples.}
\label{fig:zbkg}
\end{figure}

After correcting the data with this value of 
\deltaecem\ we determine \alphaecem\ so that the position of the \zb\ peak
predicted by the fast Monte Carlo agrees with the data. To determine the scale
factor that best fits the data, we perform a maximum likelihood fit to 
the \mee\
spectrum between 70~GeV and 110~GeV. In the resolution function we allow for 
 background shapes determined from  samples of events with two EM
clusters that fail the electron quality cuts (Fig.~\ref{fig:zbkg}).
The background normalization is obtained from the sidebands of the $Z$ peak.

\begin{figure}[htpb!]
\vspace{0.1in}
\epsfxsize=3.0in
\centerline{\epsfbox{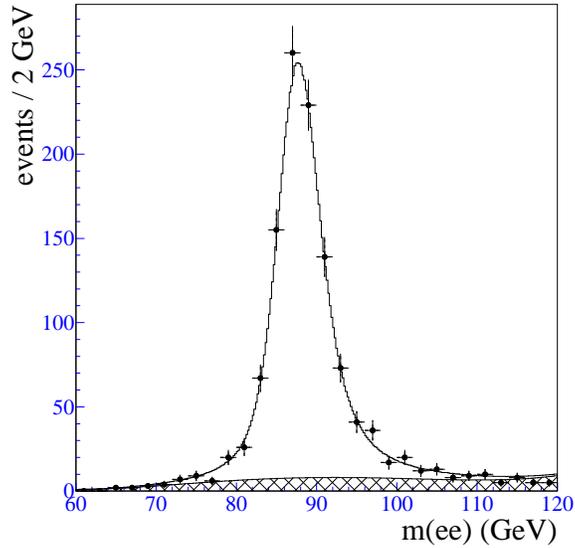}}
\vspace{-0.02in}
\caption{ The dielectron mass spectrum from the CC/EC \zb\ sample.
The superimposed curve shows the maximum likelihood fit and the shaded region
the fitted background.}
\label{ccec_z_mass}
\end{figure}

Figure~\ref{ccec_z_mass} shows the \mee\ spectrum for the CC/EC \zb\ sample 
 and the
Monte Carlo spectrum that best fits the data for $\deltaecem = - 0.1$~GeV.
The $\chi^2$ for the best fit to the CC/EC \mee\ spectrum is 14 for 19
degrees of freedom. For
$\alphaecem = 0.95143 \pm 0.00259$,
the \zb\ peak position of the CC/EC sample
 is consistent with the known \zb\ boson mass.
The error reflects the statistical uncertainty. The background
 has no measurable effect on the
result.

\begin{figure}[htpb!]
\vspace{-0.0in}
\epsfxsize=3.0in
\centerline{\epsfbox{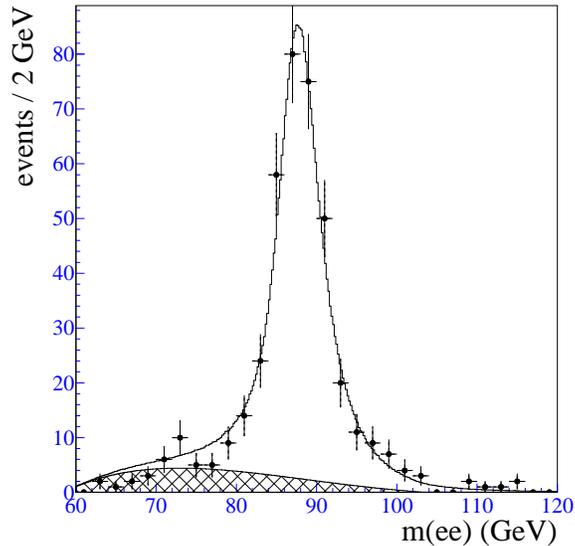}}
\vspace{-0.02in}
\caption{ The dielectron mass spectrum from the EC/EC \zb\ sample.
The superimposed curve shows the maximum likelihood fit and the shaded region
the fitted background.}
\label{ecec_z_mass}
\end{figure}

Figure~\ref{ecec_z_mass} shows the \mee\ spectrum for the EC/EC \zb\ sample 
 and the
Monte Carlo spectrum that best fits the data for $\deltaecem = - 0.1$~GeV.
The $\chi^2$ for the best fit to the EC/EC \mee\ spectrum is 12 for 17
degrees of freedom. For $\alphaecem = 0.95230 \pm 0.00231$,  
the \zb\ peak position of the EC/EC sample
 is consistent with the known \zb\ boson mass.
The error reflects the statistical uncertainty and the uncertainty in the
 background. 

Combining the \alphaecem\ measurements from the CC/EC and the EC/EC \zb\
 samples, we obtain the ECEM energy scale
\begin{equation}
\alphaecem = 0.95179 \pm 0.00187 \quad .
\label{eq:EMscale}
\end{equation}
The difference between the ECEM scales measured separately in the north and
 south calorimeters is $0.0040 \pm 0.0037$, consistent with the calorimeters
 having the same EM response. 

\subsection { Electron Energy Resolution }
\label{sec-elec-res}

\begin{figure}[htpb!]
\vspace{0.3in}
\epsfxsize=3.0in
\centerline{\epsfbox{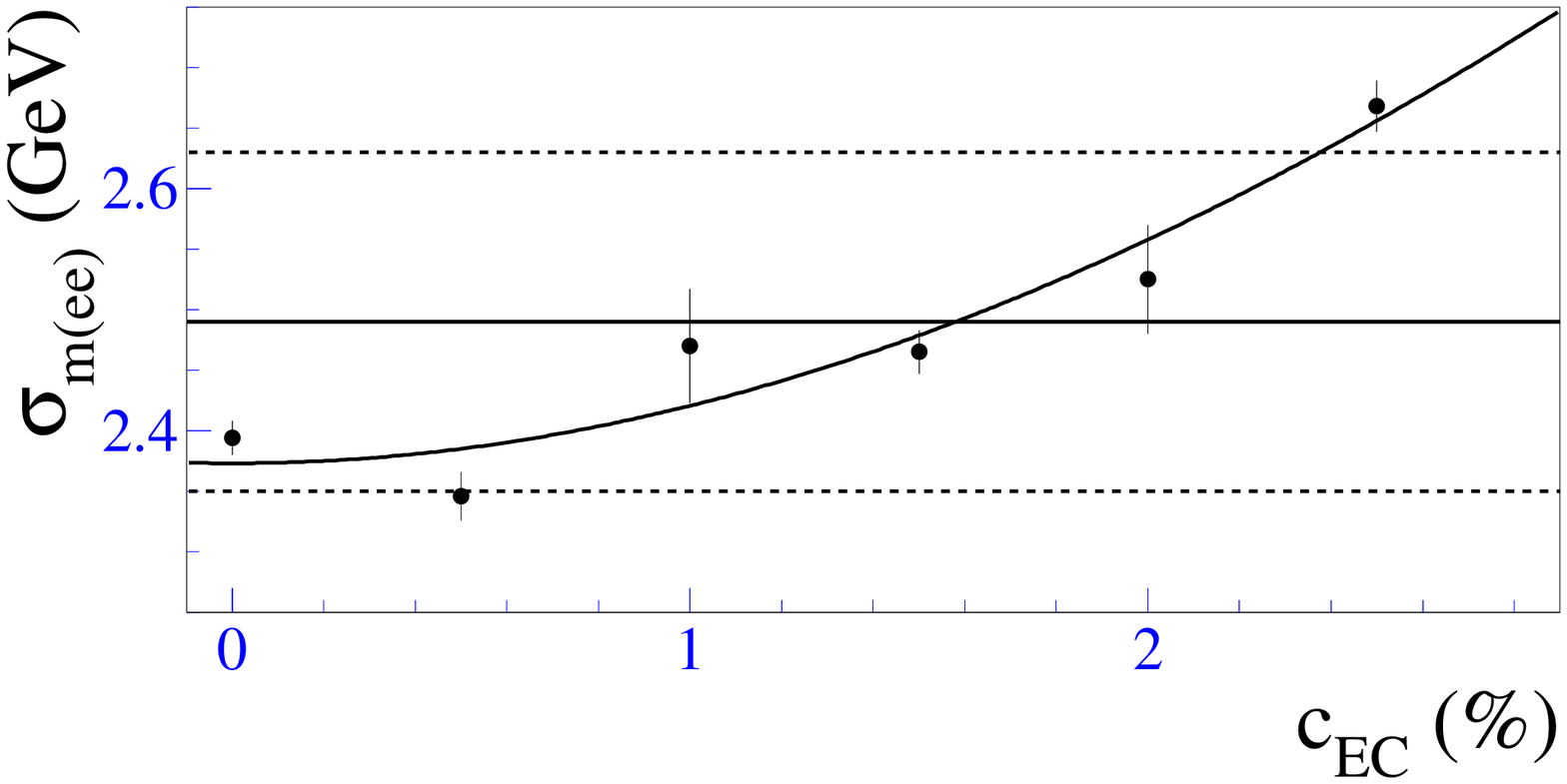}}
\vspace{0.0in}
\epsfxsize=3.0in
\centerline{\epsfbox{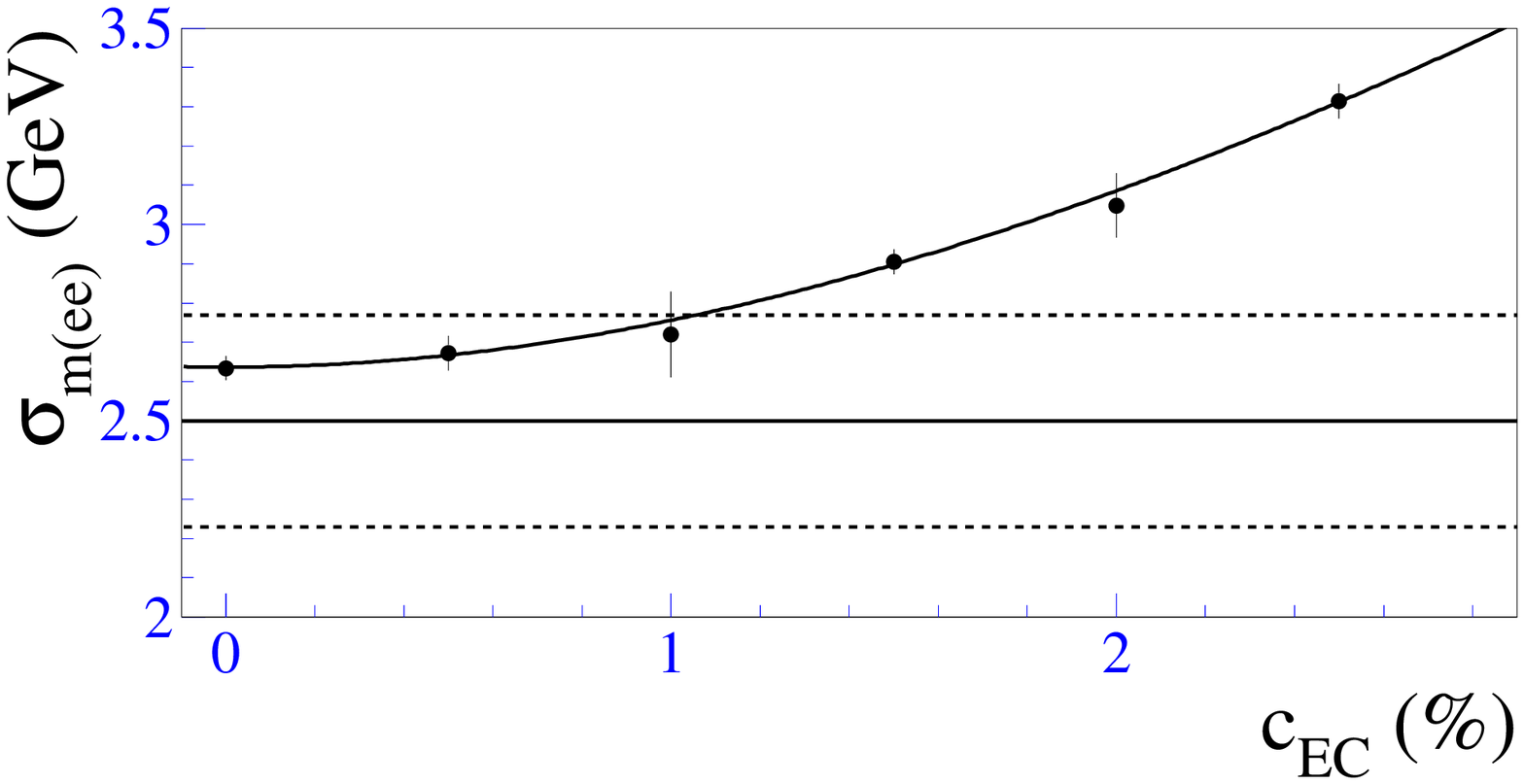}}
\vspace{-0.02in}
\caption{ The dielectron mass resolution versus
the constant term \cem. The top plot is for the
 CC/EC \zb\ events and the bottom plot is for the 
 EC/EC \zb\ events. }
\label{fig:cem}
\end{figure}

Equation~\ref{eq:emresolution} gives the functional form of the electron energy
resolution. We take the intrinsic resolution of the end
 calorimeter, which is given
by the sampling term \sem, from the test beam measurements. The noise term
\nem\ is represented by the width of the electron underlying event
 energy distribution
(Fig.~\ref{fig:deltaupar}).
We measure the constant term \cem\ from the \zb\ line
shape of the data. We fit a
Breit-Wigner convoluted with a Gaussian, whose width characterizes the
dielectron mass resolution, to the \zb\ peaks for the CC/EC and EC/EC samples
 separately. Figure~\ref{fig:cem} shows the
width $\sigma_{\mee}$ of the Gaussian fitted to the \zb\ peak predicted by the
fast Monte Carlo as a function of \cem. The horizontal lines indicate the
width of the Gaussian fitted to the \zb\ samples and its uncertainties. 
For the data measurements of 
\begin{eqnarray}
\sigma_m = 2.47 \; \pm \; 0.05 \; {\rm GeV} \; {\rm (CC/EC)} \nonumber \\
\sigma_m = 2.72 \; \pm \; 0.11 \; {\rm GeV} \; {\rm (EC/EC)} 
\end{eqnarray}
we extract  
 from the CC/EC \zb\ boson events 
 $ c_{\rm EC} = 1.6^{+0.8}_{-1.6}$\% and from the EC/EC \zb\ events we 
 extract $c_{\rm EC} = 0.0^{+1.0}_{-0.0}$\%. 
We take the combined measurement to
 be 
\begin{equation}
c_{\rm EC} = 1.0 ^ {+0.6} _ {-1.0} \; {\rm \%}.
\label{cecem}
\end{equation} 
The measured \zb\ boson mass does not depend on $c_{\rm EC}$.
\section{ Recoil Measurement }
\label{sec-recoil}
\subsection { Recoil Momentum Response }

The detector response and resolution for particles recoiling against a \wb\
boson should be the same as for particles recoiling against a \zb\ boson. For
\zee\ events, we can measure the transverse momentum of the \zb\ boson
 from the
$e^+e^-$ pair, \ptee,  into
which it decays,
  and from the recoil momentum \ut\ in the same way as for \wev\
events. By comparing \ptee\ and \ut, we calibrate the recoil response relative
to the electron response.

The recoil momentum is carried by many particles, mostly hadrons, with a wide
momentum spectrum. Since the response of the calorimeter to hadrons is slightly
nonlinear at low energies, 
 and the recoil particles see a reduced response at module boundaries,
  we expect a momentum-dependent response function with values below unity.  
 To fix the functional form of the recoil momentum response, we studied
 \cite{wmass1bcc} 
 the response predicted by a Monte Carlo \zee\ sample obtained using the \HERW\
program and a \GEAN-based detector simulation. We projected the
reconstructed transverse recoil momentum onto the transverse
 direction of motion of the
\zb\ boson and define the response as
\begin{eqnarray}
    \rrec = \frac{\left| \utv\cdot\hat{q}_T \right|}{\left| \qt \right|},
\end{eqnarray}
where \qt\ is the generated transverse momentum of the \zb\ boson.
 A response function of the form
\begin{equation}
\rrec = \alpharec + \betarec \ln \left(\qt/\hbox{GeV}\right)
\end {equation}
fits the response predicted by \GEAN\ with $\alpharec =0.713\pm0.006$ and
$\betarec =0.046\pm0.002$.
This functional form also describes the jet energy response~\cite{jetscale}
  of the \Dzero\
calorimeter.

\begin{figure}[htpb!]
\vspace{-0.0in}
\epsfxsize=3.0in
\centerline{\epsfbox{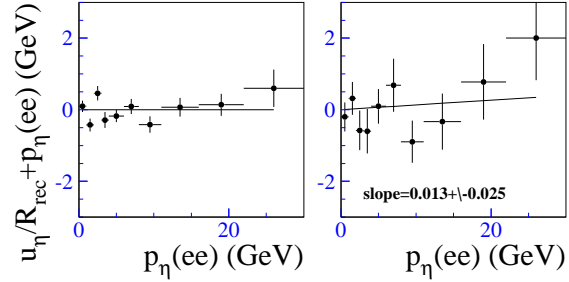}}
\vspace{-0.in}
\caption{ The recoil momentum response in the CC/CC + CC/EC (left)
  and the EC/EC 
 (right)   \zb\ samples as a function of $p_\eta (ee)$.}
\label{rec_scale_data}
\end{figure}

 The recoil response for data was calibrated against the electron response by
 requiring $p_T$ balance in $Z \rightarrow ee$ decays for our published
 CC analysis \cite{wmass1bcc}. The \zb\ boson $p_T$
 measured with the electrons and the recoil are projected on the $\eta$ axis,
 defined as the bisector of the two electron directions in the transverse
 plane. From the CC/CC + CC/EC 
 \zb\ boson events, we measured $\alpharec =0.693\pm0.060$ 
 and
 $\betarec =0.040\pm0.021$, in good agreement with the Monte Carlo prediction.
 To compare the
 recoil response measured with \zb\ events of different topologies, we scale
 the recoil measurement with the inverse of the response parametrization
\begin{equation}
R_{\rm rec} = 0.693 \; + \; 0.04 \cdot \ln \; (p_T(ee)/{\rm GeV})
\label{rrec}
\end{equation}
 and plot the sum of the projections versus $p_\eta (ee)$, as shown in 
 Fig.~\ref{rec_scale_data}. We see no $p_\eta (ee)$ dependence to 
 the $p_\eta$ balance measured using the \zb\ boson
  events with at least one central
 electron, since this sample was used to derive the values of these parameters.
 The EC/EC \zb\ boson  events give a recoil response measurement 
 statistically consistent with the above. Hence we use the same recoil 
 response for the EC and  the CC \wb\ boson events \cite{wmass1bcc}.

\subsection { Recoil Momentum Resolution }

\begin{figure}[htpb!]
\vspace{-0.0in}
\epsfxsize=3.0in
\centerline{\epsfbox{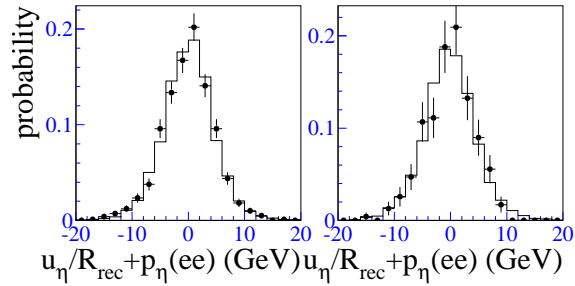}}
\vspace{0.1in}
\caption{ The $\eta$-balance distribution for the \zb\ boson
 data ($\bullet$) and the
fast Monte Carlo simulation (-----). The  plot on the left is for the
 CC/CC + CC/EC \zb\ events and the plot on the right is for the 
 EC/EC \zb\ events. }
\label{fig:eta_balance}
\end{figure}

\begin{figure}[htpb!]
\vspace{-0.0in}
\epsfxsize=3.0in
\centerline{\epsfbox{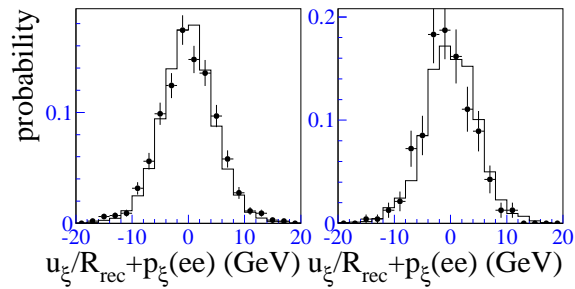}}
\vspace{0.1in}
\caption{ The $\xi$-balance  distribution for the \zb\ boson 
 data ($\bullet$) and
the fast Monte Carlo simulation (-----). The  plot on the left is for the
 CC/CC + CC/EC \zb\ events and the plot on the right is for the 
 EC/EC \zb\ events. }
\label{fig:xi_balance}
\end{figure}

 The widths of the $p_\eta$ balance and the $p_\xi$ balance (where the $\xi$
 axis is perpendicular to the $\eta$ axis) are sensitive to the recoil 
 resolution.  Figures~\ref{fig:eta_balance}--\ref{fig:xi_balance}
 show the  comparison between 
 the data and Monte Carlo for the recoil resolution determined in our CC \wb\ 
 mass analysis \cite{wmass1bcc}. The $p_\eta$ balance
 width is in good agreement between data and Monte Carlo for all \zb\ 
 boson  topologies. 
 Hence we use the same recoil resolution for EC \wb\ boson 
 events as for the CC \wb\
  boson events
 \cite{wmass1bcc}. 

\subsection {Comparison with \boldmath \wb\ \unboldmath Boson Data}

\begin{figure}[htpb!]
\vspace{-0.0in}
\epsfxsize=3.0in
\centerline{\epsfbox{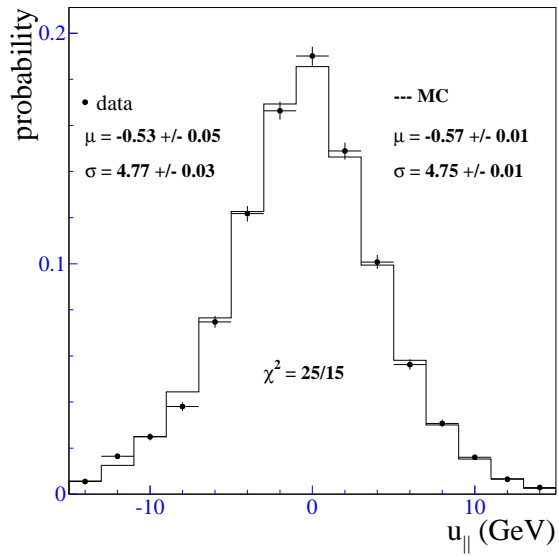}}
\vspace{0.1in}
\caption{ The \upar\ spectrum for the \wb\ data
($\bullet$) and the Monte Carlo simulation (-----). The mean ($\mu$) and r.m.s.
 ($\sigma$) of the distributions and the $\chi^2$/dof is also shown. }
\label{fig:uparproj}
\end{figure}

\begin{figure}[htpb!]
\vspace{-0.0in}
\epsfxsize=3.0in
\centerline{\epsfbox{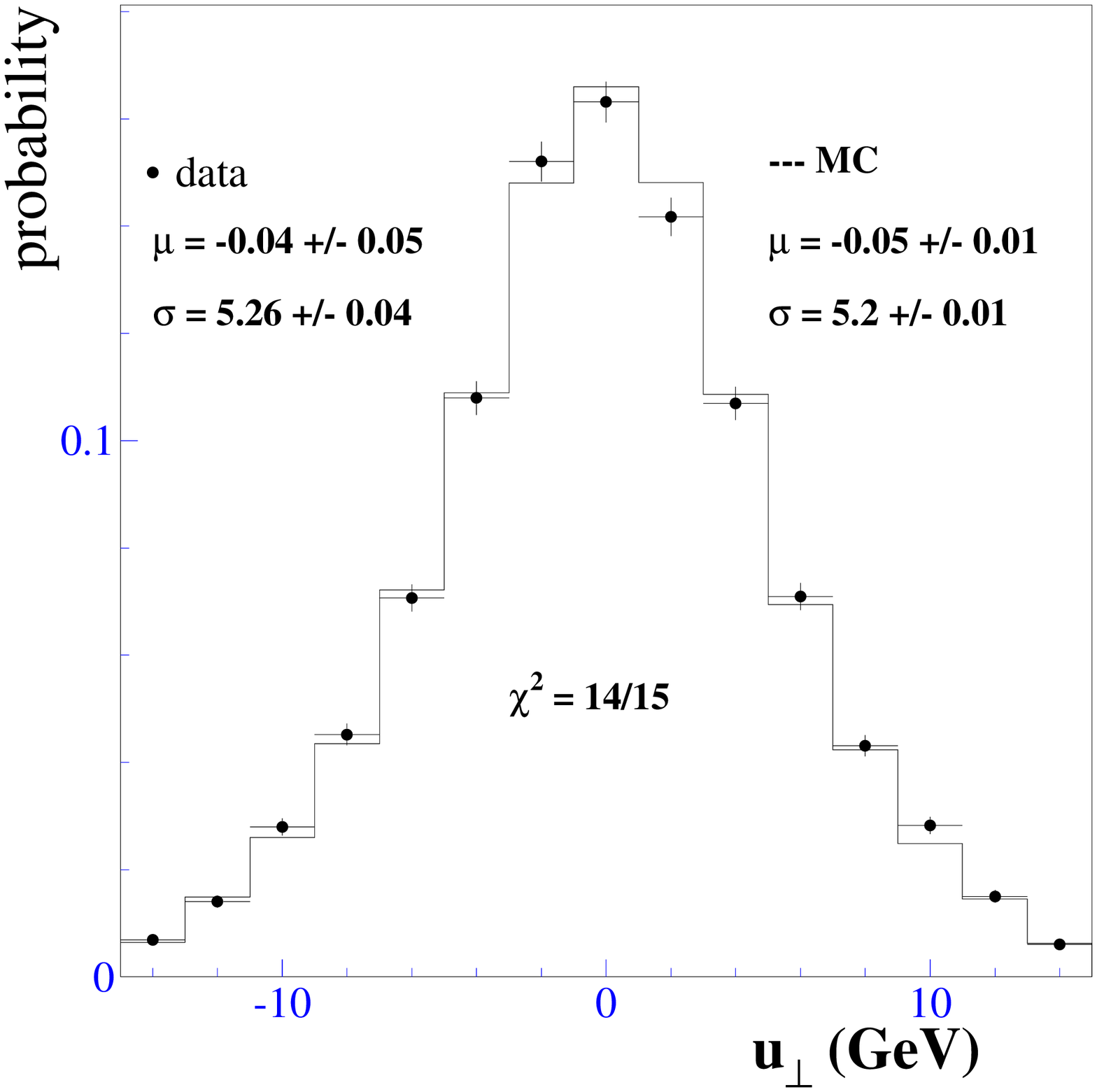}}
\vspace{0.1in}
\caption{ The \uper\ spectrum for the \wb\ data
($\bullet$) and the Monte Carlo simulation (-----). The mean ($\mu$) and r.m.s.
 ($\sigma$) of the distributions and the $\chi^2$/dof is also shown. }
\label{fig:uperproj}
\end{figure}

\begin{figure}[htpb!]
\vspace{-0.0in}
\epsfxsize=3.0in
\centerline{\epsfbox{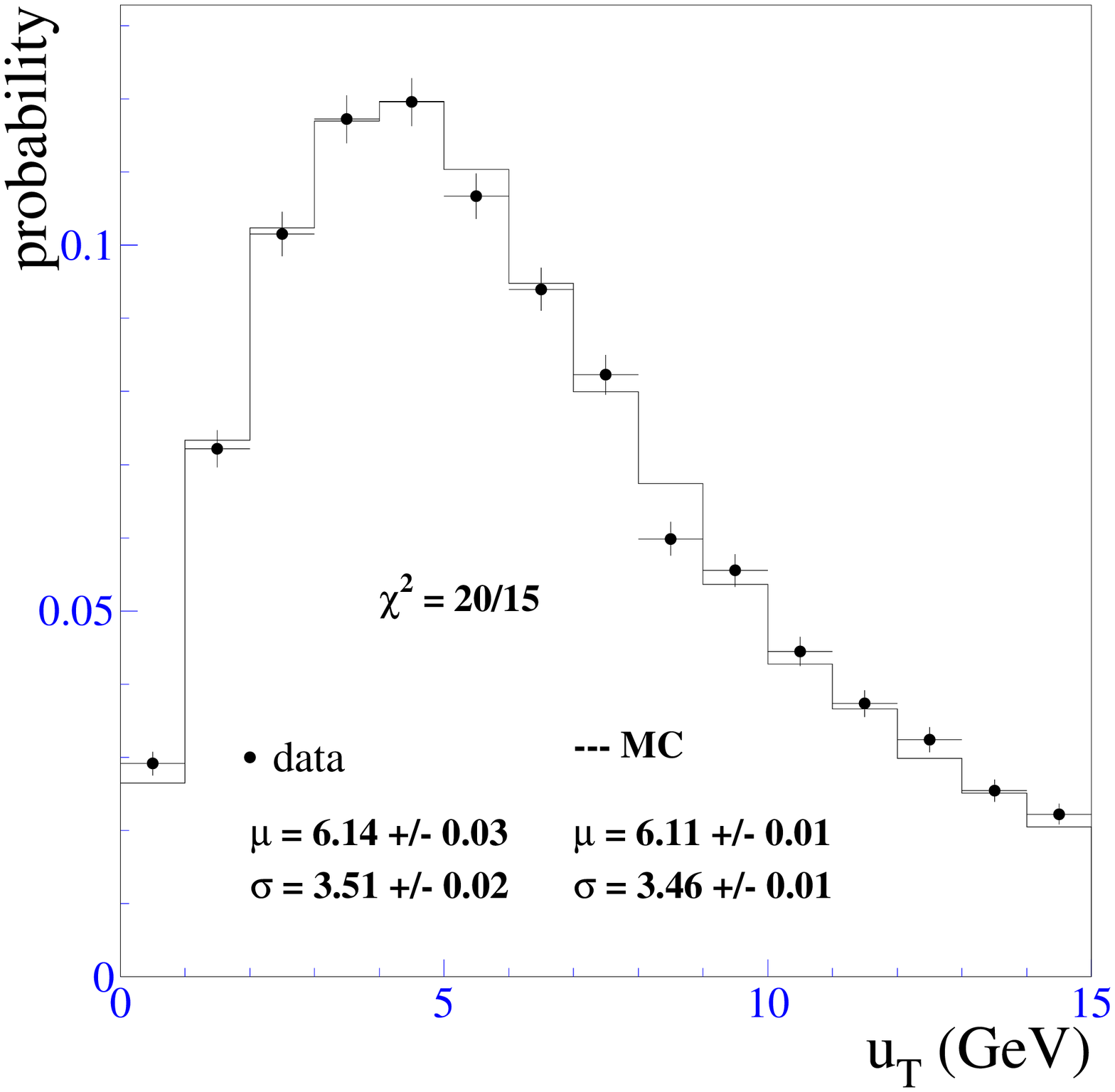}}
\vspace{0.1in}
\caption{ The recoil momentum (\ut) spectrum for the \wb\ data
($\bullet$) and the Monte Carlo simulation (-----). The mean ($\mu$) and r.m.s.
 ($\sigma$) of the distributions and the $\chi^2$/dof is also shown. }
\label{fig:ut}
\end{figure}

\begin{figure}[htpb!]
\vspace{-0.1in}
\epsfxsize=3.0in
\centerline{\epsfbox{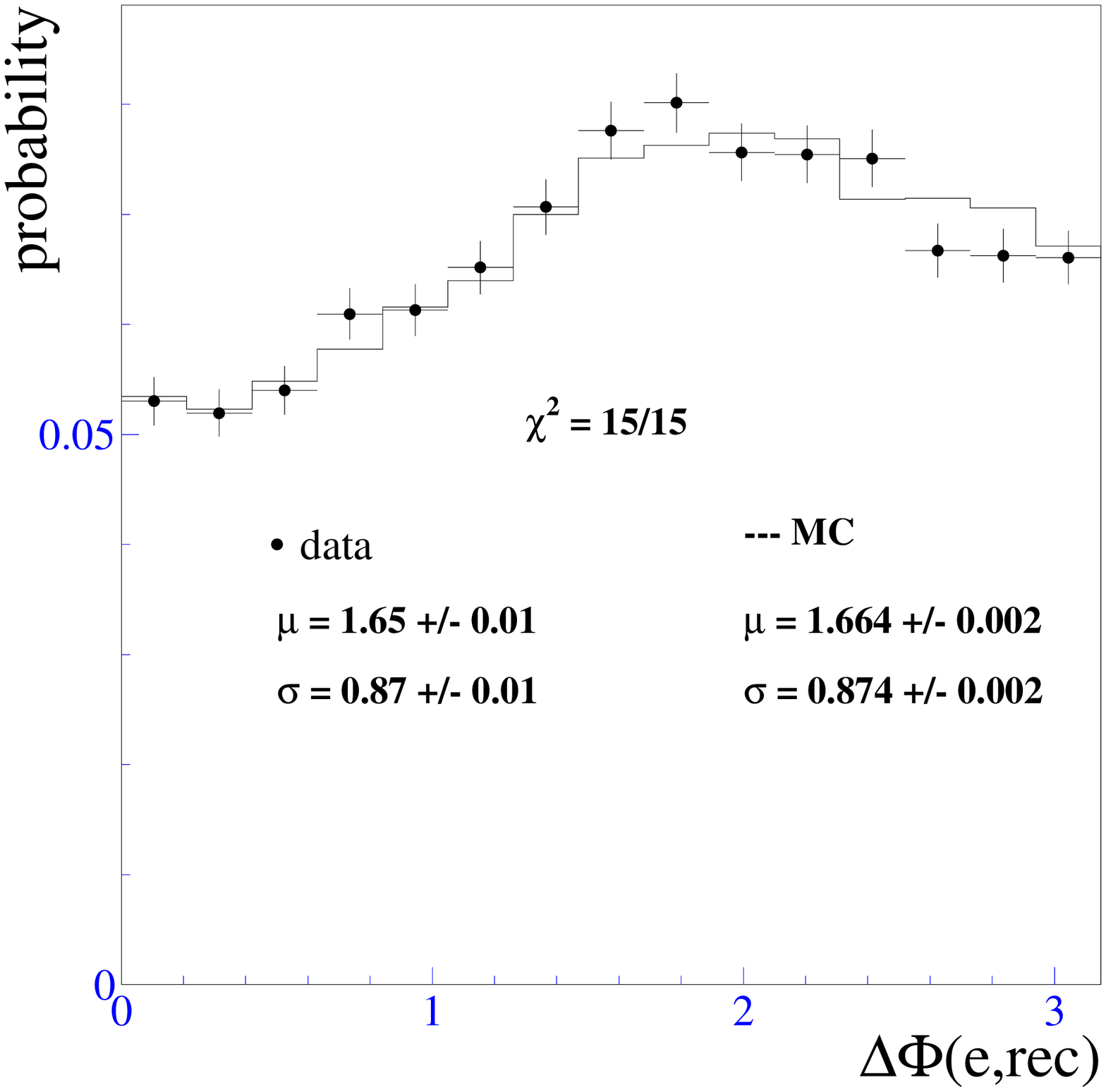}}
\vspace{0.1in}
\caption{ The azimuthal difference between electron and recoil directions
for the \wb\ data ($\bullet$) and the Monte Carlo simulation (-----).
The mean ($\mu$) and r.m.s.
 ($\sigma$) of the distributions and the $\chi^2$/dof is also shown. }
\label{fig:deltaphi}
\end{figure}

We compare the recoil momentum
distributions in the \wb\ boson
 data to the predictions of the fast Monte Carlo,
which includes the parameters described in this section and
Sec.~\ref{sec-elec}.
Figure~\ref{fig:uparproj} shows the \upar\ spectra from Monte Carlo and \wb\
data. The agreement means that the recoil momentum
response and resolution and the \upar\ efficiency parameterization
describe the data well. Figures \ref{fig:uperproj}--\ref{fig:deltaphi} show
\uper, \ut, and the azimuthal difference between electron and recoil directions
from Monte Carlo and \wb\ boson 
 data. The figures also show the mean and r.m.s. of
 the data and Monte Carlo distributions and the $\chi^2$ over the number of
 degrees of freedom (dof). 

\section{ Constraints on the \boldmath \wb\ \unboldmath 
 Boson Rapidity Spectrum}
\label{sec-constraints}
 In principle,
 if the acceptance for the $W \rightarrow e \nu$ decays were complete, 
 the transverse mass distribution or the lepton $p_T$ distributions would
 be independent of the $W$ rapidity. However, cuts on the electron angle in
 the laboratory frame cause the observed distributions of the transverse
 momenta to depend on the $W$ rapidity. Hence a constraint on the $W$ rapidity
 distribution is useful in constraining the production model uncertainty on 
 the \wb\ mass. 
\par
 The pseudorapidity distribution of the electron from $W \rightarrow e \nu$ 
 decays is correlated with the rapidity distribution of the $W$ boson.
 Therefore  we  can
 compare the electron $\eta$ distribution between data and Monte Carlo. 
\par
 To compare the data with the Monte Carlo, we need to correct for the
  jet background in the data and the electron identification efficiency 
 as a function of $\eta$. We obtain the jet background fraction as a 
 function of $\eta$ by counting the number of $W$ events 
 that fail electron cuts (see Sec.~\ref{hadbkg}) in bins of $\eta$, 
 subtracting the small contamination due to true electrons, and normalizing
 the entire distribution to the total background fraction (separately in the
 CC and EC). The normalized background $\eta$ distribution is subtracted from
 the $\eta$ distribution of the data.

\begin{figure}[tbhp]
\vspace{-0.0in}
\epsfxsize=3.0in
\centerline{\epsfbox{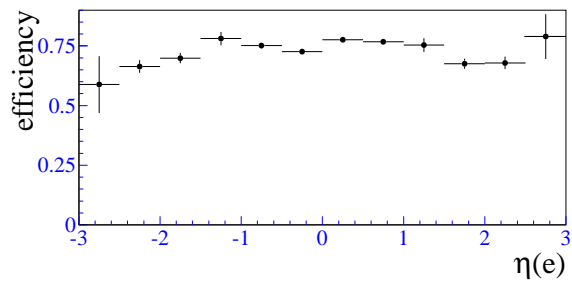}}
\vspace{0.1in}
\caption {Dependence of electron identification efficiency on electron 
 pseudorapidity. Statistical errors are shown. }
\label{eleeff_vs_eta}
\end{figure}

 The electron identification efficiency (after fiducial and kinematic cuts)
 is measured using the CC/CC and CC/EC
 $Z \rightarrow ee$ events. All the electron identification cuts are used
 to identify one electron to tag the event. Candidates are selected in the 
 mass range $81 < m_{ee} < 101$ GeV. Sidebands in the mass range $60 < m_{ee}
 < 70$ GeV and  $110 < m_{ee} < 120$ GeV are used for background subtraction. 
 The number of events in which the second electron also satisfies all the
 electron identification cuts is used to calculate the efficiency. The 
 efficiency measured in bins of the  $\eta$ of the second electron 
 is shown in Fig.~\ref{eleeff_vs_eta}. 

\begin{figure}[tbhp]
\epsfxsize=3.0in
\centerline{\epsfbox{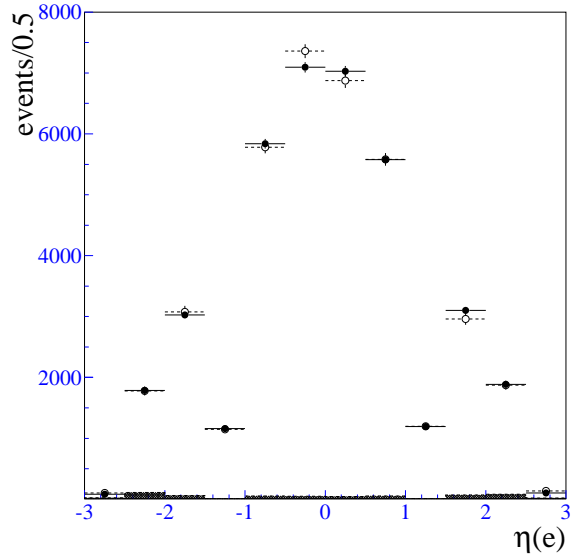}}
\caption { $\eta$ distribution of the electron from $W\rightarrow e \nu$
 decays from background-subtracted data ($\bullet$), efficiency-corrected
 Monte Carlo ($\circ$) and the jet background (shaded histogram). The
 distributions drop near $\mid \! \eta \! \mid =1.2$ because there is no EM 
 calorimetry in the range $1.1 < \mid \! \eta_{\rm det} \! \mid < 1.4$.}
\label{w_etae_datamc_1}
\end{figure}

\begin{figure}[tbhp]
\epsfxsize=3.0in
\centerline{\epsfbox{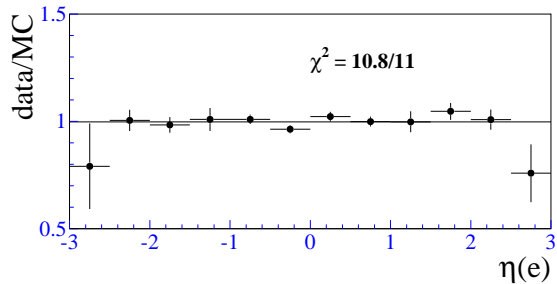}}
\vspace{0.1in}
\caption {The ratio of the background-subtracted data and efficiency-corrected
 Monte Carlo. The Monte Carlo has been normalized to the data. The $\chi^2$/dof
 is with respect to unity.  }
\label{w_etae_datamc_2}
\end{figure}

 We scale the electron $\eta$ distribution predicted by the Monte Carlo 
 by the $\eta$-dependent efficiency, and compare to the background-subtracted
  data in Fig.~\ref{w_etae_datamc_1}. 
 The errors on the Monte Carlo points include the
 statistical errors on the Monte Carlo sample and the statistical errors
 on the efficiency measurements. The errors on the data points include the
 statistical errors on the number of candidate events and the statistical
 errors on the background estimate which has been subtracted. 
 Figure~\ref{w_etae_datamc_2} shows the ratio between the 
 background-subtracted data and the efficiency-corrected Monte Carlo, with 
 the uncertainties
 mentioned above added in quadrature. The Monte Carlo has been normalized to 
 the data. The $\chi^2$/dof shown is with respect to unity. There is good
 agreement between the data and the Monte Carlo.  


\begin{figure}[tbhp]
\epsfxsize=3.0in
\centerline{\epsfbox{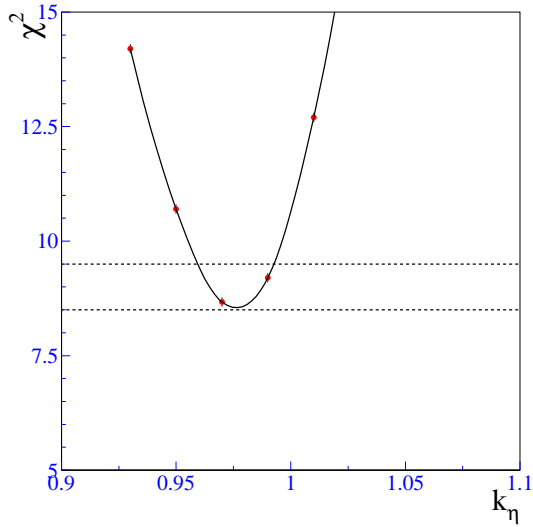}}
\vspace{0.1in}
\caption {$\chi^2$ of the electron $\eta$ distribution ratio between data and
 Monte Carlo from unity, as a function of the $W$ rapidity scale factor 
 $k_\eta$. There are 11 degrees of freedom. The Monte Carlo uses the 
 MRS(A$^\prime$) 
 parton distribution functions.
 The horizontal lines indicate $\chi^2_{\rm min}$
 and $\chi^2_{\rm min} + 1$.}
\label{chi2_vs_keta_mrsa}
\end{figure}

 To extract a constraint on the $y$ 
 distribution of the $W$ boson,
  we introduce in the Monte Carlo a scale factor
 as follows:
\begin{equation}
 y_W \rightarrow k_\eta \cdot y_W
\end{equation}
 i.e. the rapidity of the $W$ is scaled by the factor $k_\eta$. We then
 compute the $\chi^2$ between the data and Monte Carlo $\eta(e)$ distributions
 for different $k_\eta$. The result is shown in Fig.~\ref{chi2_vs_keta_mrsa}
  for the MRS(A$^ \prime$) \cite{mrsa}
  parton distribution functions.  Table~\ref{ketatable} shows
 the values of $k_\eta$ at which the $\chi^2$ is minimized for the different
 pdf's. 

 The uncertainty in $k_\eta$ is 1.6\%, which is the change in $k_\eta$ that
 causes the 
  $\chi^2$ to rise by one unit above the minimum. 
 We generate Monte Carlo events with different values of $k_\eta$ and fit them
 with templates generated with $k_\eta$ set to unity.
 For a $k_\eta$ variation of 1.6\%, 
 the variation of the fitted $W$
 mass in the EC is shown in Table~\ref{ketaerror}. 

\begin{table}[hbtp]
\caption{Value of $k_\eta$ giving the minimum $\chi^2$ 
for different pdf's.}
\medskip
\begin{center}
\begin{tabular}{cccc} 
\hline
  MRS(A$^\prime$) \cite{mrsa} & CTEQ3M \cite{cteq3m} & CTEQ2M \cite{cteq2m} & MRSD$- ^\prime$  \cite{mrsd} \\
\hline
0.975 & 0.98 & 0.985 & 0.99  \\
\hline
\end{tabular}
\end{center}
\label{ketatable}
\end{table} 

\begin{table}[hbtp]
\caption{Variation in fitted EC $W$ mass due to a 1.6\% variation in $k_\eta$.}
\medskip
\begin{center}
\begin{tabular}{cccc} 
\hline
  &$m_T$ fit & $p_T(e)$ fit & $p_T(\nu)$ fit \\
\hline
$\delta M_W$ (MeV) & 34 & 48 & 25  \\
\hline
\end{tabular}
\end{center}
\label{ketaerror}
\end{table}

 The comparison of the electron $\eta$ distribution between the data and the
 Monte Carlo provides a consistency check of the predicted
 $W$ rapidity distribution, and hence of 
 the pdf's. The measured $k_\eta$
 being consistent with unity\footnote{We have used $k_\eta = 1$ in the mass
 analysis.}
 sets an upper bound on the pdf uncertainty. 
 While this constraint can potentially be much more powerful with higher 
 statistics obtained in future data-taking, it is presently weaker than 
 the uncertainty in the modern pdf's. Therefore we do not use this constraint
 to set our final $W$ mass uncertainty due to pdf's. However, since
 our data used for this constraint are independent of the world data used to 
 derive the pdf's, we have additional evidence that the uncertainty on the $W$
 mass due to the pdf's is not being underestimated. 

\section{ Backgrounds }
\label{sec-back}
\subsection {\boldmath \wte \unboldmath }

The decay \wte\ is topologically indistinguishable from \wev. It is included in
the fast Monte Carlo simulation (Sec.~\ref{sec-mc}). This decay is suppressed 
by the branching fraction for 
\tev\ $\left( 17.83\pm0.08 \right)$\% \cite{PDG}, and by the
lepton \pt\ cuts. It accounts for  
 1\% of the events in the \wb\ sample. 

\subsection {Hadronic Background}
\label{hadbkg}

QCD processes can fake the signature of a \wev\ decay if a hadronic jet fakes
the electron signature and the transverse momentum balance is mismeasured. 

\begin{figure}[htpb!]
\epsfxsize=3.0in
\centerline{\epsfbox{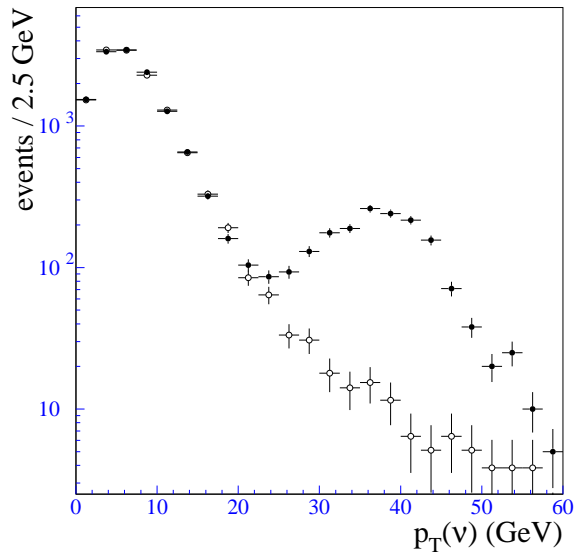}}
\vspace{0.1in}
\caption{The \mpt\ spectra of a sample of events passing
electron identification cuts ($\bullet$) and a sample of events failing
the cuts ($\circ$).}
\label{fig:bkgmet}
\end{figure}

We estimate this background from the \mpt\ spectrum of data events with an
electromagnetic cluster. Electromagnetic clusters in events with
low \mpt\ are almost all due to jets. Some of these clusters
  satisfy our electron
selection criteria and fake an electron. From the shape of the \mpt\ spectrum
for these events we determine how likely it is for these events to have
sufficient \mpt\ to enter our \wb\ sample. 

\begin{figure}[htpb!]
\vspace{-0.0in}
\epsfxsize=3.0in
\centerline{\epsfbox{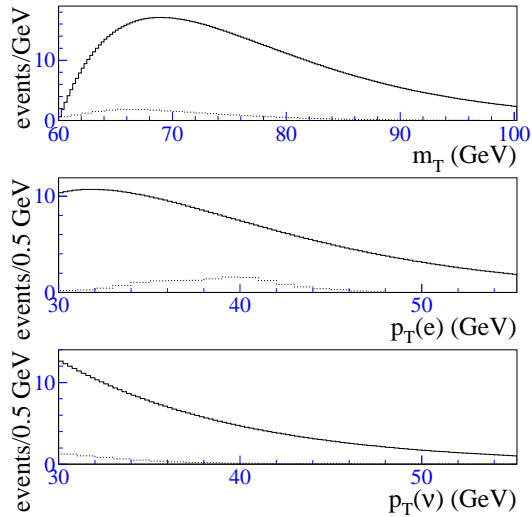}}
\vspace{0.1in}
\caption{ Shapes of \mt, \pte, and \ptnu\ spectra from hadron
(------) and \zb\ boson (- - -) backgrounds with the
proper relative normalization.}
\label{fig:bkg}
\end{figure}

We determine this shape by selecting isolated 
electromagnetic clusters that  have
$\chi^2>200$ and the 4-variable likelihood $\lambda_4 > 30$. 
 Nearly  all electrons fail this cut, so that
the remaining sample consists almost entirely of hadrons. We use data 
 collected using 
a trigger without the \mpt\ requirement to study the efficiency of this cut
for jets. If we normalize the background
spectrum after correcting for residual electrons to the electron sample, we
obtain an estimate of the hadronic background in an electron candidate sample.
Figure~\ref{fig:bkgmet} shows the \mpt\ spectra of both  samples,
normalized for $\mpt<10$~GeV. We find the hadronic background
fraction of the total \wb\ sample after all cuts 
 to be  $f_{\rm had}  =  \left( 3.64 \pm 0.78 \right)$\%.
 The error receives contributions  from the uncertainty in the 
 relative normalization of the two samples 
at low \mpt, the statistics of the failed electron sample, and the uncertainty
 in the residual contamination of the failed electron sample by true electrons.
 We fit the distributions of the background events with 
$\mpt > 30$~GeV to estimate the shape of the background contributions to the
\pte, \ptnu, and \mt\ spectra (Fig.~\ref{fig:bkg}). We use the
  statistical error of
 the fits to estimate the uncertainty in the background shapes. 

\subsection{\boldmath  \zee \unboldmath }

To estimate the fraction of \zee\ events that satisfy the \wb\ boson
 event selection, we
use a Monte Carlo sample of approximately 100{,}000 \zee\ events generated with
the \HERW\ program and a detector simulation based on \GEAN. The boson \pt\
spectrum generated by \HERW\ agrees reasonably well with the calculation in 
Ref.~\cite{LY} and with our \zb\ boson \pt\ measurement~\cite{zptprd}.
 \zee\ decays typically enter the \wb\ sample when one electron
satisfies the \wb\ cuts and the second electron is lost or mismeasured,
causing the event to have large \mpt.

An electron is most frequently mismeasured when it goes into the regions
between the CC and one of the ECs, which are covered only by the hadronic
section of the calorimeter. These electrons therefore cannot
 be identified, and their
energy is measured in the hadronic calorimeter. Large \mpt\ is more likely for
these events than when both electrons hit the EM calorimeters. 

We make the \wb\ and \zb\ selection cuts on the Monte Carlo events, and 
 normalize the number of events passing the \wb\ cuts to the number of \wb\
 data events, scaled by the ratio of selected \zb\ data and Monte Carlo events.
 We estimate the fraction of 
\zb\ events in the \wb\ sample to be   
$f_Z = \left( 0.26 \pm 0.02 \right)\%$. 
The uncertainties quoted
include systematic uncertainties in the matching of momentum scales between
Monte Carlo and collider data. Figure~\ref{fig:bkg} shows the distributions of
\pte, \ptnu, and \mt\ for the \zb\ events with one lost or mismeasured
 electron that satisfy
the \wb\ selection.
\section{ Mass Fits}
\label{sec-fit}
\subsection { Maximum Likelihood Fitting Procedure }

We use a binned maximum likelihood fit to extract the \wb\ mass.
Using the fast Monte Carlo program, we compute the \mt, \pte, and \ptnu\
spectra for 200 hypothesized values of the \wb\ mass between 79.7 and 81.7~GeV.
For the 
spectra we use 250~MeV bins. 
The statistical precision of the spectra for the \wb\ mass fit corresponds
to about 8 million \wb\ decays. When fitting
the collider data spectra, we add the background contributions with the shapes
and normalizations described in Sec.~\ref{sec-back} to the signal spectra. We
normalize the spectra within the fit interval and interpret them as probability
density functions to compute the likelihood
\begin{equation}
L(m) = \prod_{i=1}^N p_i^{n_i}(m),
\end{equation}
where $p_i(m)$ is the probability density for bin $i$, assuming $\mw=m$, and
$n_i$ is the number of data entries in bin $i$. The product runs over all $N$
bins inside the fit interval. We fit $-\ln[L(m)]$ with a quadratic function of
$m$. The value of $m$ at which the function assumes its minimum is the fitted
value of the \wb\ mass and the 68\% confidence level interval is the
interval in $m$ for which $-\ln[L(m)]$ is within half a unit of its minimum.

\subsection { Electron \boldmath \pt\ \unboldmath Spectrum }

\begin{figure}[htpb!]
\vspace{0.2in}
\epsfxsize=3.0in
\centerline{\epsfbox{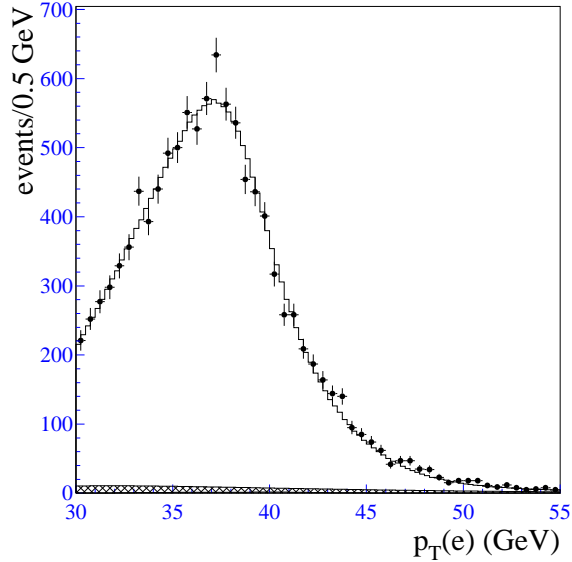}}
\vspace{0.1in}
\caption{ Spectrum of \pte\ from the \wb\ data. 
The superimposed curve shows the
maximum likelihood fit and the shaded region the estimated background.}
\label{fig:eet}
\end{figure}

\begin{figure}[htpb!]
\vspace{0.2in}
\epsfxsize=3.0in
\centerline{\epsfbox{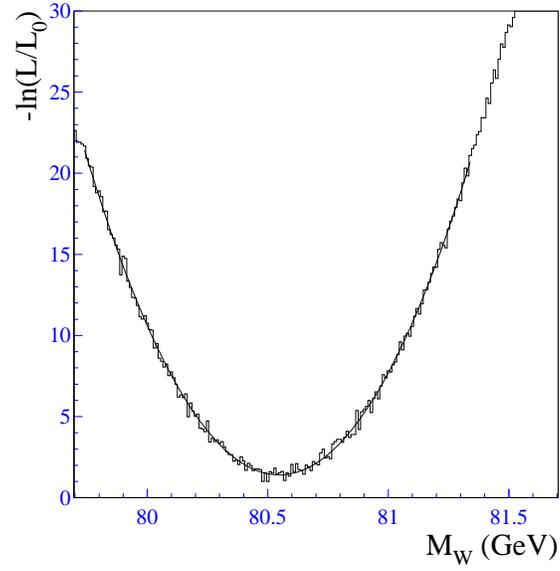}}
\vspace{0.1in}
\caption{ The likelihood function for the \pte\ fit.}
\label{fig:likepte}
\end{figure}

We fit the \pte\ spectrum in the region $32<\pte<50$~GeV. The interval is
 chosen to span the Jacobian peak. 
 The data points in Fig.~\ref{fig:eet} represent the
\pte\ spectrum from the \wb\ sample. The solid line shows the sum of the
simulated \wb\ signal and the estimated background for the best fit, and the
shaded region indicates the
sum of the estimated hadronic and \zee\ backgrounds.
The maximum likelihood fit gives
\begin{equation}
\mw = 80.547 \PM 0.128\ \hbox{GeV}
\label{eq:mwpte}
\end{equation}
for the \wb\ mass. Figure~\ref{fig:likepte} shows
$-\ln(L(m)/L_0)$ for this fit, where $L_0$ is an arbitrary number.

\begin{figure}[htpb!]
\vspace{-0.3in}
\epsfxsize=3.0in
\centerline{\epsfbox{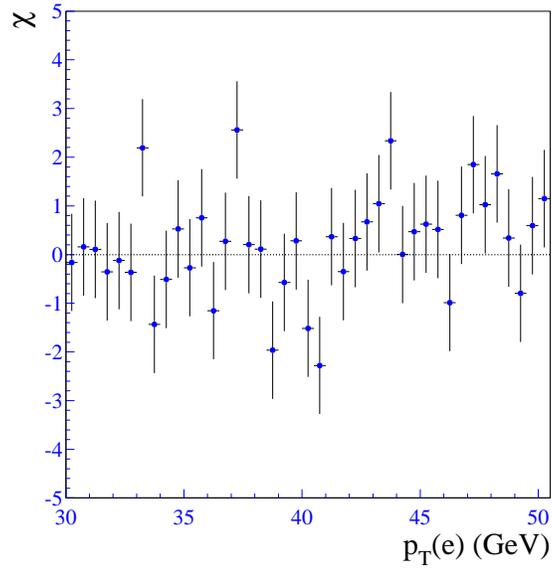}}
\vspace{0.1in}
\caption{ The $\chi$ distribution for the fit to the \pte\ spectrum.}
\label{fig:chieet}
\end{figure}

As a goodness-of-fit test, we divide the fit interval into 0.5~GeV bins,
normalize the integral of the probability density function 
to the number of events in the fit interval, and compute
$\chi^2=\sum_{i=1}^N(y_i-P_i)^2/y_i$. The sum runs over all $N$ bins,
$y_i$ is the observed number of events in bin $i$, and $P_i$ is the
integral of the normalized probability density 
function over bin $i$. The parent
distribution is the $\chi^2$ distribution for $N-2$ degrees of
freedom. For the spectrum in Fig.~\ref{fig:eet} we compute $\chi^2=46$. For
36 bins there is a 8\% 
 probability for $\chi^2\ge46$. Figure~\ref{fig:chieet}
shows the contributions
$\chi_i=(y_i-P_i)/\sqrt{y_i}$ to $\chi^2$ for the 36 bins in the fit
interval.

\begin{figure}[htpb!]
\vspace{-0.3in}
\epsfxsize=3.0in
\centerline{\epsfbox{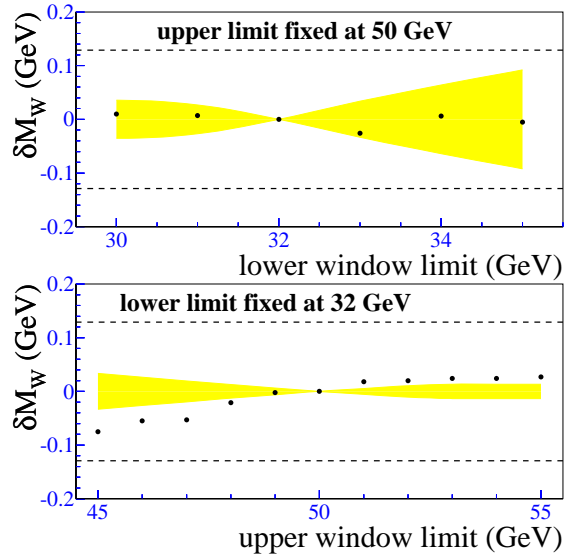}}
\vspace{0.1in}
\caption{ Variation of the fitted mass with the \pte\ fit window limits. See
text for details. }
\label{fig:vareet}
\end{figure}

Figure~\ref{fig:vareet} shows the sensitivity of the
fitted mass value to the choice of fit interval. The points in the two plots
indicate the observed deviation of the fitted mass from the value given in
Eq.~\ref{eq:mwpte}. We expect some variation due to statistical
fluctuations in the spectrum and systematic uncertainties in the
probability density functions. We estimate the effect due to statistical
fluctuations using Monte Carlo  ensembles. We expect the
fitted values
to be inside the shaded regions indicated in the two plots with 68\%
probability. The dashed lines indicate the statistical error for the nominal
fit. Figure~\ref{fig:vareet} 
 shows that the probability density function provides a good description
of the observed spectrum.

\subsection { Transverse Mass Spectrum }

\begin{figure}[htpb!]
\vspace{0.2in}
\epsfxsize=3.0in
\centerline{\epsfbox{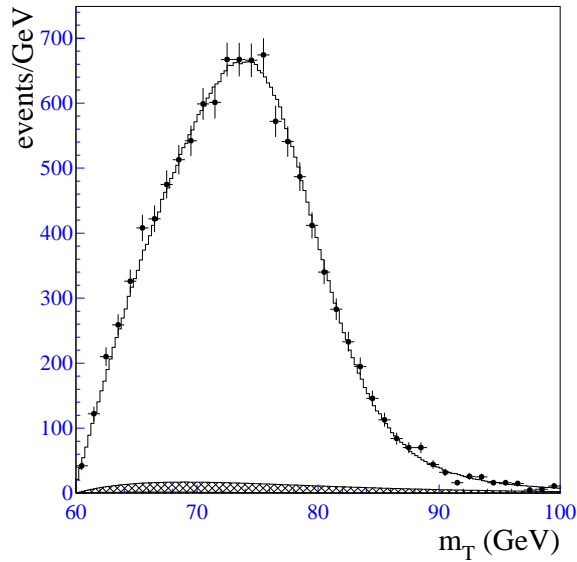}}
\vspace{0.1in}
\caption{ Spectrum of \mt\ from the \wb\ data. The superimposed curve shows the
maximum likelihood fit and the shaded region shows the estimated background.}
\label{fig:mtw}
\end{figure}

\begin{figure}[htpb!]
\vspace{-0.0in}
\epsfxsize=3.0in
\centerline{\epsfbox{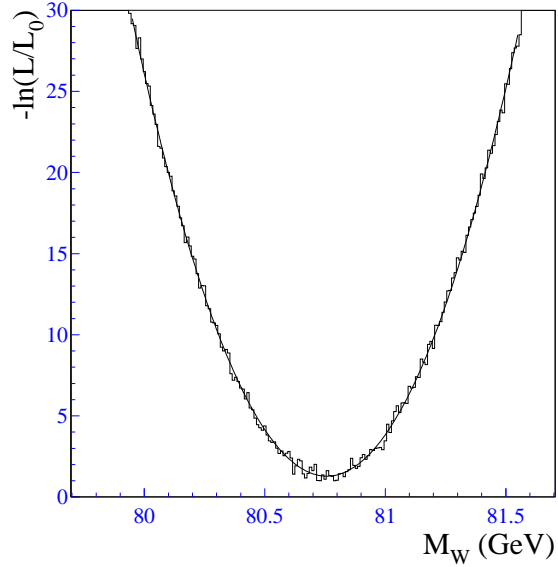}}
\vspace{0.1in}
\caption{ The likelihood function for the \mt\ fit.}
\label{fig:likemt}
\end{figure}

The \mt\ spectrum is shown in Fig.~\ref{fig:mtw}. The points are the observed
spectrum, the solid line shows signal plus background for the best fit, and
the shaded region indicates the estimated background contamination. We fit in
the interval
$65<\mt<90$~GeV. 
Figure~\ref{fig:likemt} shows
$-\ln(L(m)/L_0)$ for this fit where $L_0$ is an arbitrary number.
The best fit occurs for
\begin{equation}
\mw = 80.757 \PM 0.107\ \hbox{GeV}.
\label{eq:mwmt}
\end{equation}

\begin{figure}[htpb!]
\vspace{-0.0in}
\epsfxsize=3.0in
\centerline{\epsfbox{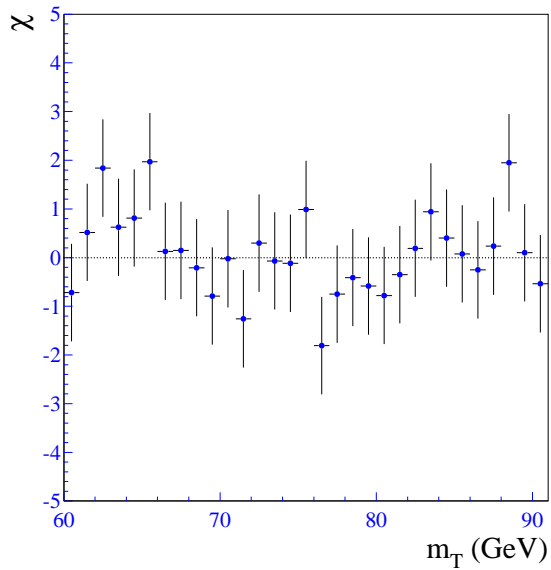}}
\vspace{0.1in}
\caption{ The $\chi$ distribution for the fit to the \mt\ spectrum.}
\label{fig:chimt}
\end{figure}

\begin{figure}[htpb!]
\vspace{-0.0in}
\epsfxsize=3.0in
\centerline{\epsfbox{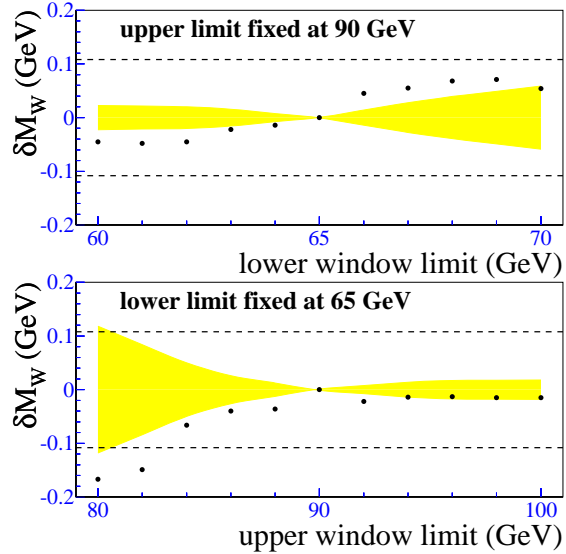}}
\vspace{0.1in}
\caption{ Variation of the fitted mass with the \mt\ fit window limits. See
text for details. }
\label{fig:varmt}
\end{figure}

Figure~\ref{fig:chimt} shows the deviations of the data from the fit.
Summing over all bins in the fitting window, we get $\chi^2 =17$ for 25 bins.
For 25 bins there is a 81\% probability to obtain a larger value. 
Figure~\ref{fig:varmt} shows the sensitivity of the fitted mass to 
the choice of
fit interval.

\subsection { Neutrino \boldmath \pt\ \unboldmath Spectrum }

\begin{figure}[htpb!]
\vspace{-0.0in}
\epsfxsize=3.0in
\centerline{\epsfbox{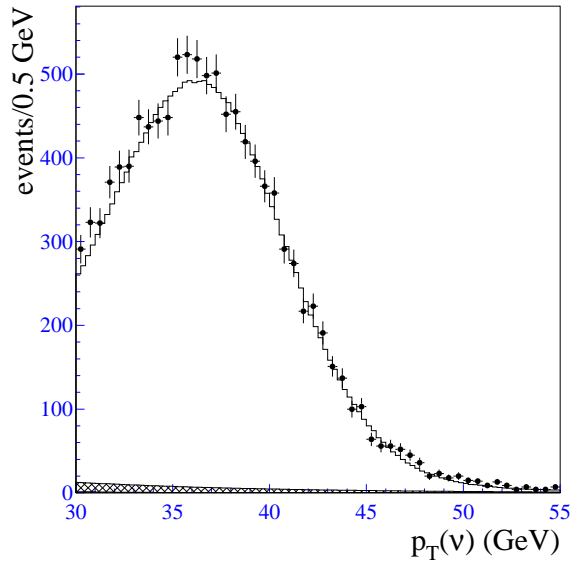}}
\vspace{0.1in}
\caption{ Spectrum of \ptnu\ from the \wb\ data. The superimposed curve shows
the maximum likelihood fit and the shaded region shows the estimated
background.}
\label{fig:met}
\end{figure}

\begin{figure}[htpb!]
\vspace{-0.0in}
\epsfxsize=3.0in
\centerline{\epsfbox{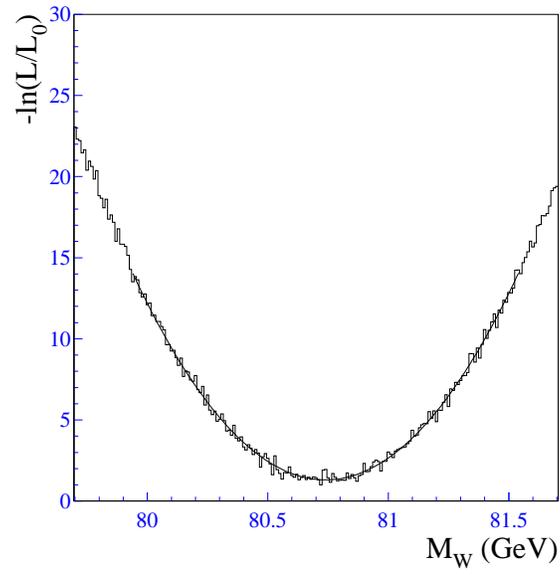}}
\vspace{0.1in}
\caption{ The likelihood function for the \ptnu\ fit.}
\label{fig:likeptnu}
\end{figure}

Figure~\ref{fig:met} shows the neutrino \pt\ spectrum.
  The points are the observed
spectrum, the solid line shows signal plus background for the best fit, and
the shaded region indicates the estimated background contamination. We fit in
the interval
$32<\ptnu<50$~GeV. 
Figure~\ref{fig:likeptnu} shows
$-\ln(L(m)/L_0)$ for this fit where $L_0$ is an arbitrary number.
The best fit occurs for
\begin{equation}
\mw = 80.740 \PM 0.159\ \hbox{GeV}.
\label{eq:mwptnu}
\end{equation}

\begin{figure}[htpb!]
\vspace{-0.3in}
\epsfxsize=3.0in
\centerline{\epsfbox{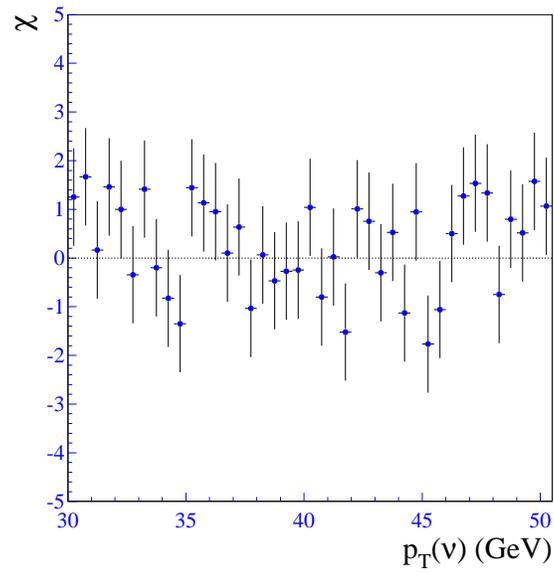}}
\vspace{0.1in}
\caption{ The $\chi$ distribution for the fit to the \ptnu\ spectrum.}
\label{fig:chimet}
\end{figure}

\begin{figure}[htpb!]
\vspace{-0.3in}
\epsfxsize=3.0in
\centerline{\epsfbox{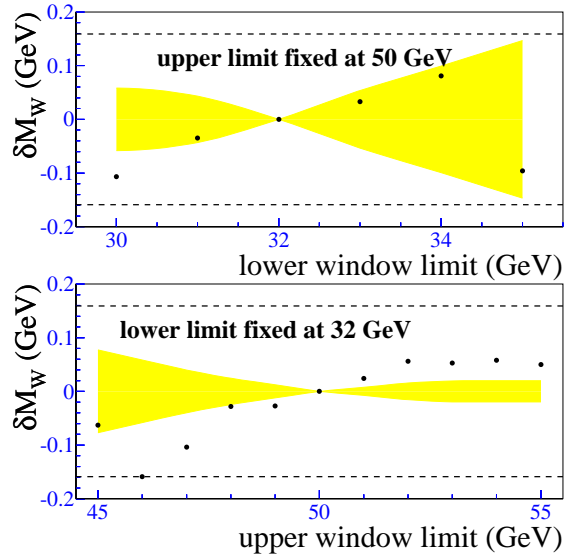}}
\vspace{0.1in}
\caption{ Variation of the fitted mass with the \ptnu\ fit window limits. See
text for details. }
\label{fig:varmet}
\end{figure}

Figure~\ref{fig:chimet} shows the deviations of the data from the fit.
Summing over all bins in the fitting window, we get $\chi^2 =37$ for 36 bins.
For 36 bins there is a 33\% probability to obtain a larger value. 
Figure~\ref{fig:varmet} shows the sensitivity of the fitted mass to 
the choice of
fit interval.

\section{ Consistency Checks}
\label{sec-checks}
\subsection {North vs South Calorimeters}

Since the detector is north-south symmetric, we expect the measurements
 made with the north and south calorimeters separately to be consistent.
We find
\begin{eqnarray}
M_W^{\rm ECN} - M_W^{\rm ECS} & = & \: \; \; \; \;  88 \;  \pm  \; 215 \; {\rm MeV} \; 
 (m_T \; {\rm fit)} \nonumber \\
M_W^{\rm ECN} - M_W^{\rm ECS} & = &  -116 \;  \pm  \; 258 \; {\rm MeV} \; 
 (p_T^e \; {\rm fit)} \nonumber \\
M_W^{\rm ECN} - M_W^{\rm ECS} & = & \: \; \; 107 \;  \pm  \; 318 \; {\rm MeV} \; 
 (p_T^\nu \; {\rm fit)} 
\end{eqnarray} 
 where the uncertainty is statistical only. 
\subsection {Time Dependence}

\begin{figure}[htpb!]
\vspace{0.2in}
\epsfxsize=3.0in
\centerline{\epsfbox{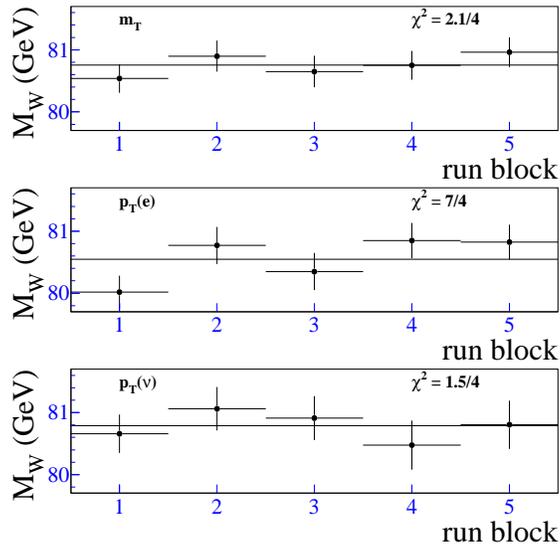}}
\vspace{0.1in}
\caption{ The fitted \wb\ boson masses in bins of run blocks from
the \mt, \pte, and \ptnu\ fits.  The solid
line is the central value for the respective fit over
the entire sample. The \wb\ fit statistical error for each subsample is shown.
The average instantaneous luminosity in the bins is 4.2, 6.1, 7.1, 9.3 and
 10.1 respectively, in units of $10^{30}$/cm$^2$/s.}
\label{fig:mwtime}
\end{figure}

We divide the \wb\ boson data sample into five sequential calender
 time intervals
  such that 
 the subsamples have equal number of events.
We generate resolution functions for
the luminosity distribution of these five subsamples.
We fit the transverse mass and lepton \pt\ spectra from the \wb\
samples in each
 time bin. 
 The fitted masses are plotted in Fig.~\ref{fig:mwtime} where the time bins
 are labelled by run blocks. The errors shown are
statistical only.
We compute the $\chi^2$ with respect to the \wb\ mass fit to the
 entire data sample.
The $\chi^2$ per degree of freedom (dof) for the \pte\ fit is 7.0/4 and for the
\ptnu\ fit is 1.5/4. The \mt\ fit has a $\chi^2$/dof of 2.1/4. 

 Since the luminosity was increasing with time throughout the run, the time
 slices correspond roughly to luminosity bins. 

\subsection { Dependence on \boldmath $ u_T$ \unboldmath  Cut }

We change the cuts on the recoil momentum \ut\ and study how well the fast
 Monte
Carlo simulation reproduces the variations in the spectra. We split the \wb\
sample into subsamples with $\upar\gt 0~\GeV$ and $\upar\lt 0~\GeV$, 
and fit the subsamples with corresponding Monte Carlo spectra generated with
 the same cuts. 
The difference in the fitted masses from the two subsamples corresponds to
  0.3$\sigma$, 0.8$\sigma$ and 1.3$\sigma$ for the \mt, 
 \pte, and \ptnu\ fits respectively, based on the statistical uncertainty
 alone.  Although there is
significant variation among the shapes of the spectra for the different cuts,
the fast Monte Carlo models them well.

\subsection {Dependence on Fiducial Cuts }

\begin{figure}[htpb!]
\epsfxsize=3.0in
\centerline{\epsfbox{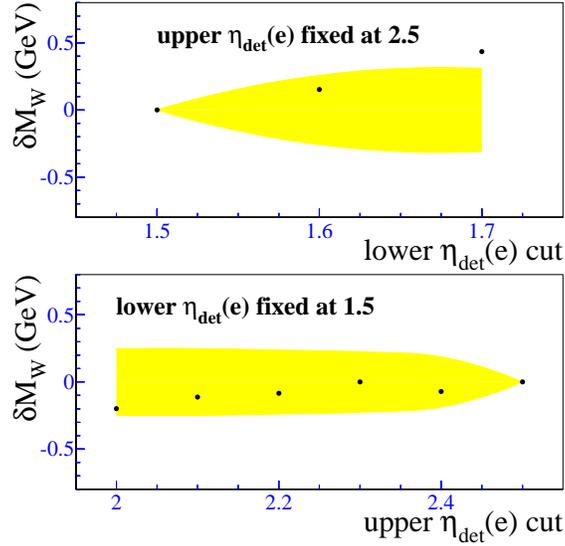}}
\vspace{0.1in}
\caption{The variation in the \wb\ mass from the \pte\ fit
 versus the $\eta_{\rm det}(e)$ cut.  The shaded
region is the expected statistical variation.  }
\label{fig:eta_cut_eet}
\end{figure}

\begin{figure}[htpb!]
\epsfxsize=3.0in
\centerline{\epsfbox{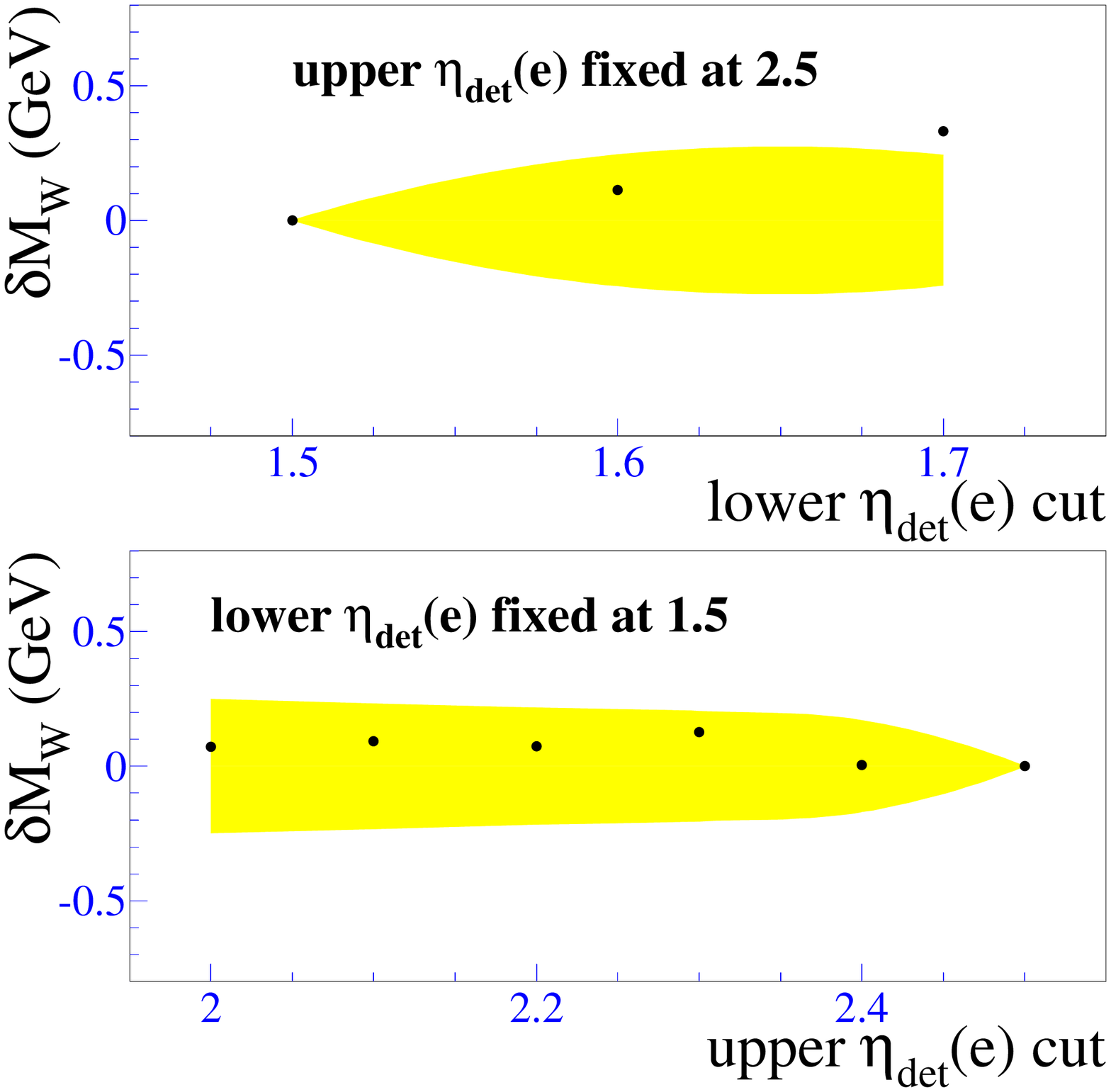}}
\vspace{0.1in}
\caption{The variation in the \wb\ mass from the \mt\ fit
 versus the $\eta_{\rm det}(e)$ cut.  The shaded
region is the expected statistical variation.  }
\label{fig:eta_cut_mtw}
\end{figure}

\begin{figure}[htpb!]
\epsfxsize=3.0in
\centerline{\epsfbox{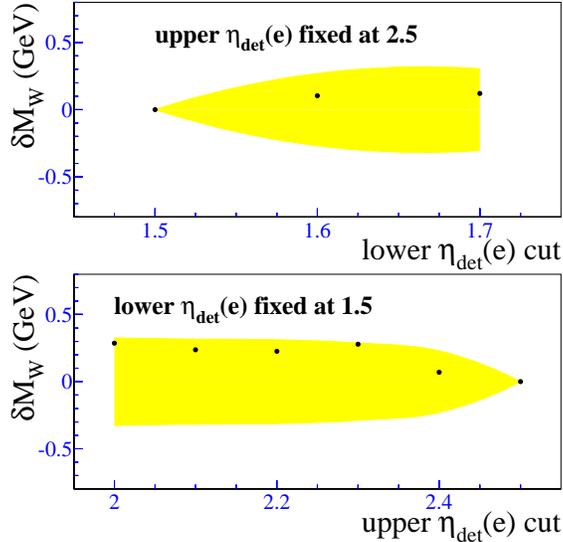}}
\vspace{0.1in}
\caption{The variation in the \wb\ mass from the \ptnu\ fit
 versus the $\eta_{\rm det}(e)$ cut.  The shaded
region is the expected statistical variation.  }
\label{fig:eta_cut_met}
\end{figure}

We fit the \mt\ spectrum from the \wb\ sample and the \mee\ spectrum
from the \zb\ sample for different pseudorapidity cuts on the electron
direction.
Keeping the upper $\mid \! \! 
 \eta_{\rm det} (e) \! \! \mid$ cut fixed at 2.5, we vary the lower
 $\mid \! \! \eta_{\rm det} (e) \! \! \mid$ 
 cut from 1.5 to 1.7. Similarly, we vary the upper
 $\mid \! \! \eta_{\rm det} (e) \! \! 
 \mid$ cut from 2.0 to 2.5, keeping the lower
  $\mid \! \! \eta_{\rm det} (e) \! \! \mid$ cut
 fixed at 1.5. 
 Figures~\ref{fig:eta_cut_eet}--\ref{fig:eta_cut_met}  show the
change in the \wb\ mass versus the $\eta_{\rm det} (e)$
  cut using the electron energy scale
calibration from the corresponding \zb\ sample.  The shaded 
 region indicates the
statistical error. Within the uncertainties, the mass is independent
of the $\eta_{\rm det} (e)$ cut.

\subsection { \boldmath \zb\ \unboldmath Boson Transverse Mass Fits}

\begin{figure}[htpb!]
\epsfxsize=3.0in
\centerline{\epsfbox{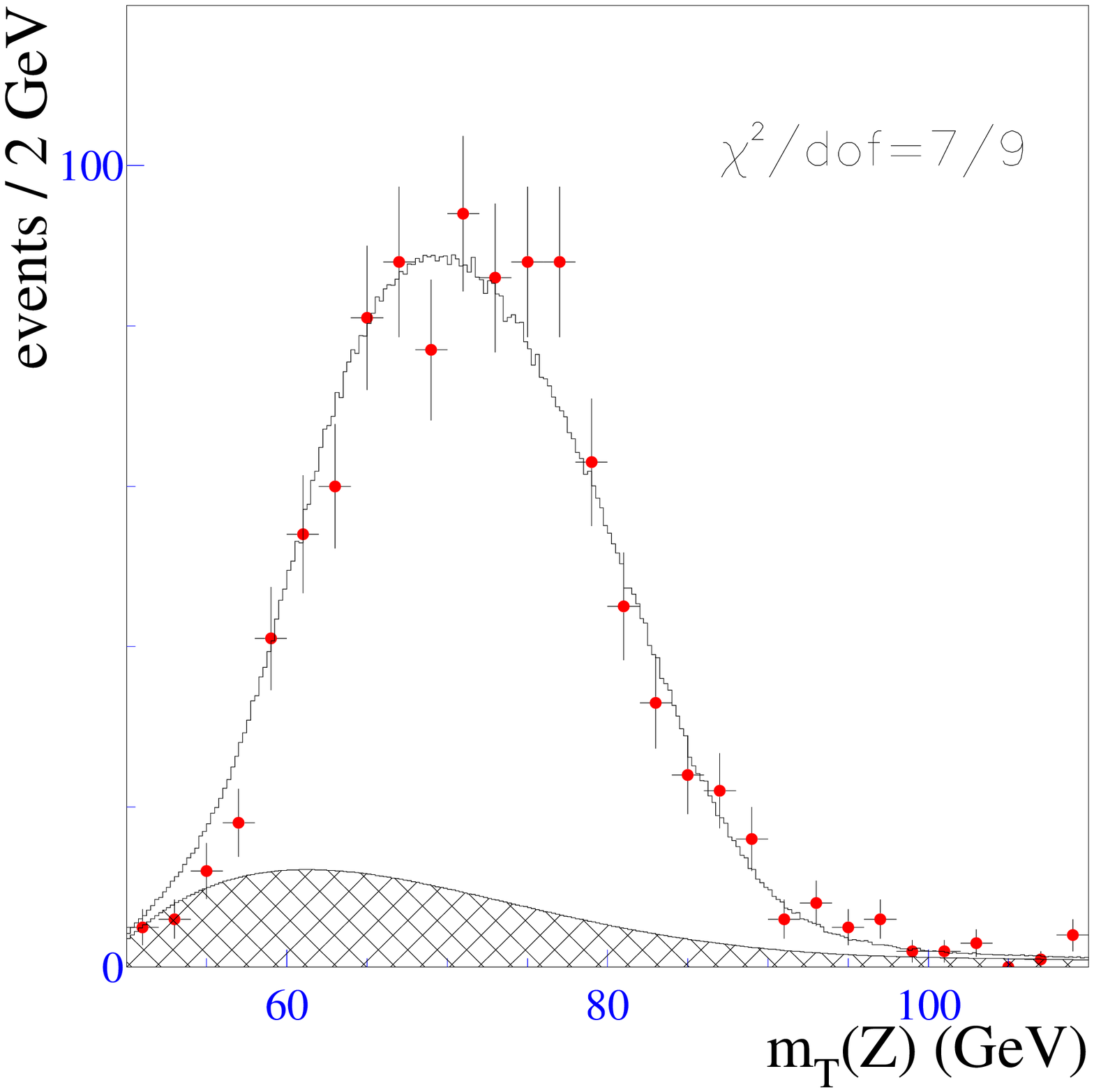}}
\vspace{0.0in}
\epsfxsize=3.0in
\centerline{\epsfbox{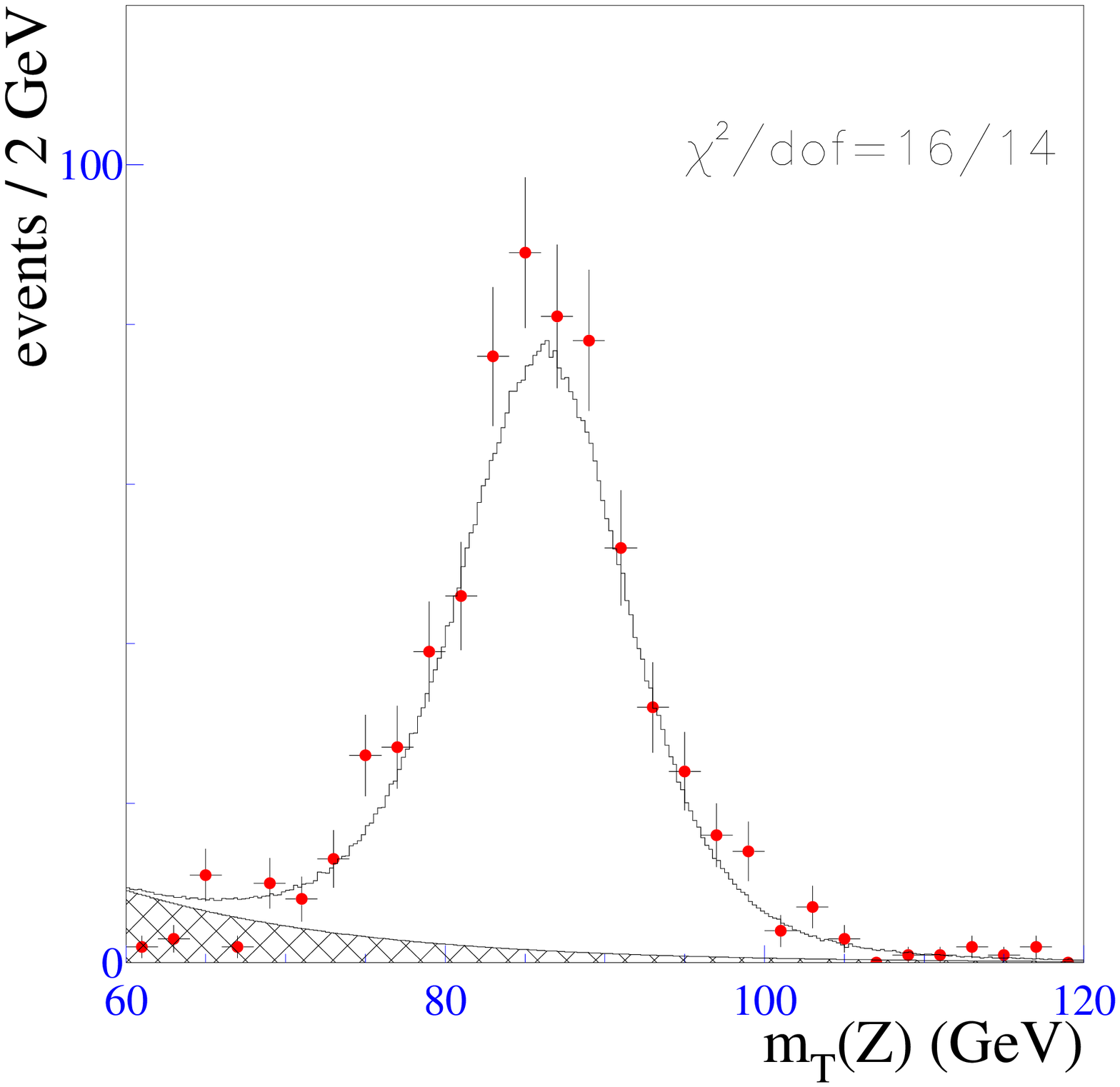}}
\vspace{-0.02in}
\caption{ Spectra of the \zb\ boson
 transverse mass, from the CC/EC data (top) and
 the EC/EC data (bottom). The second electron in the \zb\ boson decay is
 treated like the neutrino in \wb\ boson decay. The superimposed curves show
the maximum likelihood fits and the shaded regions show the estimated
backgrounds. The $\chi^2$/dof between the data and the Monte Carlo are also
 shown.}
\label{mtz_fit_ec}
\end{figure}

 As a consistency check, we fit the transverse mass distribution of the 
 $Z \rightarrow ee$ events, 
 reconstructed using each electron and the recoil. The measured energy of the
 second electron is ignored, both in the data and in the Monte Carlo used
 to obtain the templates. Each \zb\ event is treated (twice) as a \wb\ event,
 where the neutrino transverse momentum is recomputed using the first electron
 and the 
 recoil. One of the two electrons is required to
 be in the EC. The
 fitting range is $70 < m_T < 90$ GeV  for the CC/EC events and
 $70 < m_T < 100$ GeV for the EC/EC events.
 Figure~\ref{mtz_fit_ec} shows the results. The CC/EC fit yields $M_Z = 92.004 
 \pm 0.895$(stat) GeV with  $\chi^2$/dof = 7/9. 
 The EC/EC fit yields $M_Z = 91.074 
 \pm 0.299$(stat) GeV with  $\chi^2$/dof = 16/14. The average fitted mass is 
 $M_Z = 91.167 \pm 0.284$(stat) GeV. 
 The fits are
 good and  the fitted masses are
 consistent with the input $Z$ mass.

\section{ Uncertainties in the Measurement }
\label{sec-syst}

 Apart from the statistical error in the fitted \wb\ mass, 
 uncertainties in the various inputs needed for the measurement lead to 
 uncertainties in the final result. Some of these inputs are
 discrete (such as the choice of the parton distribution function set) and 
 others are parameterized by continuous variables. For a different choice of
 pdf set, or a shift in the value of an input parameter by one
 standard deviation, the expected shift in the fitted \wb\ mass is computed
 by using the fast Monte Carlo to generate spectra with the changed parameter
 and fitting the spectra with the default templates. The expected shifts due
 to various input parameter uncertainties (given in Table~\ref{parerror})
 or choice of pdf set are discussed
 in detail below, and are summarized in Tables~\ref{systematicsec}
 and~\ref{systematicscc}. The shifts in the fitted mass obtained from
 the different
 kinematic spectra may be in opposite
 directions, in which case they are indicated with opposite signs.

\begin{table}[hbtp]
\caption{Errors on the parameters in the \wb\ mass analysis. The correlation
 coefficient between \alpharec\ and \betarec\ is $-0.98$; that
 between \srec\ and
 \alphamb\ is $-0.60$. }
\medskip
\begin{center}
\begin{tabular}{cc} 
\hline
parameter & error  \\
\hline
parton luminosity $\beta$ & 0.001 GeV $^{-1}$\\
\hline
photon coalescing radius $R_0$ & 7 cm \\
\hline
$W$ width & 59 MeV \\
\hline
ECEM offset \deltaecem\ & 0.7 GeV  \\
\hline
ECEM scale \alphaecem\ & 0.00187   \\
\hline
FDC radial scale \betafdc\ & 0.00054  \\
\hline
FDC-EC radial scale \betaec\ & 0.0003  \\
\hline
ECEM constant term $c_{\rm EC}$ & $^{+0.006}_{-0.01}$  \\
\hline
 recoil response (\alpharec, \betarec) & (0.06, 0.02)  \\
\hline
 recoil resolution (\srec, \alphamb) & (0.14 GeV$^{1/2}$, 0.028) \\ 
& $\oplus$ (0.0,0.01) \\
\hline
\upar\ correction \dupar/$\delta \eta \delta \phi$ & 28 MeV  \\
\hline
\upar\ efficiency slope $s$ & 0.0012 GeV$^{-1}$ \\
\hline
\end{tabular}
\end{center}
\label{parerror}
\end{table}

\begin{table}[hbtp]
\caption{Variation in the fitted $M_W$ and $M_Z$ (in MeV) for the forward 
 electron sample  
 due to variation in the model input parameters
  by the respective uncertainties.}
\medskip
\begin{center}
\begin{tabular}{cccccc} 
\hline
Source&$\delta M_Z$&$\delta M_Z$&$\delta M_W$&$\delta M_W$&$\delta M_W$\\
&(CC/EC)&(EC/EC)&$(m_T)$&$(p_T^e)$&$(p_T^\nu)$\\
\hline
statistics & 124 & 221 & 107 & 128 & 159 \\
\hline
$p_T(W)$ spectrum & & & 22 & 37 & 44 \\
\hline
MRSR2 \cite{mrsr2}  &  &  & $-$11 & $-$21 & $-$43 \\
\hline
MRS(A$^ \prime$) \cite{mrsa}   &  &  &  $-7$ & $-43$ & $-19$ \\
\hline
CTEQ5M \cite{cteq5m}  &  &  &  14 &   9 & $-17$ \\
\hline
CTEQ4M \cite{cteq4m}  &  &  &   1 & $-21$ &  22 \\
\hline
CTEQ3M \cite{cteq3m}  &  &  &  13 &  30 &  28 \\
\hline
parton &  &  &  &  &  \\
luminosity $\beta$ & 8 & 7 & 9 & 11 & 18 \\
\hline
$R_0$ & 10 & 13 & 9 & 17 & 12 \\
\hline
$2 \gamma$ & 5 & 10 & 5 & 10 & 0 \\
\hline
$W$ width & & & 10 & 10 & 10 \\
\hline
ECEM offset & 284 & 421 & 437 & 433 & 386 \\
\hline
ECEM scale &  &  &  &  &  \\
variation 0.0025 & 114 & 228 & 201 & 201 & 201 \\
\hline
CCEM scale &  &  &  &  &  \\
variation 0.0008 & 37 & 0 & 0 & 0 & 0 \\
\hline
FDC radial scale & 8 & 36 & 43 & 37 & 28 \\
\hline
FDC-EC radial scale & 10 & 52 & 57 & 54 & 48 \\
\hline
ECEM constant &  &  &  &  &  \\
term \cecem\ & 0 & 0 & 45 & 29 & 78 \\
\hline
 hadronic &  &  &  &  &  \\
response &  &  & 11 & 20 & $-50$ \\
\hline
 hadronic &  &  &  &  &  \\
resolution &  &  & 40 & 4 & 203 \\
\hline
\upar\ correction & 20 & 30 & 18 & 34 & $-6$ \\
\hline
\upar\ efficiency &   &   & 4 & $-22$ & 40 \\
\hline
 background &  &  &  &  &  \\
normalization & 0 & 11 & 12 & 15 & 25 \\
\hline
 background &  &  &  &  &  \\
shape & 0 & 5 & 16 & 23 & 78 \\
\hline
\end{tabular}
\end{center}
\label{systematicsec}
\end{table}

\begin{table}[hbtp]
\caption{Variation in the fitted $M_W$ and $M_Z$ (in MeV) for the central 
 electron sample  
 due to variation in the  model input parameters
  by the respective uncertainties.}
\medskip
\begin{center}
\begin{tabular}{cccccc} 
\hline
Source&$\delta M_Z$&$\delta M_Z$&$\delta M_W$&$\delta M_W$&$\delta M_W$\\
&(CC/CC)&(CC/EC)&$(m_T)$&$(p_T^e)$&$(p_T^\nu)$\\
\hline
statistics & 75 & 124 & 70 & 85 & 105 \\
\hline
$p_T(W)$ spectrum & & & 10 & 50 & 25 \\
\hline
MRSR2  \cite{mrsr2}  & &  &   5 &  26 &   3   \\
\hline
MRS(A$^ \prime$) \cite{mrsa}   & &  &  $-5$ &  16 & $-31$   \\
\hline
CTEQ5M \cite{cteq5m}  & &  &  $-8$ &   6 & $-22$   \\
\hline
CTEQ4M \cite{cteq4m} & &  &  10 &  11 & $-18$   \\
\hline
CTEQ3M \cite{cteq3m}  & &  &   0 &  64 &  $-9$   \\
\hline
parton &  &  &  &  &  \\
luminosity $\beta$ & 4 & 8 & 9 & 11 & 9 \\
\hline
$R_0$ & 19 & 10 & 3 & 6  & 0 \\
\hline
$2 \gamma$ & 10 & 5 & 3 & 6 & 0 \\
\hline
$W$ width & & & 10 & 10 & 10 \\
\hline
 CC EM offset & 387 & 467 & 367 & 359 & 374 \\
\hline
 CDC  scale & 29 & 33 & 38 & 40 & 52 \\
\hline
 uniformity &  &  & 10 & 10 & 10 \\
\hline
CCEM constant &  &  &  &  &  \\
term \cccem\ &  &  & 23 & 14 & 27 \\
\hline
 hadronic &  &  &  &  &  \\
response &  &  & 20 & 16 & $-46$ \\
\hline
 hadronic &  &  &  &  &  \\
resolution &  &  & 25  & 10 & 90 \\
\hline
\upar\ correction &  &  & 15 & 15 & 20 \\
\hline
\upar\ efficiency &   &   & 2 & $-9$ & 20 \\
\hline
 backgrounds &  &  & 10 & 20 & 20 \\
\hline
\end{tabular}
\end{center}
\label{systematicscc}
\end{table}

 Since the most important
 parameter, the EM energy scale, is measured by calibrating to the \zb\ mass,
 we are  measuring the ratio of the \wb\ and \zb\ boson 
 masses. There can
 be significant cancellation in uncertainties between the \wb\ and \zb\ masses
 if their variation due to an input parameter change is very similar. 
 For those parameters that affect the fitted \zb\ mass, 
 Tables~\ref{systematicsec} and~\ref{systematicscc} also show the expected
 shift in the fitted \zb\ mass. The signed \wb\ and \zb\ mass shifts are
 used to construct a covariance matrix between the various fitted \wb\ mass
 results, which is used to obtain the final \wb\ mass value and uncertainty; 
 thus simple combination of the uncertainties in Tables~\ref{systematicsec}
 and~\ref{systematicscc} is inappropriate.  
 This is discussed in detail in Section~\ref{erroranalysis}. 

\subsection { Statistical Uncertainties }

Tables~\ref{systematicsec} and~\ref{systematicscc}  list the uncertainties
 in the \wb\ mass measurement due to the finite sizes of the \wb\ and \zb\
samples used in the fits to the \mt, \pte, \ptnu, and \mee\ spectra. The
statistical uncertainty due to the finite \zb\ sample propagates into the \wb\
mass measurement through the electron energy scale \alphaecem.

 Since the \mt, \pte\ and \ptnu\ fits are performed using
  the same \wb\ data set,
 the results from the three fits are statistically correlated. The
 correlation coefficients between the respective statistical errors are 
 calculated using Monte Carlo ensembles, and 
 are shown in Table~\ref{correlation}.

\begin{table}[ht]
\begin{center}
\caption{\small The statistical correlation coefficients obtained from 
  Monte Carlo ensemble tests fitting the 
\wb\ boson mass for 260 samples of 11{,}089 events each.}
\medskip
\begin{tabular}{cccc}
             & \multicolumn{3}{c}{correlation matrix} \\
              & \mt   & \pte    & \ptnu    \\
             &     &         &       \\ \hline
\mt          & 1   & 0.634   & 0.601    \\
\pte         & 0.634 & 1     & 0.149    \\
\ptnu        & 0.601 & 0.149   & 1      \\
\end{tabular}
\label{correlation}
\end{center}
\end{table}

\subsection { \boldmath \wb\ \unboldmath Boson Production and Decay Model }
\label{sec-model_errors}

\subsubsection{Sources of Uncertainty}

Uncertainties in the \wb\ boson 
 production and decay model arise from the following
sources: the phenomenological parameters in the calculation of the \ptw\
spectrum, 
 the choice of parton distribution functions, radiative decays, and the
\wb\ boson width.
In the following we describe how we assess the size of the systematic
uncertainties introduced by each of these. We summarize the
size of the uncertainties in
Tables~\ref{systematicsec} and~\ref{systematicscc}.

\subsubsection{\wb\ Boson \pt\ Spectrum}

In Sec.~\ref{sec-constraints} of Ref.~\cite{wmass1bcc},
  we described our constraint on the \wb\ boson \pt\ spectrum. This constraint
 was obtained by studying the \zb\ boson \pt\ spectrum, which can be
 measured well using the two electrons in \zee\ decays. For any chosen
 parton distribution function, the parameters of the theoretical model were
 tuned so that the predicted \zb\ boson 
 \pt\ spectrum after simulating all detector
 effects agreed with the data. The precision with which the parameters could be
 tuned was limited by the statistical uncertainty and the uncertainty in the
 background. These parameter values were used to predict the \wb\ boson \pt\
 spectrum. 

  The uncertainties in the fitted \wb\ boson
 mass for the CC \wb\ sample due to the uncertainty in the \wb\ boson 
 \pt\ spectrum
 were listed in Ref.~\cite{wmass1bcc}, and are reproduced in 
 Table~\ref{systematicscc}.
  The corresponding uncertainty in the EC analysis is given
 in Table~\ref{systematicsec}. The CC and EC \wb\ mass
 uncertainties from this source
 are assumed to be fully correlated. 

\subsubsection{Parton Distribution Functions}

To quantify the \wb\ mass uncertainty due to variations in the
 input parton distribution functions, we select the MRS(A$^\prime$),
  MRSR2, CTEQ5M,
 CTEQ4M and CTEQ3M sets to compare to MRST.
  We select these sets because their predictions
 for the lepton charge asymmetry in $W$ decays and the neutron-to-proton
 Drell-Yan ratio span the range of consistency with the measurements from 
 CDF~\cite{cdfasym}
 and E866~\cite{e866dy}. 
 These measurements constrain the ratio of $u$ and $d$ quark
 distributions which have the most influence on the \wb\ rapidity spectrum. 
 
Using these parton distribution function sets as input to the fast
Monte Carlo model, we generate \mt\ and lepton \pt\ spectra. For each chosen
  parton distribution function set we use the appropriate \wb\ boson
 \pt\ spectrum
 as used in our CC \wb\ mass analysis. We then fit the 
 generated spectra in the
same way as the spectra from collider data, \ie\ using MRST parton
distribution functions. Table~\ref{systematicsec} lists the variation of the
fitted EC \wb\ mass values relative to MRST.
The CC and EC \wb\ mass uncertainty from this source
 is taken to be fully correlated, taking the relative signs of the mass shifts
 into account. 

 We find that the combination of the CC and EC \wb\ boson mass measurements
 is less sensitive to pdf variations, than for 
  the CC measurement alone. The
 pdf uncertainty on the CC measurement  is 11 MeV. The pdf uncertainty
 on the CC+EC combined measurement is 7 MeV. As expected, the larger
 combined rapidity coverage makes the observed transverse mass and 
 transverse momentum distributions less sensitive to the longitudinal 
  boost of the \wb\ boson. 

\subsubsection{Parton Luminosity}

The uncertainty of $10^{-3}\ \mbox{GeV}^{-1}$ in the parton
luminosity slope $\beta$ (Sec.~\ref{sec-mc}) translates into an uncertainty in
the fitted \wb\ and \zb\ boson masses. We
estimate the sensitivity in the fitted \wb\ and \zb\ masses
  by fitting Monte Carlo
spectra generated with different values of $\beta$.
 The uncertainty in $\beta$ is taken to be fully correlated between the CC
 and EC \wb\ mass analyses. 

\subsubsection{Radiative Decays}

We assign an error to the modeling of radiative decays based on varying
the detector parameter $R_0$ (Sec.~\ref{sec-mc}).
$R_0$ defines the maximum separation between the photon and
electron directions above which the photon energy is not included in the
electron shower. In general, radiation shifts the fitted mass down for the
transverse mass and electron fits, because for a fraction of the events the
photon energy is subtracted from the electron. Hence increasing
$R_0$ decreases the radiative shift. Both the fitted \wb\ and \zb\
masses depend on $R_0$.
To estimate the systematic error, we fit Monte Carlo
spectra generated with different values of
$R_0$. \GEAN\ detector simulations show that, for an $R_0$ variation of 
$\pm 7$ cm, the electron-photon cluster 
 overlap changes to give the maximum variation
 in the electron identification efficiency. 
 The changes in the mass fits when varying $R_0$ by $\pm 7$ cm
are listed in Table~\ref{systematicsec}. 

There are also theoretical uncertainties in the radiative decay calculation.
Initial state QED radiation is not included in the calculation of
Ref.~\cite{rad_decays_th}. However, initial state radiation does not affect the
kinematic distributions used to fit the mass in the final state.
We studied the effect of QED radiation off the initial state quarks on the
 parton luminosity by 
computing the parton luminosity including and excluding QED radiative effects 
on the quark momentum distribution. The change in the parton luminosity slope 
parameter was less then half of the quoted uncertainty on the parameter, which
was dominated by acceptance effects.

The calculation of Ref.~\cite{rad_decays_th} 
 includes only processes in which a single photon
is radiated. We use the code provided by the authors of
Ref.~\cite{baur_twophoton} to estimate the shift
introduced in the measured \wb\ and \zb\ masses
by neglecting two-photon emission.
 The estimated shifts in the \wb\ and \zb\ fitted masses due to
 two-photon radiation are shown in Table~\ref{systematicsec}.
Since this effect is an order of
magnitude smaller than the statistical uncertainty in our measurement we do not
correct for it, but add it in quadrature to the uncertainty due to radiative
corrections.
 The uncertainty in the radiative correction
  is taken to be fully correlated between the CC
 and EC \wb\ mass analyses. 

\subsubsection{\wb\ Boson Width}

The uncertainty on the fitted \wb\ mass corresponds to the uncertainty in the
measured value of the \wb\ boson width
 \wwidth\ = 2.062\PM0.059~GeV~\cite{Wwidth}.
 We take this uncertainty to be fully correlated between the CC
 and EC \wb\ mass analyses. 

Our recent measurement of the \wb\ width~\cite{newwwidth} considerably 
improves the precision of \wwidth\ and would reduce the \wb\ mass 
uncertainty from this source. However, since this is already a small source
 of uncertainty, the impact on the total \wb\ mass uncertainty is small. 

\subsection { Detector Model Parameters }

The uncertainties on the parameters of the detector model determined in
Secs.~\ref{sec-elec}--\ref{sec-recoil} translate into uncertainties in the
\wb\ mass measurement.
We study the sensitivity of the \wb\ mass measurement to the values of the
parameters by fitting the data with spectra generated by the fast Monte Carlo
with  input parameters modified by $\pm$1 standard deviation.

Table~\ref{systematicsec} lists the variation in the measured EC \wb\
mass  due to variation in the individual parameters. 
 For each item the uncertainty is determined with a typical Monte Carlo
 statistical error of 5~MeV.
  To achieve this precision,
10--20 million \wev\ decays are simulated for each item.

The residual calorimeter nonlinearity is parametrized by the offset
\deltaecem.
The electron momentum resolution is parametrized by \cem. The electron angle
calibration includes the effects of the parameters \betafdc\ and
\betaec, discussed in Section~\ref{sec-elec}. The recoil response is
 parameterized by \alpharec\
and \betarec. The recoil resolution is parameterized by \srec\ and \alphamb.
Electron removal refers to the bias \dupar\ introduced in the
\upar\ measurement by the removal of the cells occupied by the electron.
Selection bias refers to the \upar\ efficiency.

\subsection { Backgrounds }

We determine the sensitivity of the fit results to the assumed background
normalizations and shapes by repeating the fits to the data with 
background shapes and normalizations modified by $\pm$1 standard deviation.
  Table~\ref{systematicsec} lists the
uncertainties introduced in the 
EC $W$ boson mass measurement.

\section{Combined EC and CC \boldmath \wb\ \unboldmath 
 Boson Mass Error Analysis}
\label{erroranalysis}

 The measurement of the \wb\ mass requires the knowledge of many 
 parameters in our model of the \wb\ production, decay and detector 
 response. These parameters are constrained by measurements, and in some cases
 by theoretical input. The \wb\ mass error analysis involves the propagation
 of the measurement or theoretical uncertainties to the error matrix
 on the parameters, which is then propagated further to the error matrix
 on the CC and EC \wb\ mass measurements. The error matrix allows us to
 combine the fitted \wb\ mass values using the different data samples and
 techniques into a single value with a combined error.
 
 We identify the following parameters of relevance to the \wb\ mass
 measurements in the EC and CC:
\begin{itemize}
\item \wb\
  mass statistical errors $\delta \omega_{\rm CC}$ and $\delta \omega_{\rm EC}$
\item EM scales $\alphaccem$ and $\alphaecem$
\item EM offset parameters $\deltaccem$ and $\deltaecem$
\item FDC scale $\betafdc$ and FDC-EC relative scale $\betaec$
\item CDC scale $\betacdc$
\item EM resolutions (constant terms)  \cccem\ and \cecem\
\item recoil response $\vec{a}_{\rm rec}$
 representing jointly the response parameters
 $\alpharec$ and $\betarec$
\item recoil resolution $\vec{q}_{\rm rec}$ representing jointly 
 the hadronic sampling term \srec\ and the effects of the underlying event
 \alphamb\  
\item backgrounds $b_{\rm CC}$ and $b_{\rm EC}$
\item $u_{||}$ corrections $u_{\rm CC}$ and $u_{\rm EC}$ 
\item $u_{||}$ efficiencies $\varepsilon_{\rm CC}$ and $\varepsilon_{\rm EC}$ 
\item radiative corrections as a function of the photon coalescing radius $R_0$
\item parton luminosity $\beta$
\item theoretical modeling $\vec{t}$
\end{itemize}
 We take the EM scales, EM offsets, angular scales, $u_{||}$ corrections, 
 parton luminosity  and the radiative correction
  to be a set of 
  parameters that jointly determine the measured $W$ and $Z$ masses. We also 
 take the EM resolution parameters as a correlated set.
  We take the CC and EC backgrounds and $u_{||}$ efficiencies
 to be uncorrelated.
  The recoil
 modelling and the theoretical modelling (including pdf's, $p_T(W)$ spectrum,
 parton luminosity, radiative corrections
 and $W$ width) are treated as being
  common between the CC and the EC analyses. For all correlated parameters the
 sign of the $W$ mass correlation is determined by the relative sign of the
 mass shifts. 
\par
 The following measurements provide information on the values
 of these parameters
\begin{itemize}
\item The $Z$ mass measurements $M_Z^{\rm CC/CC}$, $M_Z^{\rm CC/EC}$ 
 and $M_Z^{\rm EC/EC}$
\item FDC radial calibration $\theta_{\rm FDC}$ and FDC-EC relative radial
 calibration $\theta_{\rm EC}$
\item CDC $z$ calibration $\theta_{\rm CDC}$ 
\item CC and EC EM offset measurements $o _{\rm CC}$ and $o _{\rm EC}$
\item Gaussian width fitted to $Z$ boson peak $\sigma_Z^{\rm CC/CC}$, 
$\sigma_Z^{\rm CC/EC}$
 and $\sigma_Z^{\rm EC/EC}$ 
\item $p_T$ balance in $Z$ events
\item width of $p_T$ balance in $Z$ events 
\item measurements of \upar\ correction and \upar\ efficiency
\item constraints on theoretical model (boson \pt\ from D\O\ data, \wb\
 width from world data including D\O\ data, and pdf's and parton luminosity
 from world data)  
\end{itemize}
\par
 We express the variations on the various calibration quantities (such as \zb\
 mass, EM offset, and angular scales, collectively referred to as $\vec{C}$)
 and the \zb\ width
 measurements as a linear combination
 of the variations on the parameters
\begin{eqnarray}
\delta \vec{C} & = & \Delta_C \; \delta \vec{p}  \nonumber \\
\delta \vec{ \sigma}_Z& = & \Delta_{\sigma} \; \delta \vec{ c}_{\rm EM}
\end{eqnarray}
 where
\begin{eqnarray}
 \delta \vec{C} & = & (\delta M_Z^{\rm CC/CC},\delta M_Z^{\rm CC/EC},
 \delta M_Z^{\rm EC/EC}, \delta \theta_{\rm FDC}, \delta \theta_{\rm EC}, \nonumber \\
 & & \delta \theta
_{\rm CDC}, \delta o_{\rm CC}, \delta o_{\rm EC}, \delta R_0, 
 \delta u_{\rm CC},
 \delta u_{\rm EC}, \delta \beta), \nonumber \\
 \delta \vec{p} & = & (\delta \alphaccem,\delta \alphaecem,
 \delta \betafdc, \delta \betaec, \delta \betacdc, \nonumber \\
 & & \delta \deltaccem,
 \delta \deltaecem, \delta R_0, \delta u_{\rm CC},
 \delta u_{\rm EC}, \delta \beta)
\end{eqnarray}
 and
\begin{eqnarray}
 \delta \vec{ \sigma}_Z & = &  (\delta \sigma_Z^{\rm CC/CC}, 
\delta \sigma_Z^{\rm CC/EC}, \delta \sigma_Z^{\rm EC/EC}), \nonumber \\
 \delta \vec{ c}_{\rm EM} & = & (\delta \cccem, \delta \cecem ). 
\end{eqnarray}
 The $\Delta$ matrices contain the
  partial derivatives of the observables with respect to the 
 parameters. 
\par
 Similarly, the variations on the $W$ mass are related linearly to the
 parameter variations
\begin{eqnarray}
\delta \vec{ M}_W & = & \Delta_W \; \delta \vec{ p} \nonumber \\
      & + & \Delta_{\sigma_W} \; \delta \vec{ c}_{\rm EM} \nonumber \\
      & + & \Delta_{\rm recoil \; scale} \; \delta 
\vec{ a}_{\rm rec} \nonumber \\
      & + & \Delta_{\rm recoil \; resolution} \; \delta \vec{ q}_{\rm rec} \nonumber \\
      & + & \Delta_{\rm background} \; \delta \vec{ b} \nonumber \\
      & + & \Delta_{\rm u} \; \delta \vec{ u} \nonumber \\
      & + & \Delta_{\varepsilon} \; \delta \vec{ \varepsilon} \nonumber \\
      & + & \Delta_{\rm theory} \; \delta \vec{ t} \nonumber \\        
      & + & \delta \vec{ \omega} 
\label{mwerr}
\end{eqnarray}
where $\delta \vec{ M}_W = (\delta M_W^{\rm CC}, \delta M_W^{\rm EC})$. 
\par 
 Knowing the components of
 $\delta \vec{ C}$ and  $\delta \vec{ \sigma}_Z $,
  we compute the covariance matrix for  the parameters in   $\vec{p} $ and
  $\vec{c}_{\rm EM} $. Since there are more measurements than parameters, we
 use the generalized least squares fitting procedure for this purpose.
 We then propagate the parameter covariance matrices into the covariance
 matrix for the CC and EC \wb\ mass measurements using equation~\ref{mwerr},
 by identifying the covariance matrix
  with the expected value of $\delta \vec{ M}_W 
 (\delta \vec{ M}_W)^T$, where $T$ indicates the transpose. The various
 contributions to $\delta  \vec{ M}_W$ are independent, hence they
 contribute additively to the total covariance matrix. 

 The CC \wb\ mass measurements \cite{wmass1bcc} were obtained using the 
 MRS(A$^ \prime$) parton distribution functions. We adjust these measurements
 by the estimated shifts (see Table~\ref{systematicscc}) when using the MRST
 parton distribution functions. Thus we use the following \wb\ mass values
 extracted from the CC data to combine with our EC measurements:
\begin{eqnarray}
M_W^{\rm CC} & = & 80.443 \; {\rm GeV} \; (\mt\   {\rm fit}) \nonumber \\
M_W^{\rm CC} & = & 80.459 \; {\rm GeV} \; (\pte\  {\rm fit}) \nonumber \\
M_W^{\rm CC} & = & 80.401 \; {\rm GeV} \; (\ptnu\ {\rm fit}) 
\end{eqnarray}

The combined $W$ mass $M_W$ for a set of $n$ $W$ mass measurements $m_i$
 and their covariance matrix $V$ is given by
\begin{equation}
   M_W = (\sum_{i,j=1}^{n}{H_{ij}}\,\,m_j)
  \, / \,(\,\sum_{i,j=1}^{n}{H_{ij}}\,),
\end{equation}
 where $H \equiv V^{-1}$ and $i,j$ run over the $W$ mass measurements being
 combined. 
The combined error is given by
\begin{equation}
 \sigma(M_W) = (\,\sum_{i,j=1}^{n}{H_{ij}}\,)^{-1/2},
\end{equation}
and the $\chi^2$ for the combination is given by
\begin{equation}
  \chi^2 = \sum_{i,j=1}^{n}{(m_i - M_W) \,\, H_{ij} \,\, (m_j - M_W )}.
\end{equation}
\section{ Results }
\label{sec-results}

We use the covariance matrix described above to obtain the total uncertainty
 on the EC \wb\ mass measurements and to combine 
 our CC and EC measurements. 
 We obtain the following
 results for the transverse mass fit
\begin{eqnarray}
\mw^{\rm EC} & = & 80.757 \pm 0.107 ({\rm stat}) \pm 0.204 ({\rm syst})\ \hbox{GeV} \nonumber \\
    & = & 80.757 \; \pm \; 0.230 \; {\rm GeV}  
\end{eqnarray} 
and
\begin{eqnarray}
M_W & = & 80.504 \; \pm \; 0.097 \; {\rm GeV} \; ({\rm CC \; {\rm and}
 \; EC \; combined}). 
\end{eqnarray} 
The $\chi ^2$ for the CC+EC \mt\ combination is
 1.5 for one degree of freedom, with a probability of 23\%.

 Similarly, for the $p_T(e)$ fit we obtain
\begin{eqnarray}
\mw^{\rm EC} & = & 80.547 \pm 0.128 ({\rm stat}) \pm 0.203 ({\rm syst})\ \hbox{GeV} \nonumber \\
    & = & 80.547 \; \pm \; 0.240 \; {\rm GeV}
\end{eqnarray}
and
\begin{eqnarray}
M_W & = & 80.480 \; \pm \; 0.126 \; {\rm GeV} \; ({\rm CC \; {\rm and} \; EC \; combined}). 
\end{eqnarray}
The $\chi^2$ for the CC+EC \pte\ 
 combination is 0.1 with a probability of 74\%. 

For the $p_T(\nu)$ fit we obtain
\begin{eqnarray}
\mw^{\rm EC} & = & 80.740 \pm 0.159 ({\rm stat}) \pm 0.310 ({\rm syst})\ \hbox{GeV} \nonumber \\
 & = & 80.740 \; \pm \; 0.348 \; {\rm GeV}
\end{eqnarray}
and
\begin{eqnarray}
\mw & = & 80.436 \; \pm \; 0.171 \; {\rm GeV} \; ({\rm CC \; {\rm and} \; EC \; combined}). 
\end{eqnarray}
The $\chi^2$ for the CC+EC \ptnu\ 
 combination is 1.0 with a probability of 32\%. 

The combination of the \mt, \pte\ and \ptnu\ fit values for the EC give the 
combined EC \wb\ mass result
\begin{eqnarray}
M_W  =  80.691 \; \pm \; 0.227 \; {\rm GeV}.
\end{eqnarray}
The $\chi^2$/dof is 4.0/2, with a probability of 14\%.

We combine all six measurements (CC and EC fits with the three techniques)
 to  obtain the combined 1994--1995 measurement
\begin{eqnarray}
M_W & = & 80.498 \; \pm \; 0.095 \; {\rm GeV}.
\end{eqnarray}
The $\chi^2$/dof is 5.1/5, with a probability of 41\%. 
The consistency of the six results indicates that we understand the
ingredients of our model and their uncertainties. 
 Including  the 
  measurement from the 1992--1993 data gives the 1992--1995 data measurement:
\begin{eqnarray}
M_W & = & 80.482 \; \pm \; 0.091 \; {\rm GeV}.  
\label{eq:combined}
\end{eqnarray}

Table~\ref{tab:sum} lists the \Dzero\ \wb\ mass measurement uncertainties
 from the 1994--1995 end calorimeter data alone and the combined 1994--1995
 central and end calorimeter data. 

\begin{table}[hbtp]
\caption{\wb\ mass uncertainties (in MeV) 
 in the  EC measurement  and the combined CC+EC measurement from the 1994--1995
 data.}
\medskip
\begin{center}
\begin{tabular}{ccc} 
\hline
Source                      & EC  & CC+EC \\
\hline
\wb\ statistics             & 108 & 61  \\
\hline
\zb\ statistics             & 181 & 59  \\
\hline
calorimeter linearity       &  52 & 25  \\
\hline
calorimeter uniformity      & --  &  8  \\
\hline
electron resolution         & 42  & 19  \\
\hline
electron angle calibration  & 20  & 10 \\
\hline
recoil response             & 17  & 25  \\
\hline
recoil resolution           & 42  & 25  \\
\hline
electron removal            & 4   & 12  \\
\hline
selection bias              & 5   &  3  \\
\hline
backgrounds                 & 20  &  9  \\
\hline
pdf                         & 17  &  7  \\
\hline
parton luminosity           & 2   &  4  \\
\hline
$p_T(W)$                    & 25  & 15  \\
\hline
$\Gamma (W)$                & 10  & 10  \\
\hline
radiative corrections       & 1   & 12  \\
\hline
\end{tabular}
\end{center}
\label{tab:sum}
\end{table}

\begin{figure}[htpb!]
\epsfxsize=3.0in
\centerline{\epsfbox{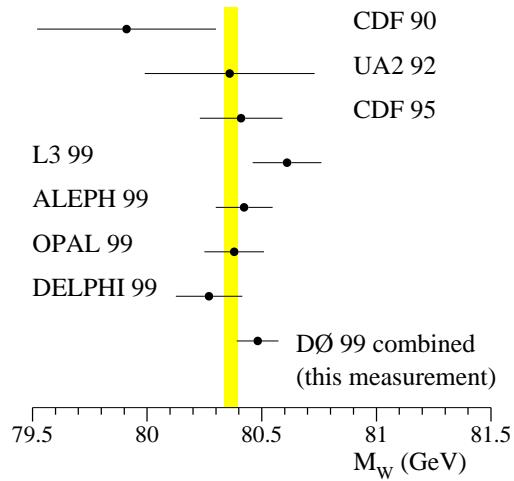}}
\vspace{0.1in}
\caption{A comparison of this measurement with previously published \wb\ boson
 mass
measurements (Table~\ref{tab:mw}). The shaded region indicates the predicted
\wb\ boson 
 mass value from global fits to all electroweak data except the \wb\ mass 
 measurements \protect\cite{mz}.}
\label{fig:mw_world}
\end{figure}

The \Dzero\ measurement is in good agreement with other measurements and is
more precise than  previously published results.
Table~\ref{tab:mw} lists previously published measurements with uncertainties
below 500~MeV, except previous D\O\ measurements which are subsumed into this
 measurement. A global fit to all electroweak measurements excluding
 the direct \wb\ mass measurements
  predicts $\mw = 80.367\pm0.029$~GeV \cite{mz}.
Figure~\ref{fig:mw_world} gives a graphical representation of these data.

\begin{table}[ht]
\begin{center}
\caption{\small Previously published measurements of the \wb\ boson mass. }
\medskip
\begin{tabular}{llc}
measurement & \mw\ (GeV)    & reference \\ \hline
CDF 90      & 79.910\PM0.390  & \protect\cite{CDF90} \\
UA2 92      & 80.360\PM0.370  & \protect\cite{UA2} \\
CDF 95      & 80.410\PM0.180  & \protect\cite{CDF} \\
L3 99       & 80.610\PM0.150  & \protect\cite{L3} \\
ALEPH 99    & 80.423\PM0.124  & \protect\cite{ALEPH}\\
OPAL 99     & 80.380\PM0.130  & \protect\cite{OPAL} \\
DELPHI 99   & 80.270\PM0.145  & \protect\cite{DELPHI} \\
\Dzero\ 99 combined (this result)  & 80.482\PM0.091  &  \\
\end{tabular}
\label{tab:mw}
\end{center}
\end{table}

\begin{figure}[htpb!]
\epsfxsize=3.0in
\centerline{\epsfbox{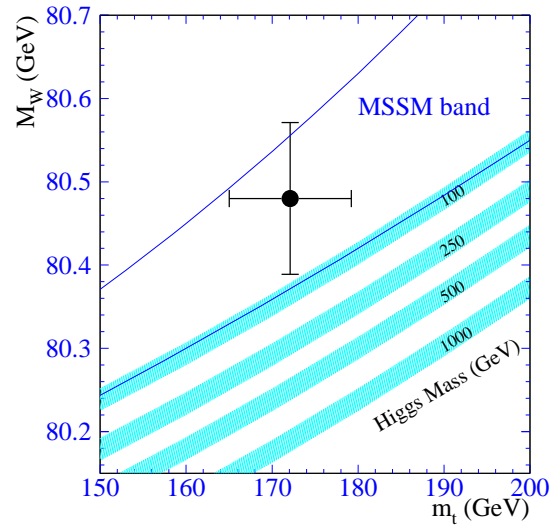}}
\vspace{0.1in}
\caption{A comparison of the \wb\ boson
 and top quark mass measurements by the
\Dzero\ collaboration with the standard model predictions for different Higgs
boson masses~\protect\cite{mw_v_mt}. The width of the bands for each Higgs
  boson
mass value indicates the uncertainty due to the error in $\alpha(\mz^2)$. Also
shown is the range allowed by the MSSM~\protect\cite{susy}.}
\label{fig:mw_mt}
\end{figure}

We evaluate the radiative corrections $\Delta r_{EW}$,
  defined in Eq.~\ref{eq:mw1}.
Our measurement of \mw\ from Eq.~\ref{eq:combined} leads to
\begin{equation}
\Delta r_{EW} = -0.0322\pm0.0059,
\end{equation}
5.5 standard deviations from the tree level value, demonstrating the 
 need for higher-order electroweak loop corrections.
  In Fig.~\ref{fig:mw_mt} we
compare the measured \wb\ boson and top quark masses~\cite{mtop_lj} from
 \Dzero\ with 
the values predicted by the standard model for a range of Higgs mass values
\cite{mw_v_mt}. Also shown is the prediction from the calculation in
Ref.~\cite{susy} for a model involving supersymmetric particles
assuming the chargino, Higgs, and left-handed selectron masses are
greater than 90~GeV.
The measured values are in agreement with the prediction
of the standard model, and in even better agreement with a supersymmetric
 extension of the standard model.

\section*{ Acknowledgements }
%
We thank the Fermilab and collaborating institution staffs for 
contributions to this work, and acknowledge support from the 
Department of Energy and National Science Foundation (USA),  
Commissariat  \` a L'Energie Atomique (France), 
Ministry for Science and Technology and Ministry for Atomic 
   Energy (Russia),
CAPES and CNPq (Brazil),
Departments of Atomic Energy and Science and Education (India),
Colciencias (Colombia),
CONACyT (Mexico),
Ministry of Education and KOSEF (Korea),
and CONICET and UBACyT (Argentina).


\begin{thebibliography}{99}
\bibitem{ecmwprl}
    B.~Abbott \etal\ (\dzero), 
    FERMILAB-Pub-99/259-E, hep-ex/9909030, to be published in \PRL\
\bibitem{d0nim}
    S.~Abachi \etal\ (\dzero), \NIM\ A {\bf338}, 185 (1994).
\bibitem{d0ec}
    H.~Aihara \etal, \NIM\ A {\bf325}, 393 (1993).
\bibitem{wmass1bcc}
    B.~Abbott \etal\ (\dzero), 
    \rm Phys. Rev. D {\bf 58}, 092003 (1998);
    B.~Abbott \etal\ (\dzero), \PRL\ {\bf 80}, 3008 (1998).
\bibitem{wmass1acc} 
    S.~Abachi \etal\ (\dzero), \PRL\ {\bf 77}, 3309 (1996); 
    B.~Abbott \etal\ (\dzero), \PR\ D {\bf 58}, 012002 (1998).
\bibitem{UA1_W_discovery} 
    G.~Arnison \etal\ (\uaone), \PL\ B {\bf122}, 103 (1983).
\bibitem{UA2_W_discovery} 
    M.~Banner \etal\ (\uatwo), \PL\ B {\bf122}, 476 (1983).
\bibitem{UA1_Z_discovery} 
    G.~Arnison \etal\ (\uaone), \PL\ B {\bf126}, 398 (1983); 
\bibitem{UA2_Z_discovery} 
    P.~Bagnaia \etal\ (\uatwo), \PL\ B {\bf129}, 130 (1983).
\bibitem{SM}
    S.L.~Glashow, \NP\ {\bf22}, 579 (1961);
    S.~Weinberg, \PRL\ {\bf19}, 1264 (1967);
    A.~Salam, in {\it Proceedings of the 8$^{th}$ Nobel Symposium}, 
    ed. N.~Svartholm (Almqvist and Wiksells, Stockholm, 1968), p. 367.
\bibitem{mz}
    The LEP collaborations, the LEP Electroweak Working Group,
    and the SLD Heavy Flavour and Electroweak Groups, 
    CERN-EP/99-15 (unpublished) and references therein.
\bibitem{UA2}
    J.~Alitti \etal\ (\uatwo), \PL\ B {\bf276}, 354 (1992). The value quoted in
    Table \protect\ref{tab:mw} uses \mz\ from Eq.~\protect\ref{eq:mz}.
\bibitem{CDF}
    F.~Abe \etal\ (\cdf), \PRL\ {\bf 75}, 11 (1995) and 
    \PR\ D {\bf52}, 4784 (1995).
\bibitem {L3}
    M.~Acciarri \etal\ (L3 Collaboration), \PL\ B {\bf 454}, 386 (1999).  
\bibitem {ALEPH}
    R.~Barate \etal\ (\alephc), \PL\ B {\bf 453} 121 (1999).  
\bibitem {OPAL}
    K.~Ackerstaff \etal\ (\opalc), \PL\ B {\bf 453} 138 (1999).
\bibitem {DELPHI}
    P.~Abreu \etal\ (DELPHI Collaboration),  CERN-EP-99-079, submitted to 
 Phys. Lett. B.
\bibitem {onshell}
    A.~Sirlin, \PR\ D {\bf22}, 971 (1980);
    W.~Marciano and A.~Sirlin, \PR\ D {\bf22}, 2695 (1980) 
    and erratum-{\it ibid.} {\bf31}, 213 (1985).
\bibitem{PDG}
    C.~Caso \etal, {\it Review of Particle Physics}, 
    Eur. Phys. J. {\bf C3} (1998). 
\bibitem{mtop_lj}
     \dzero, B.~Abbott \etal, \PRL\ {\bf80}, 2063 (1998);
     \dzero, B.~Abbott \etal, \PR\ D {\bf60}, 052001 (1999) and references
    therein.
\bibitem{cdf_mt}
    F.~Abe \etal\ (\cdf), \PRL\ {\bf 82}, 271 (1999), erratum-{\it ibid.} 
    {\bf 82},  2808 (1999) and references therein. 
\bibitem{susy} 
    P.~Chankowski \etal, \NP\  {\bf B417}, 101 (1994);
    D.~Garcia and J.~Sola, Mod. \PL\ A {\bf9}, 211 (1994); 
    A. Dabelstein, W. Hollik and W. Mosle, in {\it Perspectives for Electroweak
    Interactions in $e^+e^-$ Collisions}, ed. by B.~A.~Kniehl (World 
    Scientific,
    Singapore, 1995) p.~345;
    D. Pierce \etal, Nucl. Phys. {\bf B491}, 3 (1997).
\bibitem{Tevatron}
    H.T.~Edwards, in {\it Annual Review of Nuclear and Particle Science},
    Vol. 35, ed. J.D.~Jackson \etal, 
        (Annual Reviews, Palo Alto, 1985) p.~605.
\bibitem{calmon}
    R.D.~Schamberger, in {\em 
    Proceedings of the Fifth International Conference on 
    Calorimetry in
    High Energy Physics}, Upton, New York, 1994, edited by Howard Gordon and 
    Doris Rueger (World Scientific, Singapore, 1994);
    J. Kotcher, in {\em 
    Proceedings of the 1994 Beijing Calorimetry Symposium}, 
    Beijing, China, 1994, p. 144;
    J. A. Guida, in {\em Proceedings of the 4th International Conference on 
    Advanced Technology
    and Particle Physics}, 
    Como, Italy, 1994, Nucl. Phys. Suppl. Vol. {\bf B44},  158 (1995). 
\bibitem{testbeam}
    S.~Abachi \etal, \NIM\ A {\bf324}, 53 (1993).
\bibitem{McKinley_thesis}
    J.W.T.~McKinley, Ph.D. thesis, 
    Michigan State University, 1996 (unpublished).
\bibitem{geant}
    R.~Brun, F.~Bruyant, M.~Maire, A.C.~McPherson and P.~Zanarini, 
     \GEAN\ 3,
    CERN DD/EE/84-1 (1987); 
    F.~Carminati \etal, \GEAN\ Users Guide, 
    CERN Program Library W5013, 1991 (unpublished).
\bibitem{top_prd}
    S.~Abachi \etal\ (\dzero), \PR\ D {\bf52}, 4877 (1995).
\bibitem{LY}
    G.A.~Ladinsky and C.P.~Yuan, \PR\ D {\bf50}, 4239 (1994).
\bibitem{AK}
    P.B.~Arnold and R.P.~Kauffman, \NP\  {\bf B349}, 381 (1991).
\bibitem{AR}
    P.B.~Arnold and M.H.~Reno, \NP\ {\bf B319}, 37 (1989) and 
    erratum-{\it ibid.} {\bf B330}, 284 (1990).
\bibitem{CSS}
    J.~Collins and D.~Soper, \NP\ {\bf B193}, 381 (1981) and 
    erratum-{\it ibid.} {\bf B213}, 545 (1983); 
    J.~Collins, D.~Soper, and G.~Sterman, \NP\ {\bf B250}, 199 (1985).
\bibitem{AEMG}
    G.~Altarelli, R.K.~Ellis, M.~Greco and G.~Martinelli, 
    \NP\ {\bf B246}, 12 (1984).
\bibitem{mrst} 
    A.D.~Martin, R.G.~Roberts, W.J.~Stirling, and R.S.~Thorne,
    Eur. Phys. J. {\bf C4}, 463 (1998).
\bibitem{Wwidth}
    S.~Abachi \etal\ (\dzero), \PRL\ {\bf 75}, 1456 (1995).
\bibitem{herwig}
    G.~Marchesini \etal, Comput. Phys. Commun. {\bf 67} 
    465 (1992), release 5{.}7.
\bibitem{PDFLIB}
    H.~Plotow-Besch, CERN-PPE W5051 (1997), release 7{.}02 (unpublished).
\bibitem{mirk}
    E.~Mirkes, \NP\ {\bf B387}, 3 (1992).
\bibitem{csframe}
    J.~Collins and D.~Soper, \PR\ D {\bf16}, 2219 (1977).
\bibitem{rad_decays_th}
    F.A.~Berends and R.~Kleiss, \ZP\ C {\bf27}, 365 (1985); 
    F.A.~Berends, R.~Kleiss, J.P.~Revol, J.P.~Vialle, 
    \ZP\ C {\bf27}, 155 (1985).
\bibitem{bantly}
    J.~Bantly, Ph.D. thesis, 
    Northwestern University, 1992 (unpublished).
\bibitem{jetscale}
    B.~Abbott \etal\ (\dzero), \NIM\ A {\bf424}, 352 (1999).
\bibitem{mrsa}
    A.D.~Martin, W.J.~Stirling, and R.G.~Roberts, 
    \PR\ D {\bf50}, 6734 (1994) and \PR\ D {\bf51}, 4756 (1995).
\bibitem{cteq3m}
    H.L. Lai \etal, \PR\ D {\bf51}, 4763 (1995).
\bibitem{cteq2m}
    J.~Botts \etal, \PL\ B {\bf304}, 159 (1993).
\bibitem{mrsd}
    A.D.~Martin, W.J.~Stirling, and R.G.~Roberts, \PL\ 
    B {\bf306}, 145 (1993) and erratum-{\it ibid.} B {\bf309}, 492 (1993).
\bibitem{mrsr2}
    A.D.~Martin, R.G.~Roberts and W.J.~Stirling, \PL\ 
    B {\bf387}, 419 (1996).
\bibitem{cteq5m}
    H.L.~Lai  \etal,  hep-ph/9903282 (unpublished). 
\bibitem{cteq4m}
    H.L.~Lai  \etal, \PR\ D {\bf 55}, 1280 (1997). 
\bibitem{zptprd}
    B.~Abbott \etal\ (\dzero), 
    FERMILAB-PUB-99-197-E, hep-ex/9907009, submitted to \PR\ D.
\bibitem{cdfasym}
    F.~Abe \etal\ (\cdf), \PRL\ {\bf 81}, 5754 (1998).
\bibitem{e866dy}
    E.A.~Hawker \etal\ (Fermilab E866/NuSea Collaboration),
    \PRL\ {\bf 80}, 3715 (1998);
    J.C.~Peng \etal\ (Fermilab E866/NuSea Collaboration),
    \PR\ D {\bf 58}, 092004 (1998). 
\bibitem{baur_twophoton}
    U.~Baur \etal, \PR\ D {\bf56}, 140 (1997); U.~Baur, S.~Keller, and 
    W.K.~Sakumoto, \PR\ D {\bf 57}, 199 (1998).
\bibitem{newwwidth}
    B.~Abbott \etal\ (\dzero), 
    FERMILAB-PUB-99-171-E, hep-ex/9906025, submitted to \PR\ D.
\bibitem{CDF90}
    F. Abe \etal\ (\cdf), \PRL\ {\bf 65}, 2243 (1990) and 
    \PR\ D {\bf43}, 2070 (1991).
\bibitem{mw_v_mt}
    G.~Degrassi \etal\,  \PL\ B {\bf418}, 209 (1998);
    G.~Degrassi, P.~Gambino, and A.~Sirlin, Phys. Lett. B {\bf394}, 188 (1997).
\end{thebibliography}
\end{document}